\documentclass[a4paper,fleqn,usenatbib]{mnras}

\usepackage{newtxtext,newtxmath}

\usepackage[T1]{fontenc}
\usepackage{ae,aecompl}

\usepackage{graphicx}	

\newcommand{\kms}{km\,s$^{-1}$}

\newcommand{\figps}[3]{\resizebox{#1}{!}{\rotatebox{#2}{\includegraphics{#3}}}}
\newcommand{\fivps}[3]{\resizebox{!}{#1}{\rotatebox{#2}{\includegraphics{#3}}}}

\title[Variability of HgMn stars]{TESS survey of rotational and pulsational variability of mercury-manganese stars}

\author[O. Kochukhov et al.]
{O.\ Kochukhov$^1$\thanks{E-mail: oleg.kochukhov@physics.uu.se},
V.\ Khalack$^2$,
O.\ Kobzar$^2$,
C.\ Neiner$^3$,
E.\ Paunzen$^4$,
J.\ Labadie-Bartz$^5$,
A.\ David-Uraz$^{6,7}$
\\
$^1$Department of Physics and Astronomy, Uppsala University, Box 516, Uppsala 75120, Sweden \\
$^2$D\'epartement de Physique et d'Astronomie, Universit\'e de Moncton, Moncton, NB E1A 3E9, Canada\\
$^3$LESIA, Paris Observatory, PSL University, CNRS, Sorbonne Universit\'e, Universit\'e de Paris, 5 place Jules Janssen, 92195 Meudon, France\\
$^4$Department of Theoretical Physics and Astrophysics, Masaryk University, Kotl\'a\v{r}sk\'a 2, 611 37 Brno, Czech Republic\\
$^5$Instituto de Astronomia, Geof\'isica e Ciencias Atmosf\'ericas, Universidade de S\`ao Paulo, Rua do Mat\~ao 1226,
Cidade Universit\~aria, \\S\'ao Paulo, SP 05508-900, Brazil\\
$^6$Department of Physics and Astronomy, Howard University, Washington, DC 20059, USA\\
$^7$Center for Research and Exploration in Space Science and Technology, and X-ray Astrophysics Laboratory, NASA/GSFC, Greenbelt, MD 20771, USA
}

\date{Accepted 2021 July 19. Received 2021 July 19; in original form 2021 June 13}

\pubyear{2020}

\begin{document}
\label{firstpage}
\pagerange{\pageref{firstpage}--\pageref{lastpage}}
\maketitle

\begin{abstract}
Mercury-manganese (HgMn) stars are late-B upper main sequence chemically peculiar stars distinguished by large overabundances of heavy elements, slow rotation, and frequent membership in close binary systems. These stars lack strong magnetic fields typical of magnetic Bp stars but occasionally exhibit non-uniform surface distributions of chemical elements. The physical origin and the extent of this spot formation phenomenon remains unknown. Here we use 2-min cadence light curves of 64 HgMn stars observed by the TESS satellite during the first two years of its operation to investigate the incidence of rotational modulation and pulsations among HgMn stars. We found rotational variability with amplitudes of 0.1--3 mmag in 84 per cent of the targets, indicating ubiquitous presence of starspots on HgMn-star surfaces. Rotational period measurements reveal six fast-rotating stars with periods below 1.2~d, including one ultra-fast rotator (HD\,14228) with a 0.34~d period. We also identify several HgMn stars showing multi-periodic g-mode pulsations, tidally induced variation and eclipses in binary systems.
\end{abstract}




\section{Introduction}
\label{sec:intro}

Mercury-manganese (HgMn) stars correspond to one of the sub-classes of the heterogeneous group of chemically peculiar (CP) stars, found on the upper main sequence. These objects are generally characterised by a significant deviation of their surface chemical composition from the solar abundance pattern. HgMn stars have spectral types from B6 to A0, corresponding to the effective temperature range from 16000 to 10000~K, and are distinguished by strong lines of ionised Hg and/or Mn, discernible in low-resolution classification spectra \citep*[e.g.][]{paunzen:2021}. These line strength anomalies indicate an overabundance of these chemical elements by up to 6 orders of magnitude relative to their solar abundances. High-resolution spectroscopic analyses of HgMn stars also reveal large overabundances of Xe, Ga, Pt, Au and many other heavy elements (e.g. \citealt{castelli:2004}; \citealt{yuce:2014}; \citealt*{ghazaryan:2018}) whereas He and CNO elements are often observed to be underabundant \citep{roby:1990}. HgMn stars are frequently found in spectroscopic binaries \citep*{gerbaldi:1985} and rotate slower than normal stars of the same temperature \citep*{abt:2002}. 

HgMn stars are traditionally assigned to the non-magnetic (CP3) sub-group of chemically peculiar stars according to the classification proposed by \citet{preston:1974}. CP3 stars are qualitatively different from the magnetic Bp stars (CP2 sub-group), which occupy the same region in the Hertzsprung-Russell diagram. Magnetic Bp stars also exhibit highly unusual surface element abundances, but possess kG-strength, globally organised, stable magnetic fields. The presence of these fields leads to a stationary non-uniform distribution of chemical elements with height in stellar atmosphere and laterally across the stellar surface (e.g. \citealt*{michaud:1981}; \citealt{babel:1991,leblanc:2009,alecian:2015}). This, in turn, results in a high-amplitude spectroscopic (e.g. \citealt{kochukhov:2004e,kochukhov:2014}; \citealt*{kochukhov:2019}) and photometric \citep{jagelka:2019} rotational modulation. In contrast to this spectacular and well-studied variability of CP2 objects, HgMn stars were considered to be some of the least variable early-type stars \citep{adelman:1998b}. Most attempts to detect magnetic fields on their surfaces yielded negative results \citep[e.g.][]{shorlin:2002,wade:2006,auriere:2010a,kochukhov:2011b,makaganiuk:2011,bagnulo:2012,bagnulo:2015,martin:2018}. Occasional reports of magnetic detections in HgMn stars \citep[e.g.][]{hubrig:2010,hubrig:2012,hubrig:2020} remain inconclusive since in every such case re-analyses of the same data \citep{makaganiuk:2011a,makaganiuk:2012,kochukhov:2013a} or independent observations \citep{folsom:2010} failed to confirm the presence of the field. The most sensitive magnetic field studies of HgMn stars have reached a longitudinal field precision of a few G for bright narrow-line HgMn stars, proving that they do not host global fields comparable in strength to those found in magnetic Bp stars. This, however, does not exclude the presence of ultra-weak sub-G fields of the type found in Vega, Sirius and several hot Am stars \citep{petit:2010,petit:2011,blazere:2016}.

Despite the absence of magnetic field, the notion of constancy of HgMn stars has been challenged by high-precision spectroscopic observations. It turned out that some HgMn stars show a low-amplitude rotational modulation of spectral line profiles. This variability is consistent with the presence of low-contrast spots of, typically, those chemical elements which show the largest overabundances \citep[e.g.][]{adelman:2002,kochukhov:2005b,hubrig:2006,briquet:2010,kochukhov:2011b,makaganiuk:2011,makaganiuk:2012,korhonen:2013,strassmeier:2017}. Remarkably, it was also discovered that the geometry of chemical spots on HgMn stars evolves with time \citep{kochukhov:2007b,korhonen:2013}. In comparison, higher contrast surface inhomogeneities on magnetic Bp stars show no temporal changes on time scales of at least several decades. These observations offer an intriguing possibility that the structure formation in HgMn stellar atmospheres is governed by a hitherto unknown variety of dynamic atomic diffusion process \citep*{alecian:2011}, which is distinct from the magnetically-controlled atomic diffusion operating in magnetic Bp stars \citep*{michaud:2015}.

The incidence of spots on HgMn stars remains largely undetermined. Rotational line profile variability was detected in a relatively small number of the brightest HgMn stars that could be observed with high signal-to-noise ratio ground-based spectroscopy. Rotational periods were securely measured from these data for only four objects. High-precision photometric observations from space opened another avenue for finding variable HgMn stars. Detection of variability with periods of a few days was reported for individual HgMn stars using CoRoT \citep{morel:2014,strassmeier:2017}, Kepler \citep{balona:2011a,balona:2015}, K2 \citep{white:2017,krticka:2020}, the BRITE nanosat constellation \citep{strassmeier:2020}, and TESS \citep{balona:2019a,gonzalez:2021}. In some of these cases, interpretation of photometric variability in terms of rotational modulation was supported by the detection of line profile variations in follow-up time-resolved spectroscopy. In other cases variability could not be unambiguously ascribed to spots \citep{hummerich:2018} and interpretation in terms of g-mode pulsations was preferred \citep{alecian:2009a}. Recent studies identified another cause of variability of some HgMn stars -- the ellipsoidal and heartbeat variation associated with the orbital motion in a close binary system \citep{kochukhov:2021a,paunzen:2021a}. To summarise, variability of HgMn appears to be ubiquitous and diverse, emphasising the necessity of advancing from case studies of individual objects to statistical analyses of meaningful stellar samples.

The full-sky survey carried out by the Transiting Exoplanet Survey Satellite \citep[TESS,][]{ricker:2015} enables the first systematic unbiased investigation of the photometric variability of HgMn stars. The goal of our study is to take advantage of this unique research opportunity by performing a comprehensive assessment of the information content of the TESS light curves of HgMn stars observed during the first two years of mission operation. Using these data we identify periodic signals, associate them with rotational modulation or stellar pulsations, and assess the incidence of these variability phenomena among HgMn stars. The paper is organised as follows: Sect.~\ref{sec:obs} introduces observational data and target selection, Sect.~\ref{sec:analys} provides details of our time series analysis, Sect.~\ref{sec:res} presents results of this analysis for stars with rotational modulation (Sect.~\ref{ssec:rot}), pulsations (Sect.~\ref{ssec:puls}) and photometric variation related to binarity (Sect.~\ref{ssec:bin}). The paper is concluded by Sect.~\ref{sec:disc}, which summarises and discusses our work.

\section{Observational data and target selection}
\label{sec:obs}

This study is based upon the observations collected by the TESS satellite during the first two years of its operation. In this period, lasting from 25 July 2018 to 4 July 2020, TESS has surveyed most of the sky in 26 pointings, commonly referred to as sectors. Sectors 1--13 covered the Southern ecliptic hemisphere whereas the remaining sectors (14--26) correspond to observations of targets North of the ecliptic equator. Observations within each sector were carried out continuously for 27.4~d, except for a small gap in the middle of each sector. Depending on the ecliptic latitude, targets were observed for a minimum of 27.4~d (one sector) and up to about one year (13 sectors). Here we used the 2-min cadence PDCSAP (pre-search data conditioning simple aperture photometry) light curves provided by the TESS science team and available for download from the Mikulski Archive for Space Telescopes (MAST)\footnote{\url{https://mast.stsci.edu}}. Details of data processing steps included in the pipeline that produced these light curves can be found in \citet{jenkins:2016}. The pixel scale of TESS is 21\arcsec\ per pixel and typical apertures used by the pipeline are 3--7 pixels across, resulting in signal integration over 1--2\arcmin. With such large effective apertures contamination by nearby sources may be of concern.

To find HgMn stars with 2-min TESS light curves, we first constructed a unified catalogue of all known stars with this chemical peculiarity type. We started by extracting all objects with Hg and/or Mn spectral peculiarity from the general \textit{Catalogue of Ap, HgMn and Am stars} \citep{renson:2009}. The resulting list was complemented with HgMn stars identified by \citet{chojnowski:2020} based on SDSS/APOGEE spectra and by \citet{paunzen:2021} using LAMOST DR4 data. A compilation of HgMn abundance analyses \citep{ghazaryan:2016,ghazaryan:2018} as well as a number of recent studies reporting discovery of individual HgMn stars (e.g. \citealt{catanzaro:2010,catanzaro:2020}; \citealt*{monier:2015}; \citealt{monier:2019}; \citealt*{sikora:2020}; \citealt{gonzalez:2021}) were also taken into account. These selection steps resulted in a catalogue containing 544 definite and probable HgMn stars brighter than $V=12$~mag. This list of targets was cross-matched with the revised TESS Input Catalogue \citep[TIC,][]{stassun:2019} in order to find TIC identification numbers and stellar magnitudes. The TIC numbers were then used to query MAST. For 71 stars from our list 2-min light curves were obtained during the first two years of the TESS mission. After further examination of the information and literature on these stars, TIC~99025917 (HD~193772) and 174194250 (HD~220885) were removed since these are well-known magnetic Ap stars with `Hg' or `Mn' listed in their spectral peculiarity field in \citet{renson:2009}. Likewise, TIC~118573876 (HD~22128) has `Mn' in its peculiarity type in this catalogue, but is known to be an SB2 system containing two cool Am stars \citep*{folsom:2013}. Furthermore, TIC~29715050 (HD~179709B), TIC~868508027 (HD~82984B) and TIC~358467049 (CPD~$-60$~944B) are fainter secondary components in close visual binaries unresolved by TESS. These objects were not considered since their light curves are likely dominated by variability of the non-HgMn primaries. 

\begin{figure}
\centering
\figps{\hsize}{0}{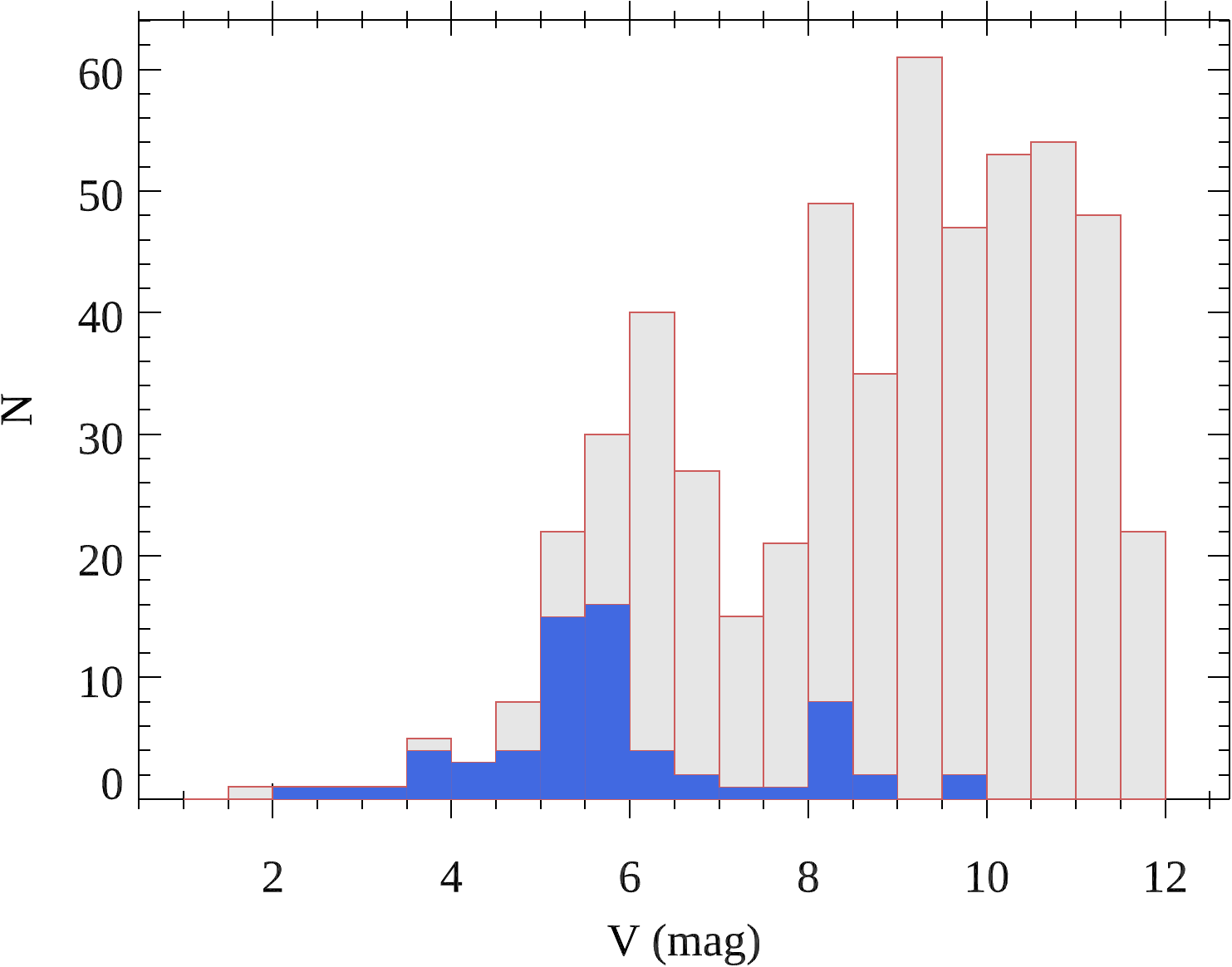}
\caption{Distribution of $V$ magnitudes of HgMn stars analysed in this study (dark histogram bars) relative to all known stars of this type (light histogram bars).}
\label{fig:targets}
\end{figure}

The final sample of targets analysed in this study comprises 65 objects listed in Table~\ref{tab:results}. Their $V$ magnitudes range between 2.1 and 9.7, with a median value of 5.6. The magnitude distribution of the studied stars relative to the initial HgMn-star sample is shown in Fig.~\ref{fig:targets}. The studied sample is reasonably representative of HgMn stars brighter than $V\approx6$ but is highly incomplete for fainter stars.

Considering a coarse pixel scale of TESS, we assessed contamination of all 65 targets by examining {\it Gaia} eDR3 \citep{gaia-collaboration:2021} sources contributing to the apertures used to derive the light curves for each studied star. Three targets -- TIC~121161014 (HD~133833), TIC~163024899 (HD~99803), TIC~307291308 (HD~71066) -- were found to have contaminating sources 2.1--3.5~mag fainter than the main target. Contamination is negligible for the remaining stars.

\section{Time series analysis}
\label{sec:analys}

Our assessment of the information content of TESS light curves of HgMn stars was carried out using standard time series analysis techniques: examination of the generalised Lomb Scargle (GLS) periodograms \citep{zechmeister:2009} and least-squares model fitting in the time domain. Prior to these analysis steps, cadences marked as low quality in the TESS data files were excluded. Then, GLS periodograms were computed for frequencies up to the Nyquist limit and examined visually. The frequency positions and heights of peaks corresponding to obvious periodic signals were read off the GLS periodogram and adopted as initial guesses for subsequent non-linear least-squares model fitting in the time domain. In this fitting, implemented with the help of the {\sc mpfit} package written in IDL \citep{markwardt:2009}, the frequencies, amplitudes and phases of all harmonic signals included in the model were adjusted simultaneously. Whenever periodogram showed evidence of higher-order harmonics (as is very common for a non-sinusoidal rotational modulation), the frequencies of these signals were fixed to multiples of the first harmonic and only amplitudes and phases were adjusted. Applying this constrained fitting to all harmonic components simultaneously provided a more precise determination of the fundamental frequency.

After initial satisfactory fit in the time domain was achieved, we subtracted the resulting model from the data and calculated GLS periodogram of the residuals in order to reveal further low-amplitude periodic signals. The significance of periodogram peaks was assessed using the canonical $S/N\ge4$ criterium \citep{breger:1993,kuschnig:1997}, with the noise floor determined as described by \citet{shultz:2021}. Any additional harmonic signals found in this way were added to the model and optimisation of the model parameters by fitting the light curve in the time domain was repeated with new frequencies taken into account. This process was continued until no further significant frequencies were evident in the periodogram of the residuals. Formal parameter errors derived by the least-squares fitting algorithm were adopted as uncertainty estimates for the amplitudes and frequencies of periodic signals included in the model. 

TESS light curves exhibit several types of instrumental artefacts, including slow drifts and occasional significant deviations at the beginning and/or end of 13.7~d continuous data segments. Strongly deviating light curve sections were manually removed prior to the GLS periodogram calculation and least-squares fitting. To account for the drifts, we approximated the background stellar brightness with low-order (typically 1--3) polynomials, fitting their coefficients simultaneously with harmonic signals. The background of each 13.7~d segment was treated independently from other segments. The high-order polynomials were typically applied in the presence of a clear short-period variation, otherwise the straight-line background was used to avoid removing variation with a period comparable to the segment's length. This piecewise polynomial background was removed before producing the final light curve and periodogram plots presented below.

A few of our targets discussed in more detail in Sect.~\ref{ssec:bin} exhibit abrupt changes of brightness due to transits or eclipses. Harmonic analysis is poorly suited for this type of variability. Since the main focus of our study is investigation of the rotational and pulsational variability of HgMn stars, we removed parts of light curves affected by transits and eclipses prior to the time series analysis described above. When multiple eclipses were observed, their times were found by computing the centroid time of each event and the period of an eclipsing binary was estimated by a linear fit to the times of individual events (separately for the primary and secondary eclipses).

\section{Results}
\label{sec:res}

\subsection{Rotational variability}
\label{ssec:rot}

Photometric rotational variability of HgMn stars is expected to occur due to low-contrast chemical abundance spots present on the surfaces of these stars \citep*[e.g.][]{prvak:2020}. Given the typical 27.4-d length of most TESS data sets and the methodology of the time series analysis adopted in our study, in particular the independent treatment of slow drifts within individual 13.7~d light curve segments, we are not sensitive to rotational modulation exceeding periods of $\approx$\,13~d. On the other end, frequencies higher than about 2~d$^{-1}$ are unlikely to be related to stellar rotation. Consequently, candidate rotational signals were searched in the 0.5--13~d period range. Rotational photometric modulation of CP stars is frequently non-sinusoidal \citep[e.g.][]{jagelka:2019}. Therefore, the presence of a low-frequency periodogram peak accompanied by harmonic frequencies was considered to be a tell-tale sign of rotational variability.

\begin{figure*}
\centering
%
\fivps{0.129\vsize}{0}{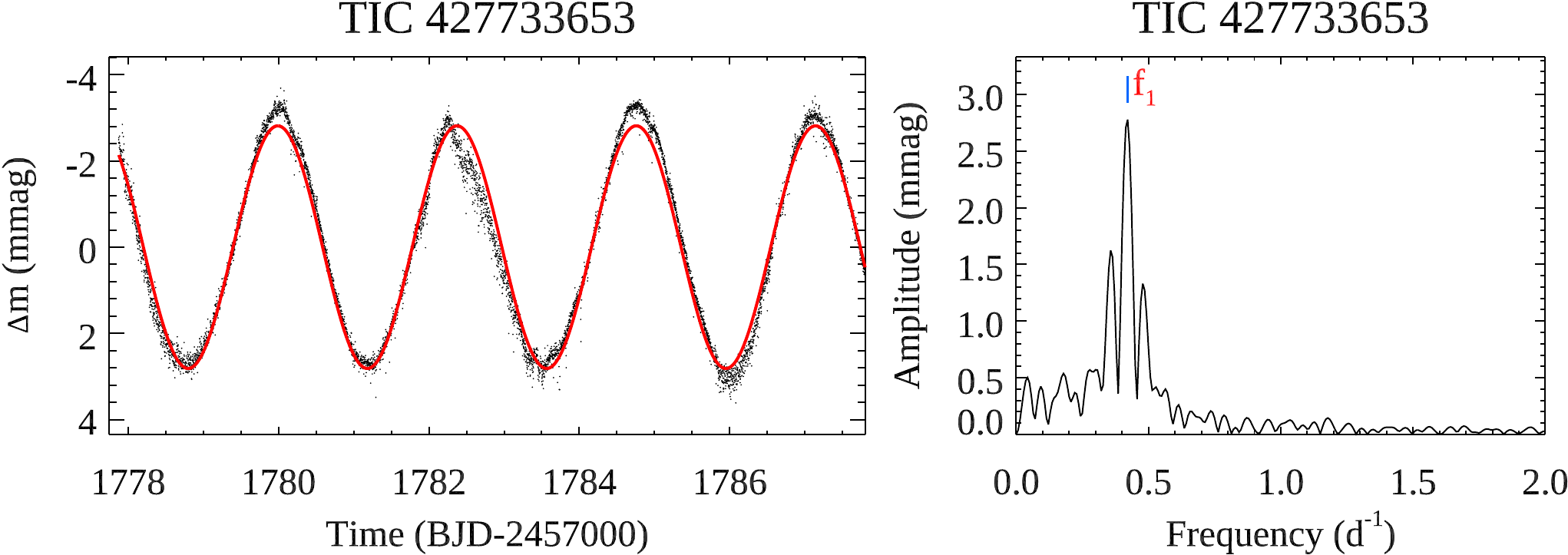}
\fivps{0.129\vsize}{0}{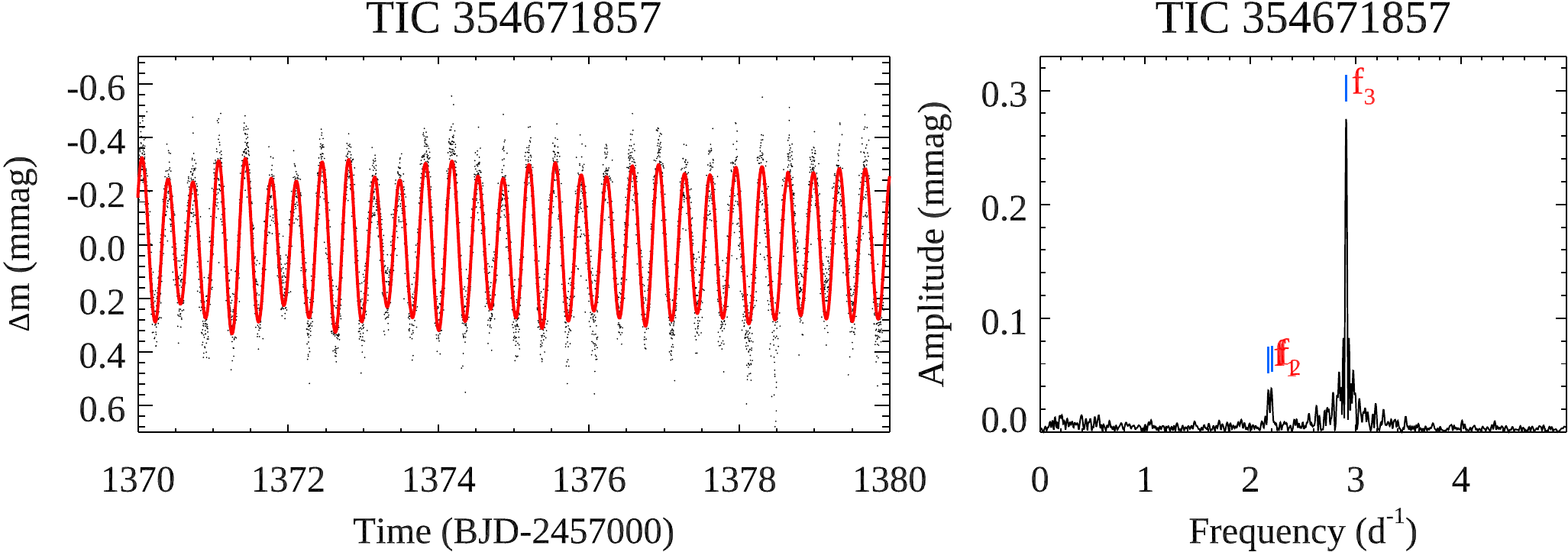}\vspace*{2mm}
\fivps{0.129\vsize}{0}{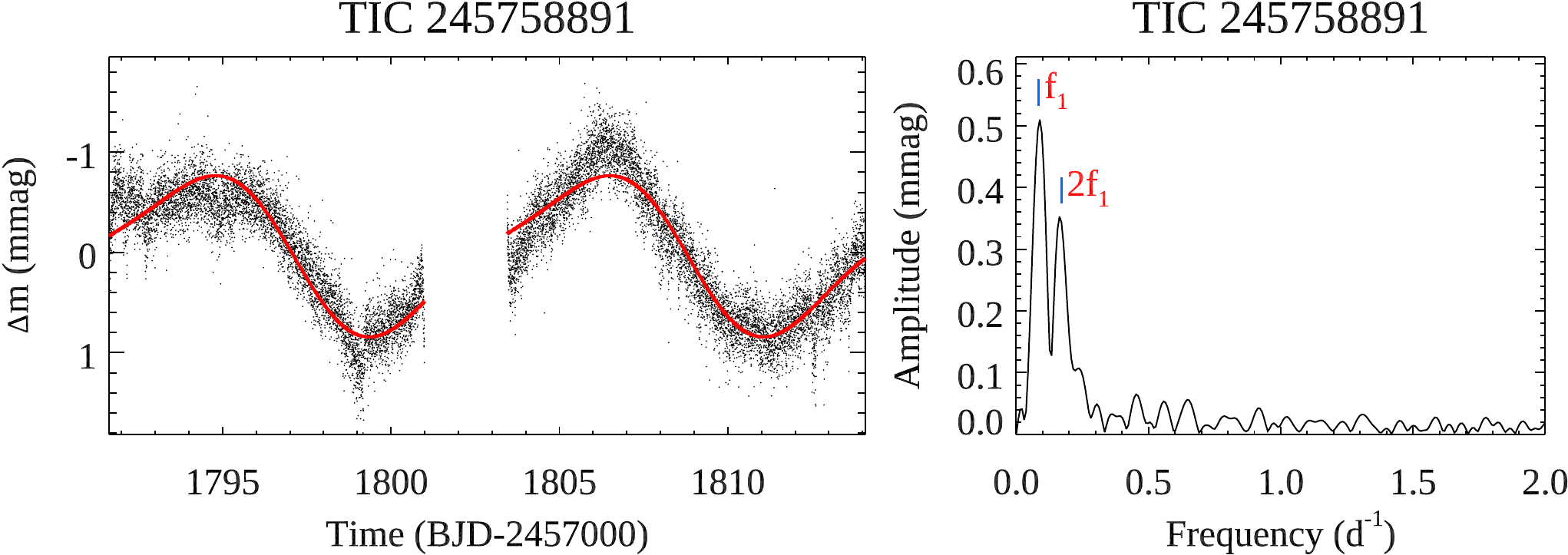}
\fivps{0.129\vsize}{0}{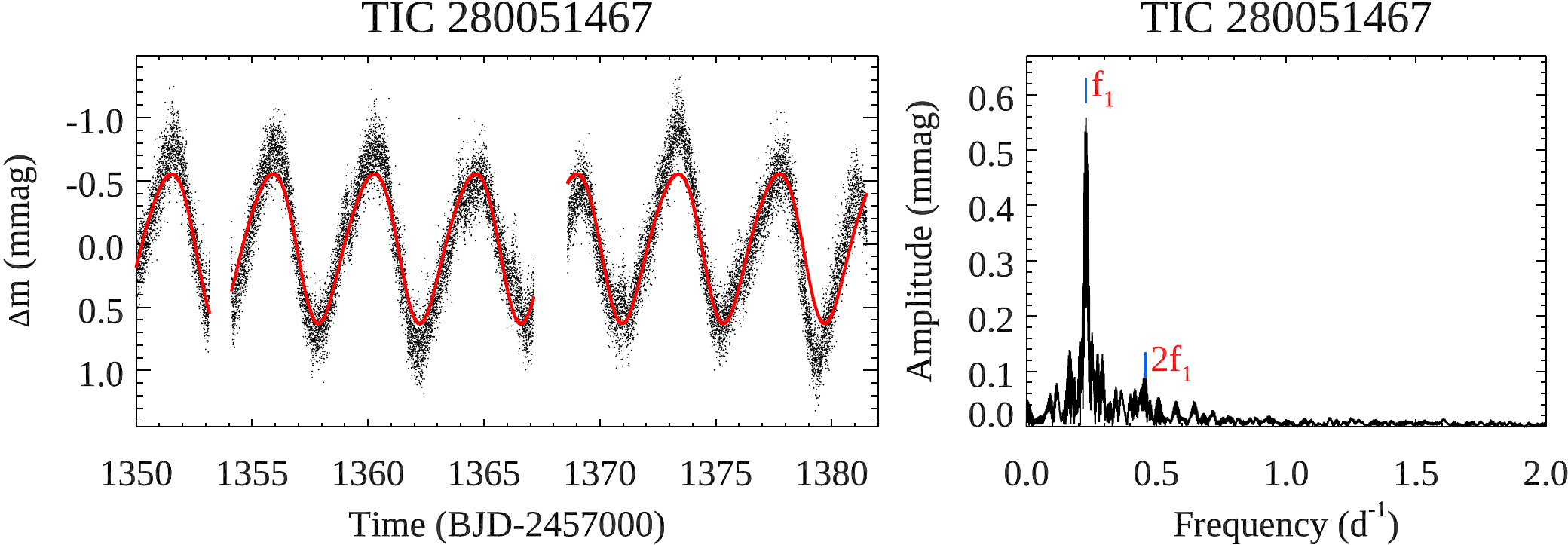}\vspace*{2mm}
\fivps{0.129\vsize}{0}{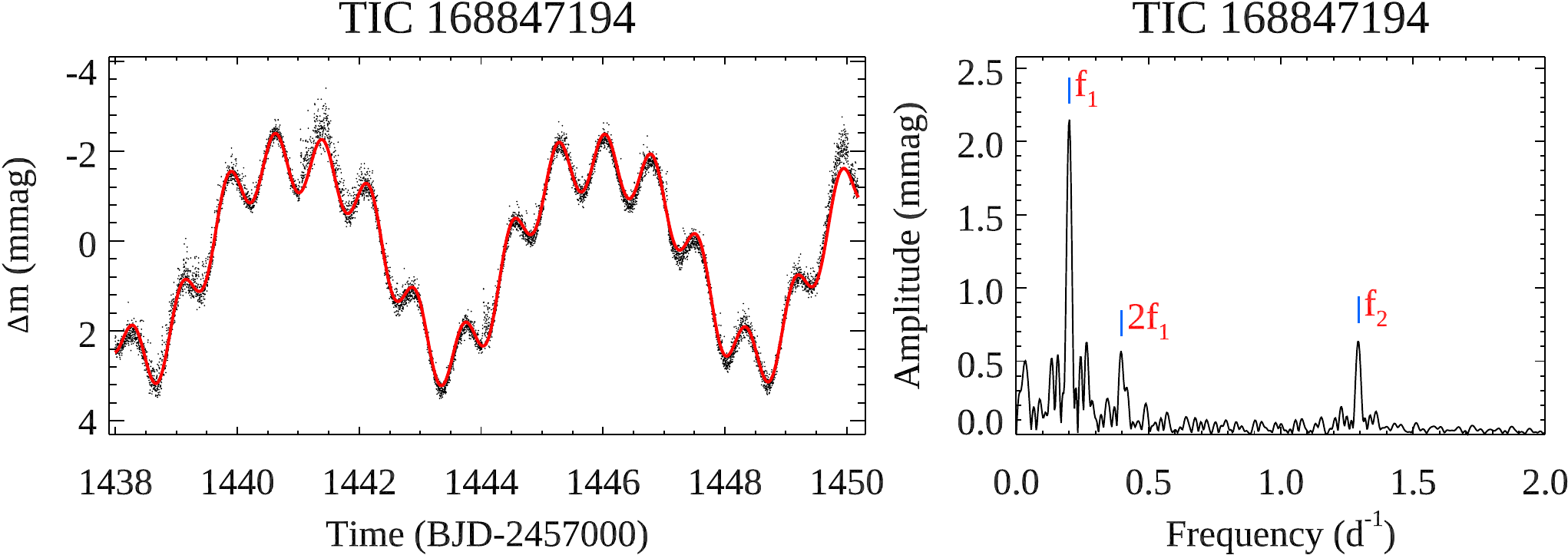}
\fivps{0.129\vsize}{0}{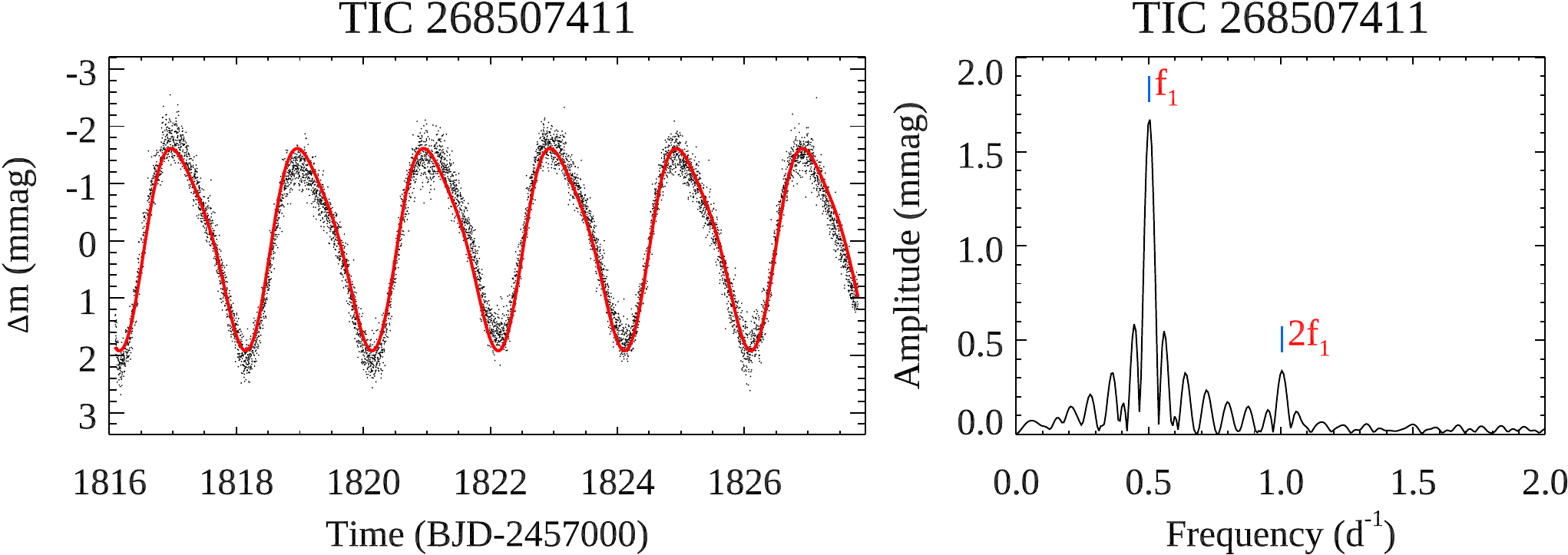}\vspace*{2mm}
\fivps{0.129\vsize}{0}{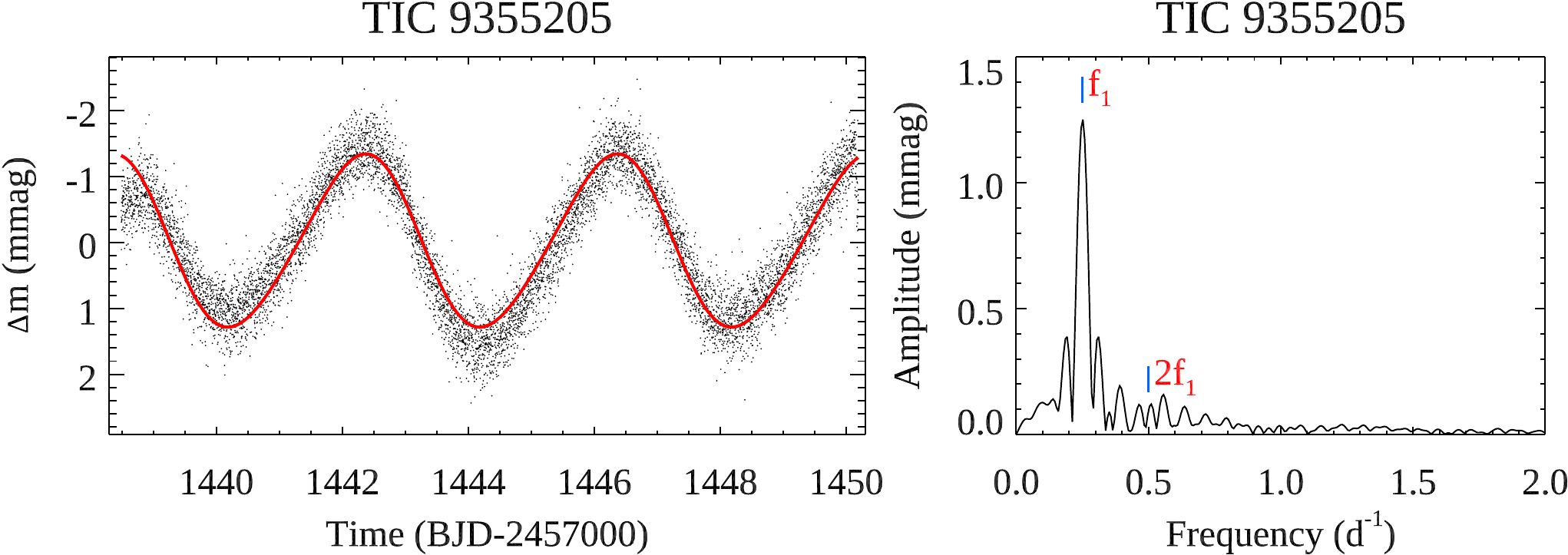}
\fivps{0.129\vsize}{0}{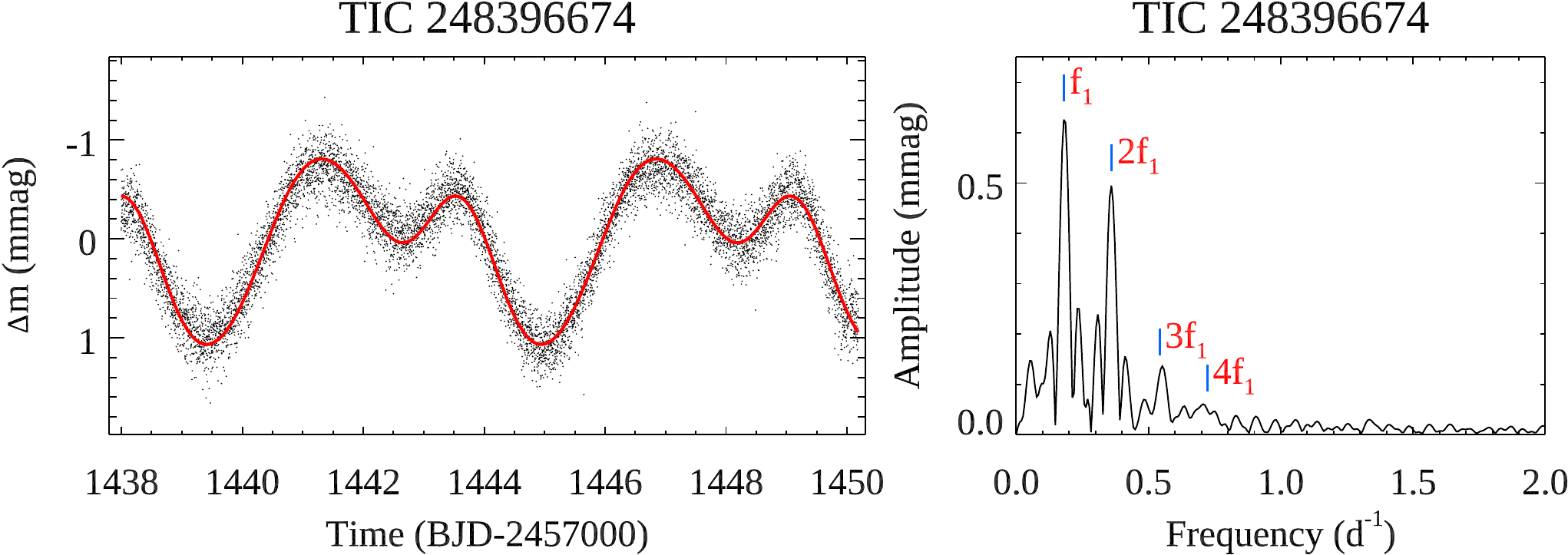}\vspace*{2mm}
\fivps{0.129\vsize}{0}{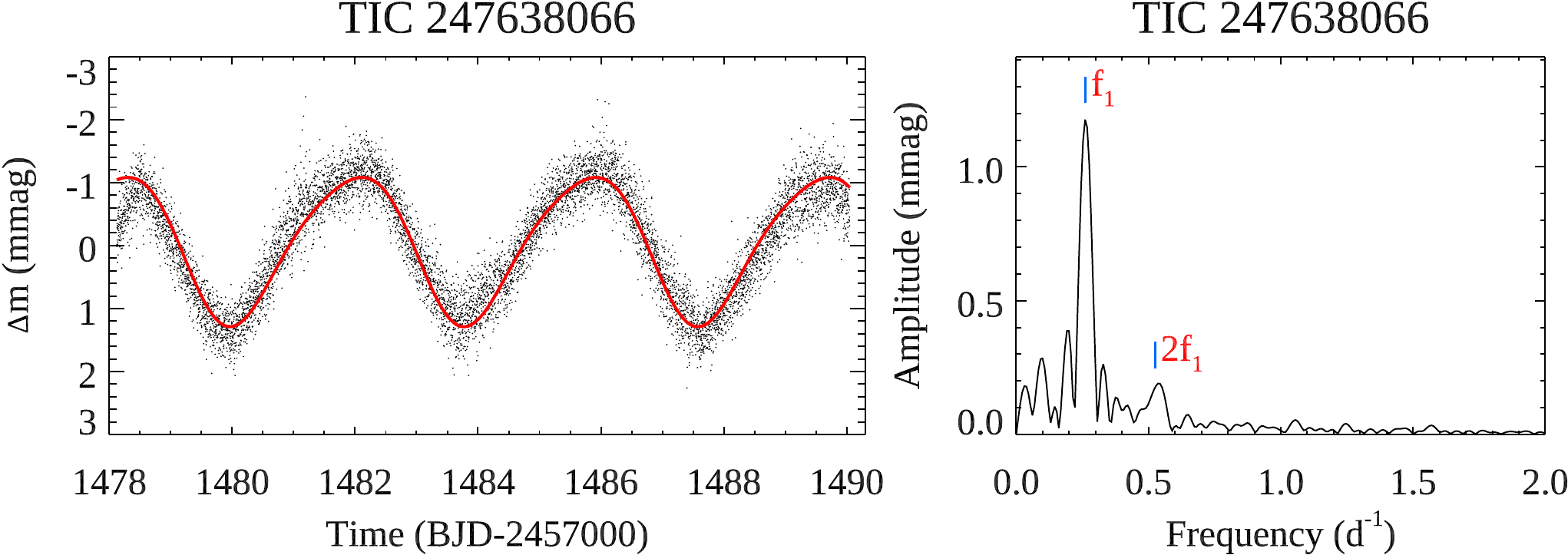}
\fivps{0.129\vsize}{0}{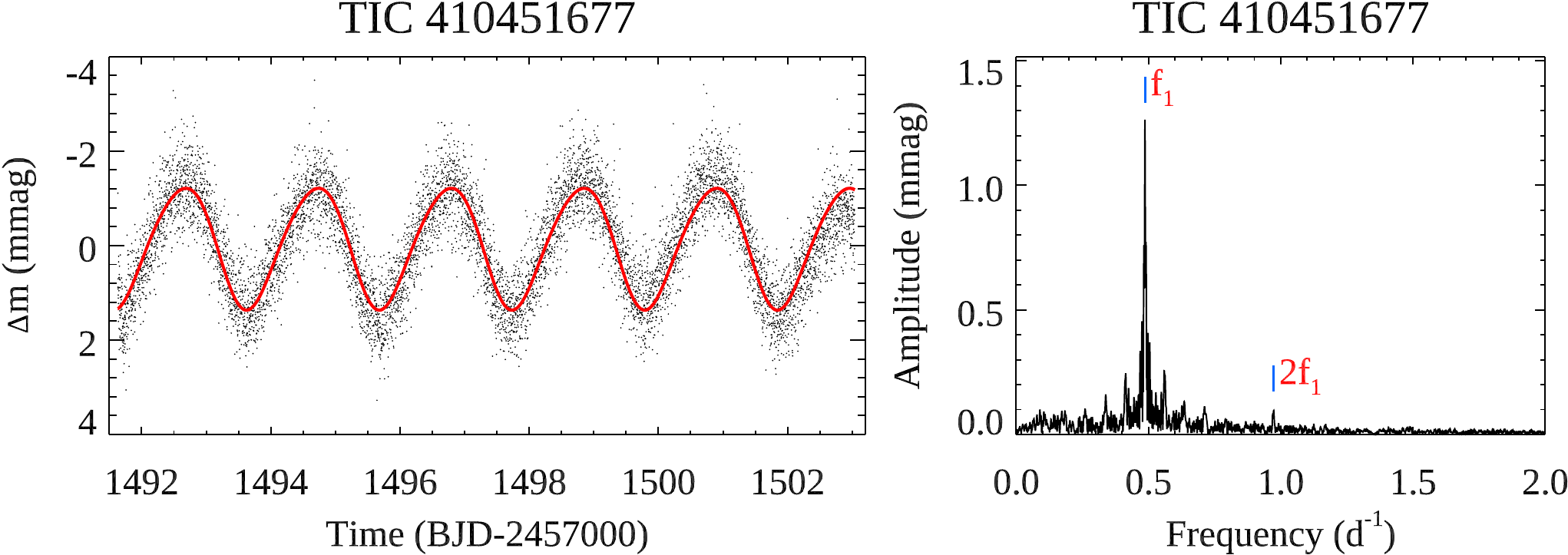}\vspace*{2mm}
\fivps{0.129\vsize}{0}{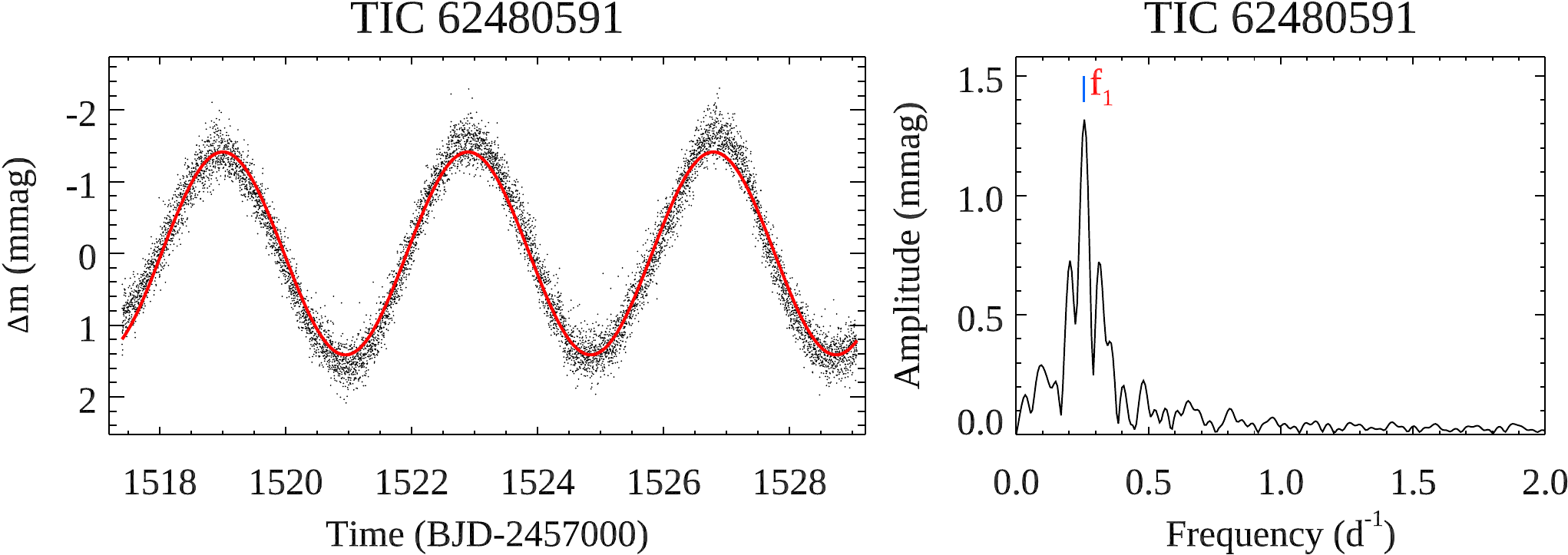}
\fivps{0.129\vsize}{0}{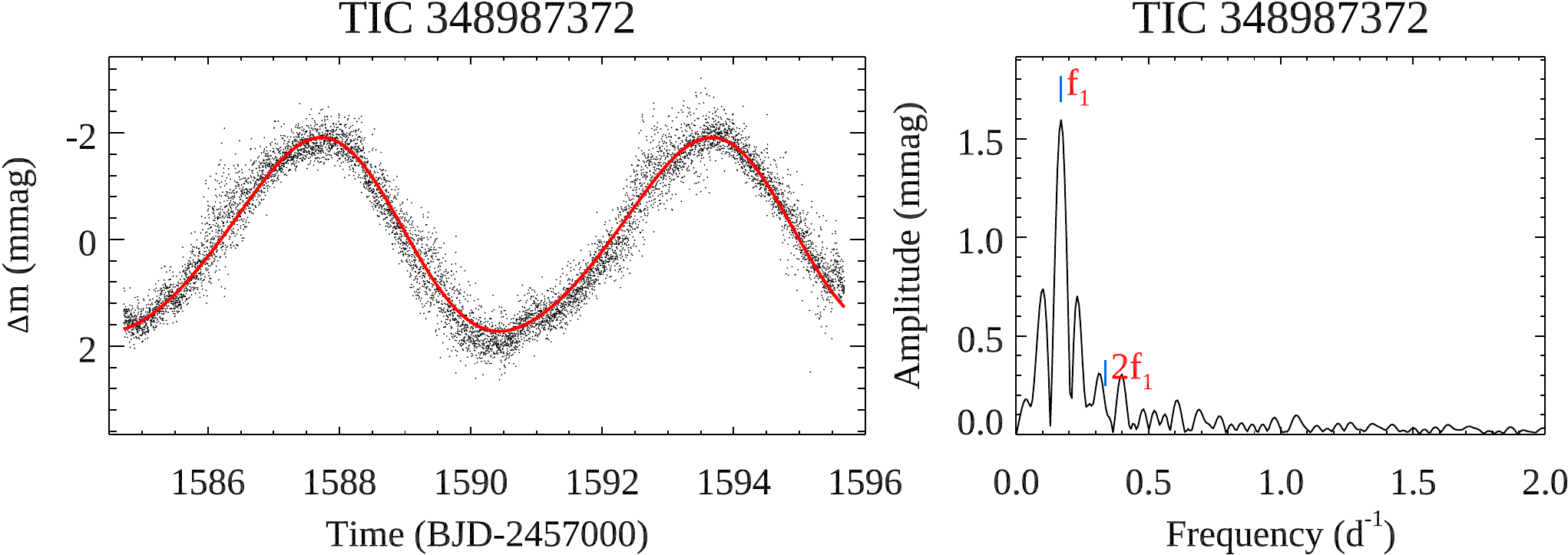}\vspace*{2mm}
\fivps{0.129\vsize}{0}{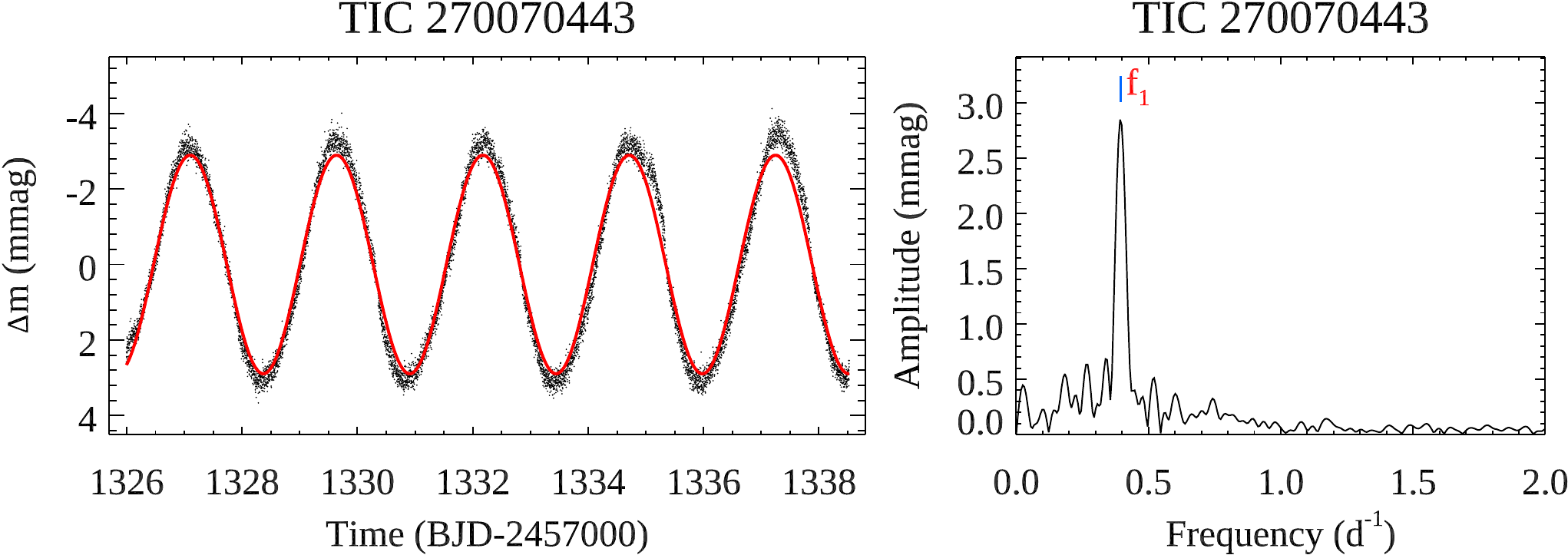}
\fivps{0.129\vsize}{0}{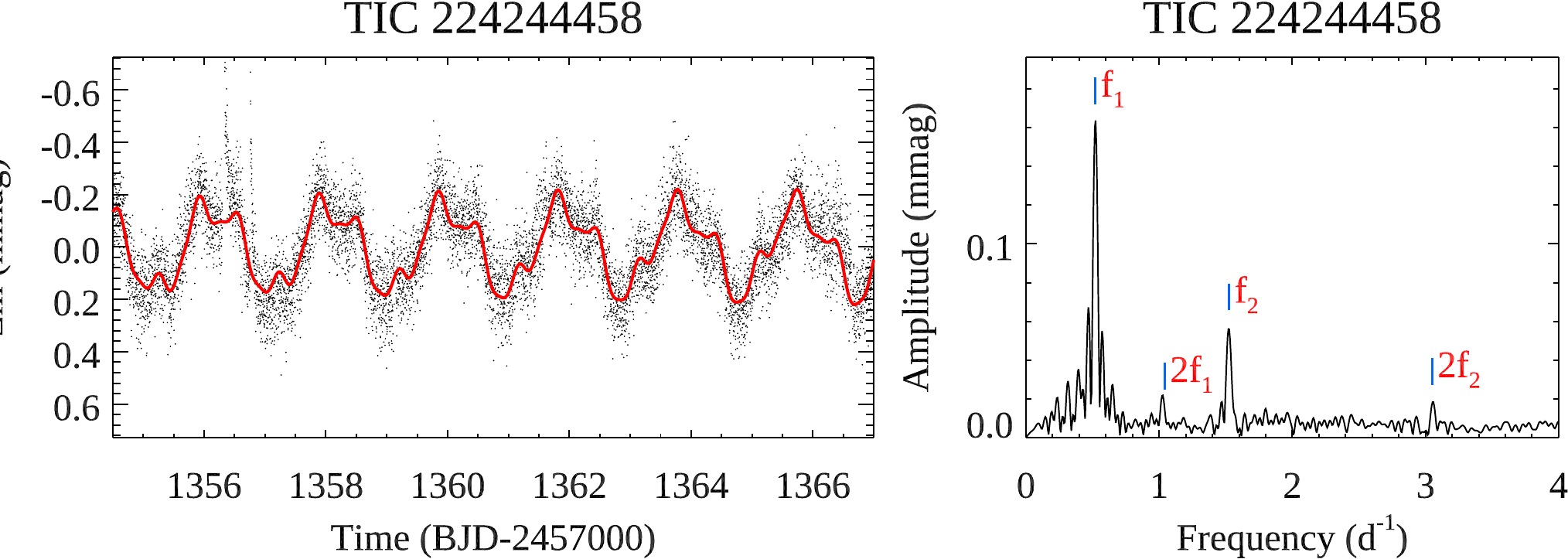}
\caption{Examples of HgMn stars showing clear rotational variability. For each target the left panels show a section of the TESS light curve (points) together with the harmonic model fit (solid line). The right panels show the associated periodogram with significant frequencies identified by short vertical bars. 
}
\label{fig:rot1}
\end{figure*}

\begin{figure*}
\centering
%
\fivps{0.129\vsize}{0}{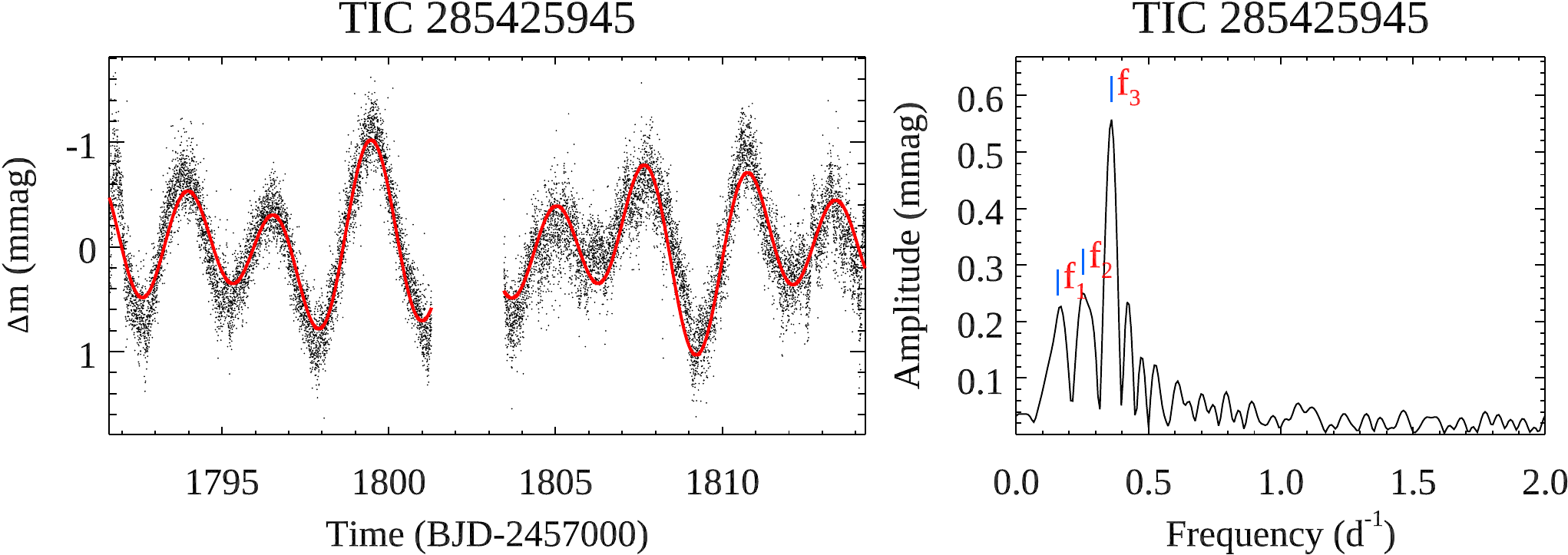}
\fivps{0.129\vsize}{0}{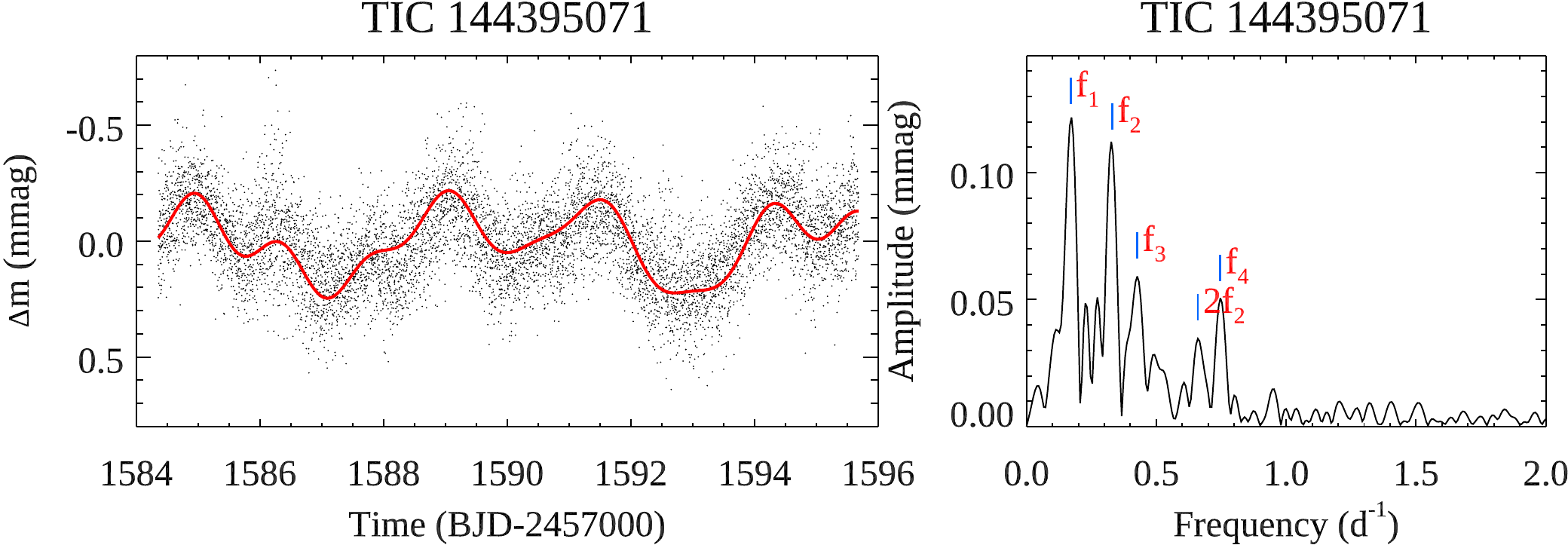}
\caption{Same as Fig.~\ref{fig:rot1}, but for the cases when rotational modulation cannot be unambiguously identified among multiple low-frequency signals present in the data.}
\label{fig:rot2}
\end{figure*}

In this study we were able to detect photometric variability compatible with rotational modulation for 55 target stars. For 44 of them a single low-frequency peak, often accompanied by harmonics, dominated the GLS periodogram. Representative examples of such photometric behaviour, corresponding to a range of light curve complexity and variability periods, are presented in Fig.~\ref{fig:rot1}. We assign these targets to the group exhibiting definite rotational modulation (identified by `ROT' in Table~\ref{tab:results}). For 11 other stars with low amplitudes of photometric variability in the TESS band and/or multiple low-frequency periodogram peaks of comparable amplitude (Fig.~\ref{fig:rot2}), our identification of rotational variability is tentative. These objects are marked by `ROT?' in Table~\ref{tab:results}. That table summarises results of our detailed frequency analysis, including derived amplitudes and periods, for all significant signals present in TESS light curves of studied stars. For five stars lacking coherent periodic signals an upper amplitude limit in the 0--2 d$^{-1}$ frequency range is provided. Below we comment on some of the most interesting rotationally variable stars included in our study. 

\smallskip

TIC~427733653 (HD~358, $\alpha$~And, Fig.~\ref{fig:rot1}) was one of the first HgMn stars for which conclusive evidence of spectroscopic rotational variability was found and a surface map of chemical abundance spots was reconstructed \citep{ryabchikova:1999b,ilyin:2000,adelman:2002,kochukhov:2007b}. The spectroscopic rotational period of 2.38195(3)~d is close to, but is formally inconsistent with, the 2.3834(2)~d period derived here from TESS observations. As demonstrated by \citet{kochukhov:2007b}, the surface spot distribution in $\alpha$~And evolves slowly with time. This may introduce additional uncertainties in the spectroscopic period determination, which have not been fully accounted for. With a semi-amplitude of 2.8~mmag, $\alpha$~And exhibits the strongest photometric rotational variability among the bona fide HgMn stars included in this study.

TIC~285425945 (HD~11291, 2 Per, Fig.~\ref{fig:rot2}). This star is known to be an SB1 system with an orbital period of 5.627~d \citep{pourbaix:2004}. The TESS light curve shows multiple low frequency signals. The highest amplitude one, corresponding to 2.784~d, is tentatively attributed to rotational modulation. Given that this signal is close to half of the orbital period, interpretation in terms of ellipsoidal variability \citep{kochukhov:2021a} cannot be excluded either.

TIC~229099027 (HD~11753, $\varphi$~Phe) is a well-known spectroscopically variable HgMn star with evolving surface distributions of chemical elements \citep{briquet:2010,makaganiuk:2012,korhonen:2013}. It is a member of a long-period, potentially eclipsing, SB1 system with $P_{\rm orb}=1126$~d \citep{pourbaix:2013}. The rotational period found in spectroscopy, $P_{\rm rot}=9.531$~d, is longer than the 9.324(1)~d period that provides the best description of the TESS light curve of this star. This discrepancy can be understood in the context of weak solar-like differential rotation reported by \citet{korhonen:2013}. The photometric rotational variability observed in the TESS band is complex with the second harmonic dominating the periodogram and up to four harmonics necessary to provide a satisfactory description of the light curve. The same TESS data set of HD~11753 was assessed by \citet{prvak:2020}. These authors demonstrated that the observed photometric amplitude is well matched by their theoretical calculations using abundance maps available for this HgMn star, although the detailed shape of the light curve could not be reproduced.

TIC~354671857 (HD~14228, $\varphi$~Eri, Fig.~\ref{fig:rot1}). Variability of this star was detected in previous studies based on TESS data \citep{balona:2019a,pedersen:2019}. The periodogram shows a prominent peak at $P=0.344$ d and weaker signals at 0.45--0.46 d. This star is an extremely rapid rotator with $v_{\rm e}\sin i=240$~\kms\ \citep*{hutchings:1979} and the 0.344~d variability is fully compatible with rotational modulation. This star is by far the fastest rotator in our study. It is consistently referred to as an HgMn star in the literature \citep{schneider:1981,renson:2009}, but its classification needs to be confirmed by a modern spectroscopic study since finding a chemical peculiarity at such extreme rotation rate is unexpected.

TIC~280051467 (HD~19400, $\theta$~Hyi, Fig.~\ref{fig:rot1}). This is a rare PGa-type CP star, which is considered to be the hotter extension of HgMn stars \citep{alonso:2003}. A subtle spectroscopic variability was reported for this object by \citet{hubrig:2014}, though no rotational period was derived. The TESS data analysed in our study shows a clear variability with $P_{\rm rot}=4.369$~d. Detection of this rotational modulation was also mentioned by \citet{balona:2019a} and \citet{pedersen:2019}.

TIC~168847194 (HD~27376, 41 Eri, Fig.~\ref{fig:rot1}) is an SB2 system with nearly equal-mass components and an orbital period of 5.010~d \citep{hummel:2017}. Both components show chemical peculiarities characteristic of HgMn stars \citep*[e.g.][]{dolk:2003} and both exhibit spectroscopic variability \citep{hubrig:2012}. Rotation of the components is expected to be synchronised with the orbital rate, which is confirmed by our measurement of $P_{\rm rot}=5.010$~d using TESS data. Interestingly, the TESS light curve of 41~Eri also shows a sinusoidal signal at 0.77~d suggesting that one of the components may exhibit pulsational variability.

TIC~373026963 (HD~28217, HR~1402). The spectroscopic variability of this star was reported by \citet{nunez:2010} and \citet{hubrig:2011b}, but the rotational period remained unknown. According to our analysis, $P_{\rm rot}=3.132$~d.

TIC~268507411 (HD~28929, HR~1445, Fig.~\ref{fig:rot1}). This HgMn star shows a clear photometric variation with $P_{\rm rot}=1.989$~d. Previous ground-based and space photometric observations yielded a somewhat shorter 1.977~d period \citep{paunzen:2013,netopil:2017}.

TIC~9355205 (HD~30963) was recently identified as an HgMn star by \citet{monier:2019}. This star shows rotational variability with a period very close to 4~d. An independent analysis of TESS data by \citet{david-uraz:2021} retrieved the same period and did not find magnetic field based on follow up spectropolarimetric observations.

TIC~436723855 (HD~31373, HR~1576) shows rotational modulation with $P_{\rm rot}=1.448$~d as well as a weaker low-frequency sinusoidal signal at 2.644~d. No photometric variability was detected for this star by \citet{adelman:1998c} and \citet{paunzen:2013} whereas \citet{pope:2019} classified this object as an SPB star based on \textit{K2} observations.

TIC~248396674 (HD~32964, 66 Eri, Fig.~\ref{fig:rot1}) is another SB2 HgMn star with a mass ratio close to one. In contrast to 41~Eri, only one of the components exhibits HgMn spectral peculiarities and line profile variability \citep{yushchenko:1999,makaganiuk:2011}. The TESS light curve of this star shows a complex double-wave shape with up to four harmonic evident in the periodogram. The most recent determination of the orbital period yields $P_{\rm orb}=5.523$~d \citep{hubrig:2012}. This is close, but not identical, to $P_{\rm rot}=5.535$~d derived from TESS photometry.

TIC~169534653 (HD~33904, $\mu$~Lep) is one of the brightest HgMn stars showing rotational line profile variability \citep{nunez:2010,kochukhov:2011b}. Here we report the first conclusive determination of the rotational period for this star, $P_{\rm rot}=2.933$~d. The period of 2~d mentioned by \citet{netopil:2017} is not supported by TESS data.

TIC~116333643 (HD~37519). No photometric variability was detected in this star by \citet{paunzen:2013} using STEREO satellite observations. On the other hand, the TESS light curve shows a clear non-sinusoidal variation with $P_{\rm rot}=0.841$~d. This is the second shortest rotational period found in our study. It cannot be excluded that this signal is coming from an unresolved companion star. However, HD~37519 is not known to be a binary and its very rapid rotation is consistent with the high projected rotational velocity, $v_{\rm e}\sin{i}=195$~\kms, reported by \citet*{abt:2002}.

TIC~247638066 (HD~38478, 129 Tau, Fig.~\ref{fig:rot1}) rotates with a period of 3.806~d. \citet{paunzen:2013} listed this object as constant.

TIC~155095401 (HD~49606, 33 Gem). Different periods, ranging between 1.4 and 3.3~d, were reported for this star (\citealt*{glagolevskij:1985}; \citealt{paunzen:2013}; \citealt{netopil:2017}). None of them corresponds to the period determined here using TESS observations, $P_{\rm rot}=8.546$~d. This star was recently found to be part of an SB1 system with an orbital period of 148.3~d \citep{catanzaro:2016}.

TIC~148109427 (HD~53244, $\gamma$~CMa) is a bright HgMn star with spectral variability investigated by several studies \citep{briquet:2010,nunez:2010,hubrig:2011b}. \citet{briquet:2010} determined $P_{\rm rot}=6.16$~d, which is somewhat shorter than the 6.214~d period measured here. \citet{burssens:2020} concluded that $\gamma$~CMa shows rotational modulation with a period of 6.250~d based on the same TESS data set.


TIC~410451677 (HD~66409, Fig.~\ref{fig:rot1}) exhibits clear rotational variability with $P_{\rm rot}=2.055$~d. The presence of this rotational modulation was also noted by \citet{balona:2019a} and \citet{pedersen:2019} based on a smaller subset of TESS observations. Weak spectroscopic variability of this HgMn star was discussed by \citet{nunez:2010}.

TIC~307291308 (HD~71066, $\kappa^2$~Vol) was not conclusively detected as variable by \citet{pedersen:2019}. However, \citet{balona:2019a} and the present study attribute the 1.296~d periodicity seen in the TESS light curve of this star to rotational modulation. This star is known to have sharp spectral lines \citep[e.g.][]{kochukhov:2013a}, which would be compatible with the short period only if the target is seen nearly pole-on. Alternatively, the observed signal could be explained by contamination by a 2.1~mag fainter nearby star HD~71046C, which is blended with HD~71066 in TESS images.

TIC~62480591 (HD~75333, 14 Hya, Fig.~\ref{fig:rot1}) shows variability with $P_{\rm rot}=3.894$~d according to our analysis. \citet{netopil:2017} reported a rotational period of 6~d, which is incompatible with our measurement.


TIC~319809753 (HD~101189, HR~4487) exhibits spectral variability that was described in several studies \citep{hubrig:2011b,hubrig:2012,hubrig:2020} but the rotational period remained undetermined. The TESS data allowed us to establish $P_{\rm rot}=2.004$~d. This is very close to an integer number of days, making interpretation of ground-based observations challenging.

TIC~348987372 (HD~106625, $\gamma$~Crv, Fig.~\ref{fig:rot1}) is the second brightest HgMn star after $\alpha$~And. Its TESS light curve exhibits a clear variation with $P_{\rm rot}=5.938$~d. No previous reports of its variability, using either photometric or spectroscopic observations, can be found in the literature.

TIC~144395071 (HD~110073, HR~4817, Fig.~\ref{fig:rot2}) shows multiple low-frequency peaks. The second highest one, corresponding to $P_{\rm rot}=3.030$~d, is tentatively attributed to rotational modulation since this signal is non-sinusoidal. Alternatively, the 5.903~d signal could also correspond to rotation. Spectroscopic variability of this star was reported by \citet{nunez:2010} and \citet{hubrig:2011b}.

TIC~442652828 (HD~141556, $\chi$~Lup) is known to be an SB2 system with $P_{\rm orb}=15.25$~d \citep{le-bouquin:2013} and with a tertiary component also present. A relatively short rotation period of 1.793~d derived from TESS data is unexpected for the sharp-line HgMn primary and may originate from one of the other two components.

TIC~51612589 (HD~145842, $\theta$~Nor). For this newly identified HgMn star \citet{gonzalez:2021} reported the 1.087~d period based on the same observational data as analysed here. Our measurement yields a close period value of 1.086~d.

TIC~323999777 (HD~168733, HR~6870) has discrepant peculiarity classifications in the literature. Most papers describe it as a magnetic CP star \citep{kochukhov:2006,briquet:2007,collado:2009}, which is supported by the definite detection of a global magnetic field \citep{mathys:1994,mathys:1997a}. A few studies group HD~168733 with HgMn stars based on its chemical abundance pattern \citep{cowley:2010,ghazaryan:2018}. Variability in the TESS light curve of HD~168733 has an unusually high amplitude for an HgMn star and a period of 6.326~d, which is compatible with $P_{\rm rot}=6.3$~d given by \citet{netopil:2017}. This is a clear outlier in our study, suggesting that this object is indeed a magnetic Bp star.

TIC~27901267 (HD~172044, HR~6997) shows a non-sinusoidal rotational modulation with $P_{\rm rot}=4.330$~d. This star is known to be an SB1 with a 1675~d period \citep{pourbaix:2004}.

TIC~347160134 (HD~174933, 112 Her) is an SB2 system with $P_{\rm orb}=6.362$~d \citep*{ryabchikova:1996a}. The period derived from TESS data, $P_{\rm rot}=12.419$~d, is approximately twice this value suggesting that the primary rotates sub-synchronously.

TIC~270070443 (HD~198174, HR~7961, Fig.~\ref{fig:rot1}) is another object with a high photometric variability amplitude in the TESS band \citep{balona:2019a,pedersen:2019} and historically a somewhat dubious relation to HgMn stars. Our measurement of $P_{\rm rot}=2.5365(2)$~d is shorter than 2.545~d reported by \citet{paunzen:2013} and \citet{netopil:2017}. The former study grouped HR~7961 with HgMn stars. However, its spectral peculiarity class is indicated as undetermined by \citet{renson:2009} and the association with HgMn stars in the literature comes only from UV spectrophotometry \citep{cucchiaro:1977}. Nevertheless, a high-resolution FEROS spectrum available in the ESO archive clearly shows the Hg~{\sc ii} 3984~\AA\ line and enhanced Mn~{\sc ii} lines, confirming the HgMn classification of this star.

TIC~129533458 (HD~216831, HR~8723) was reported to be a spectroscopically variable HgMn star \citep{kochukhov:2005b}. Our analysis establishes that this rotational modulation occurs with a period of 3.471~d, opening prospects for quantitative interpretation of ground-based spectroscopy.

TIC~224244458 (HD~221507, $\beta$\,Scl, Fig.~\ref{fig:rot1}) is one of the brightest HgMn stars showing spectroscopic variability due to surface inhomogeneities. Line profile variability in the spectra of this object was investigated by \citet{briquet:2010} and \citet{hubrig:2020}. The former study has established a rotational period of 1.93~d, which is in good agreement with $P_{\rm rot}=1.921$~d found here. Rotational variability using the sector 2 TESS light curve was reported by both \citet{balona:2019a} and \citet{pedersen:2019}. The first of these studies attributed a flare (evident at time 1356.5 in Fig.~\ref{fig:rot1}) to the HgMn star. However, our detailed time series analysis reveals the presence of another rotational signal with $P_{\rm rot}=0.656$~d, indicating that the TESS light curve of $\beta$\,Scl is contaminated by the rotational modulation of a companion star. This signal may be coming from the visual companion detected using diffraction-limited imaging \citep{scholler:2010,kammerer:2019}. A late-type star with such a short rotational period is expected to be very active and is, most likely, responsible for the observed flare.

TIC~359033491 (HD~225289, HR~9110) shows variability with $P_{\rm rot}=3.254$~d. This period is shorter by about a factor of two compared to $P_{\rm rot}=6.432$~d given by \citet{netopil:2017}. Considering that in the TESS periodogram the 3.254~d period is accompanied by the second and third harmonics, it is the more likely rotational period.

\subsection{Pulsations}
\label{ssec:puls}

\begin{figure*}
\centering
\fivps{0.129\vsize}{0}{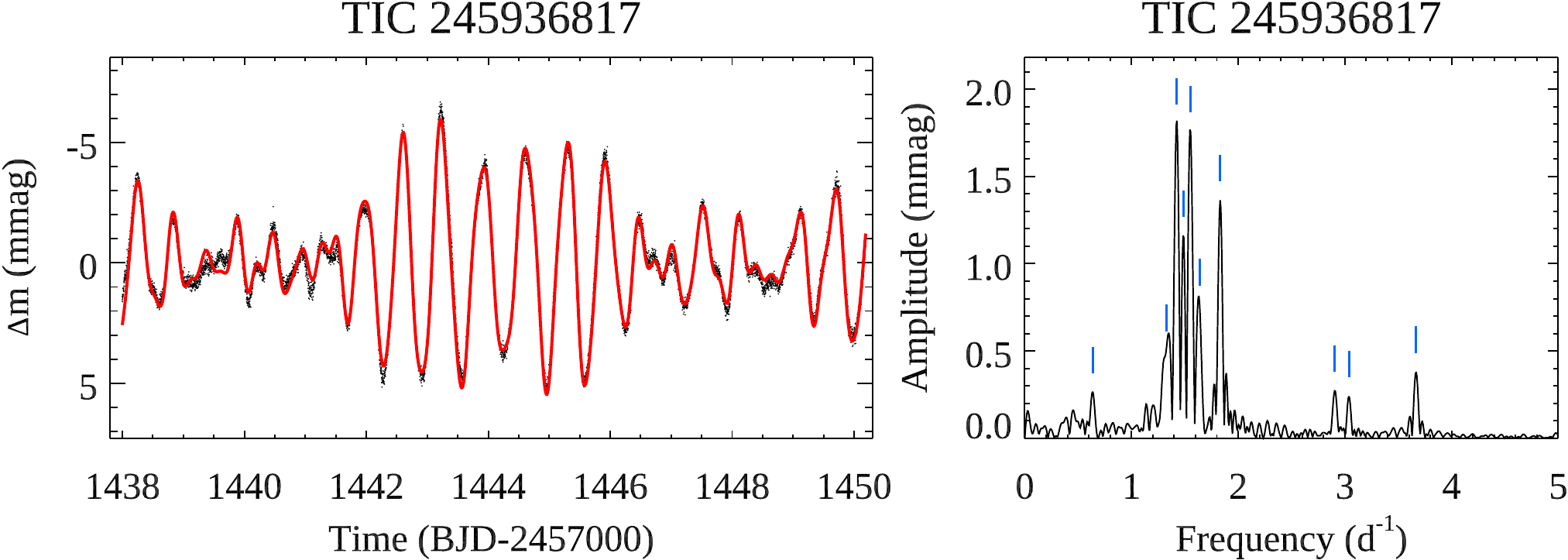}
\fivps{0.129\vsize}{0}{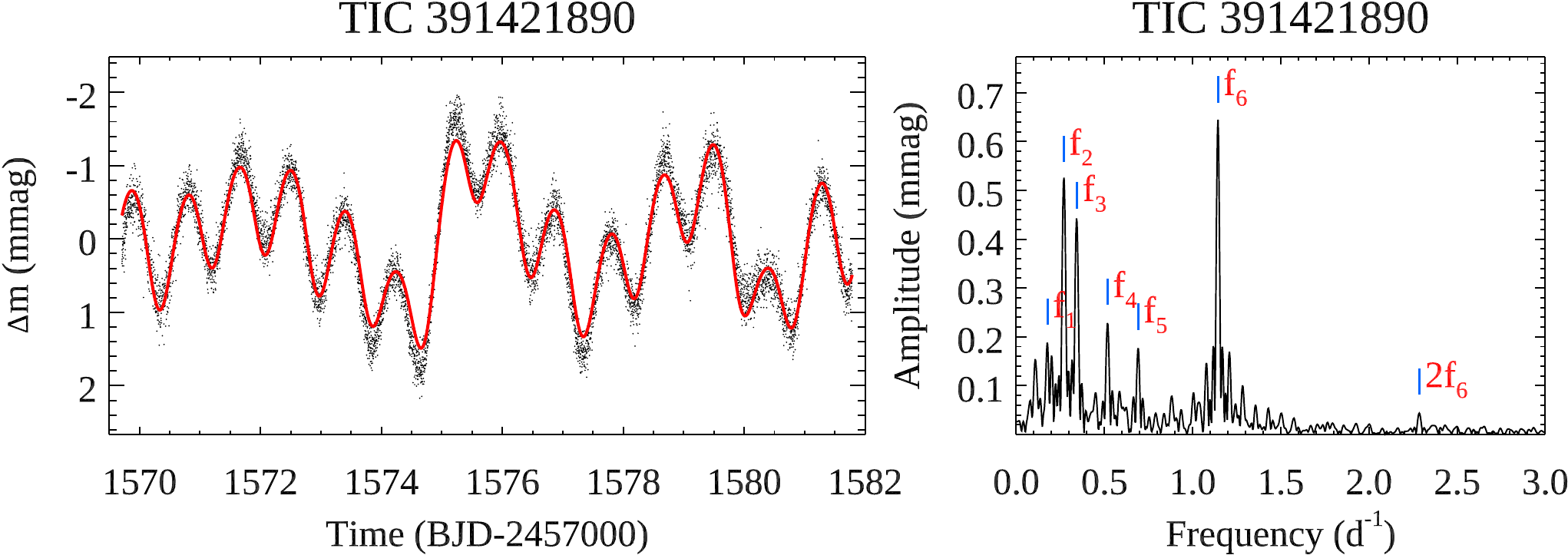}\vspace*{2mm}
\fivps{0.129\vsize}{0}{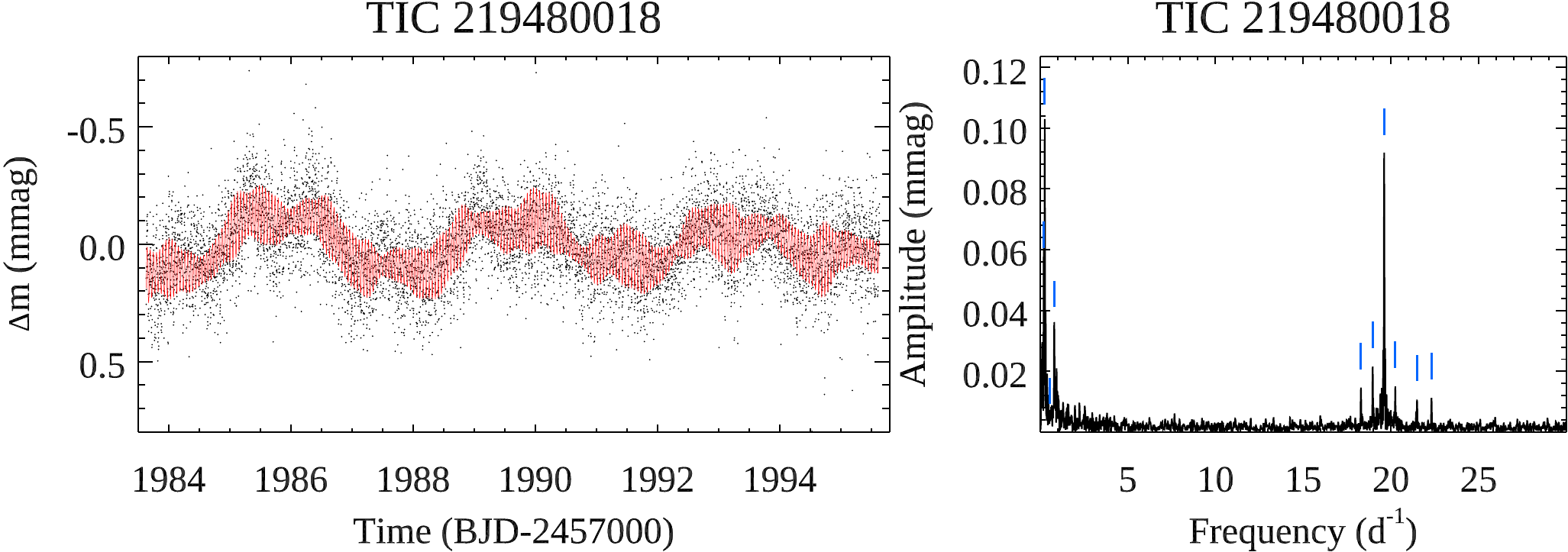}\hspace*{3mm}
\fivps{0.129\vsize}{0}{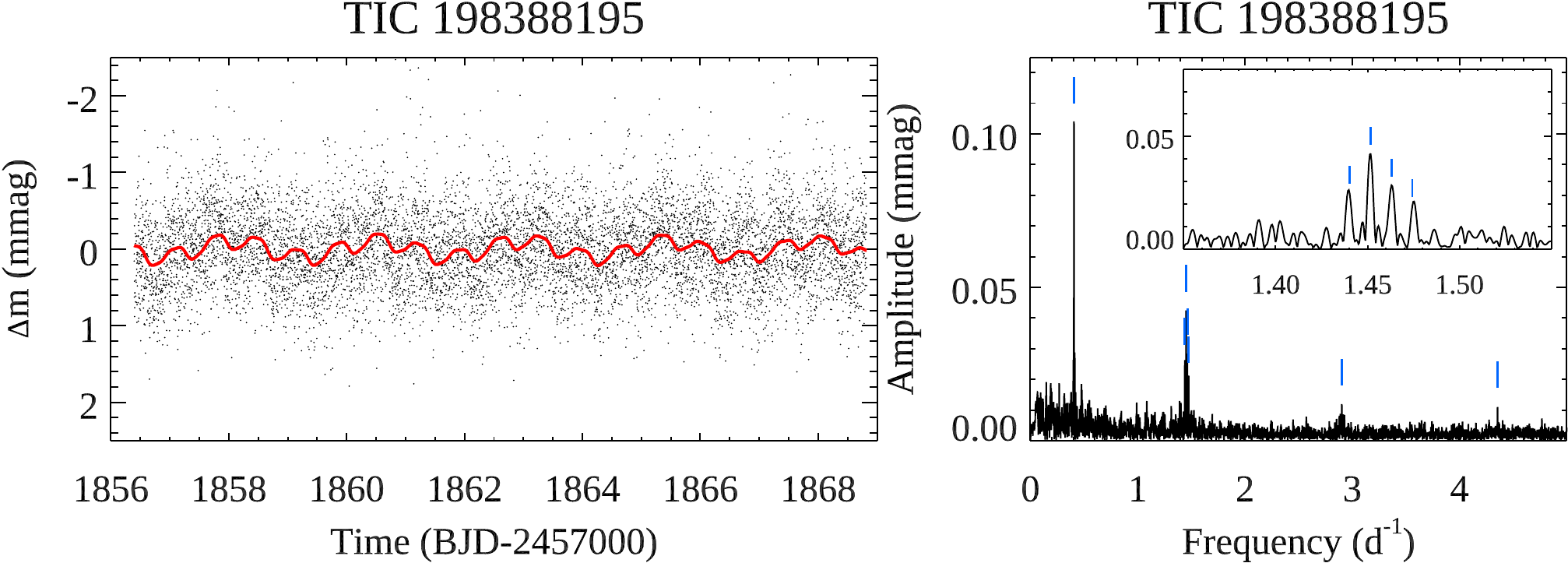}\vspace*{2mm}
\fivps{0.129\vsize}{0}{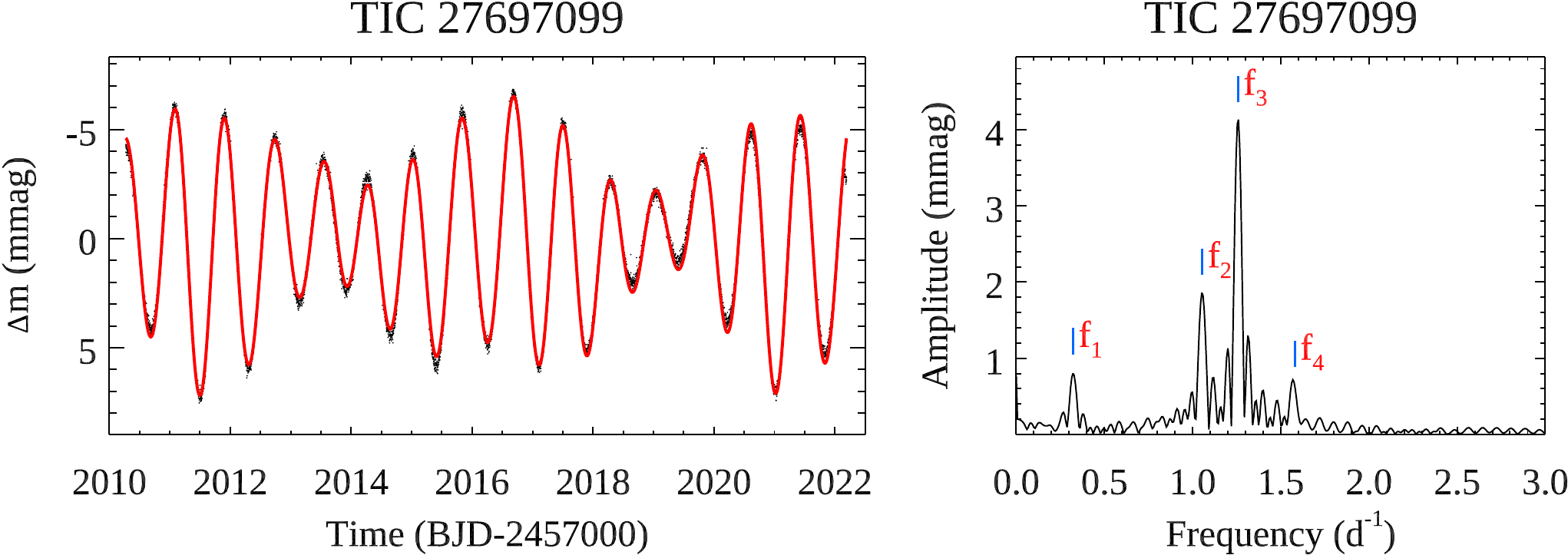}
\fivps{0.129\vsize}{0}{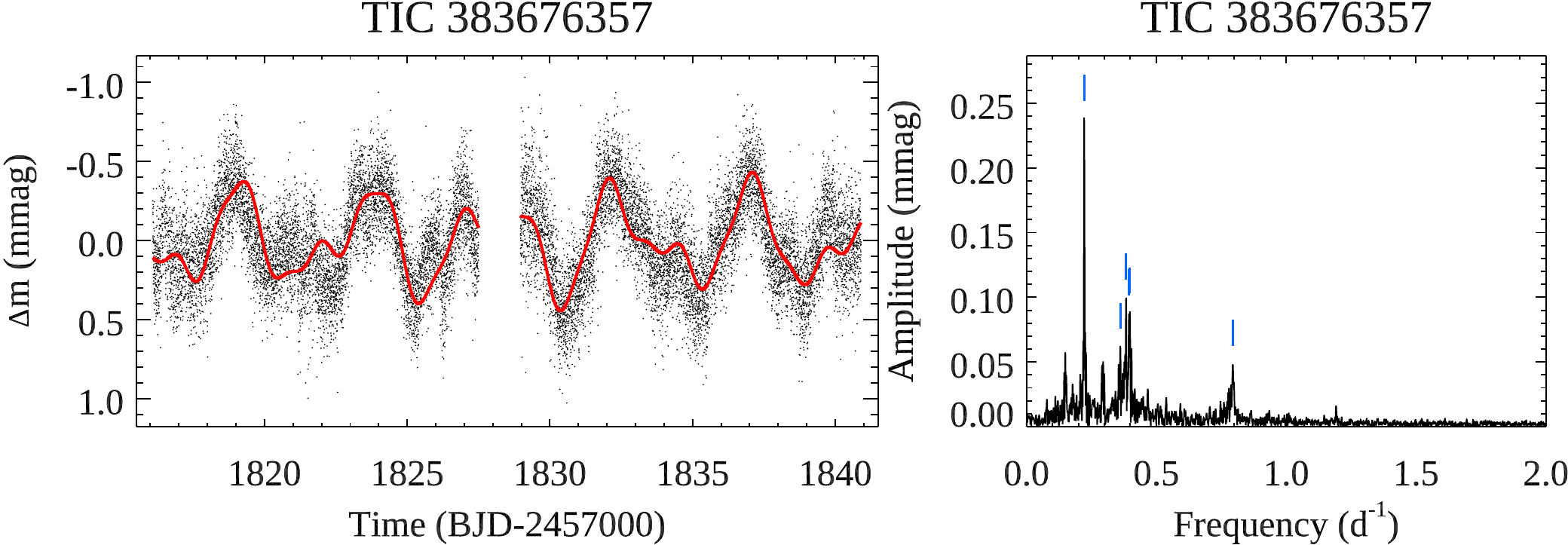}\vspace*{2mm}
\fivps{0.129\vsize}{0}{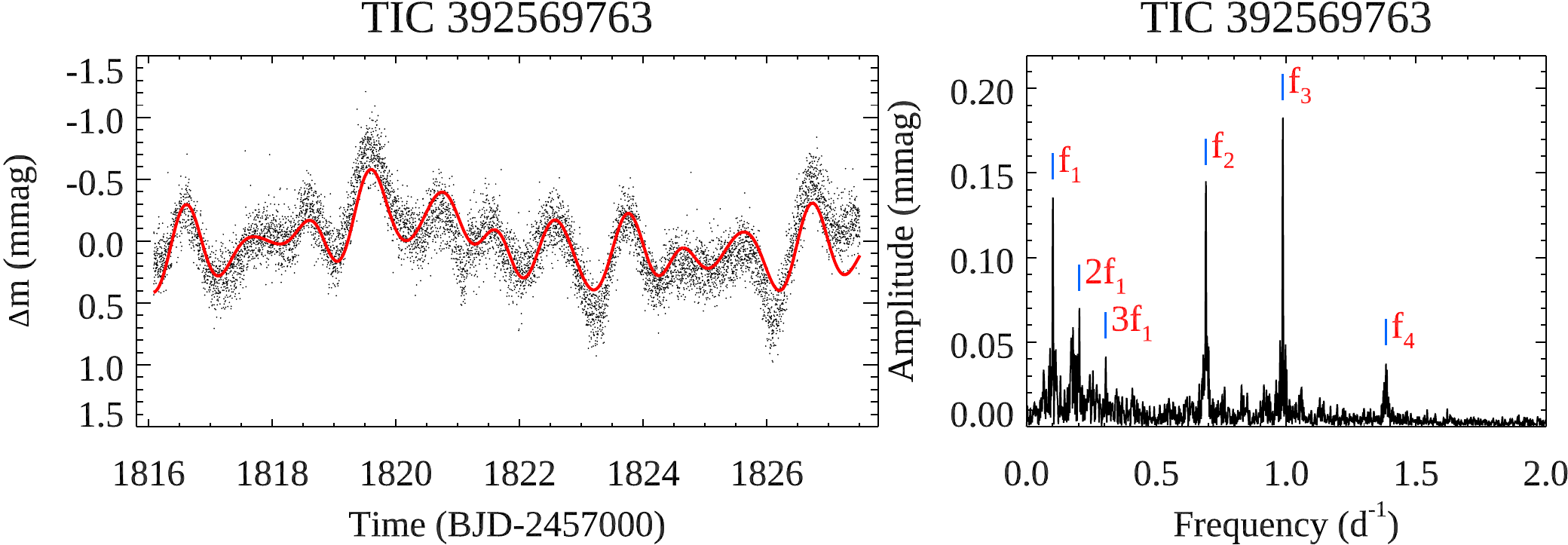}
%
%
\caption{Same as Fig.~\ref{fig:rot1}, but for HgMn stars exhibiting significant multi-periodic pulsational variability. For TIC\,198388195 the inset shows a non-sinusoidal signal at $\approx$\,1.45~d$^{-1}$ comprised of a resolved frequency quadruplet.}
\label{fig:puls}
\end{figure*}

Several classes of pulsating variables are known among B-type stars. The two best studied types are the $\beta$~Cep p-mode pulsators and the slowly pulsating B (SPB) g-mode variables \citep*[e.g.][]{aerts:2010}. The former group of stars typically pulsates with frequencies above $\sim$\,5 d$^{-1}$; the latter have frequencies below this limit. There are hybrid pulsators exhibiting both p- and g-modes simultaneously \citep[e.g.][]{buysschaert:2017}. The classical short-period $\beta$~Cep pulsations are found in early-B stars that are significantly hotter than the HgMn stars included in our study. Nevertheless, it is known that late-B stars occasionally also exhibit pulsations with periods typical of $\beta$~Cep stars. These late-B variable stars are sometimes called `Maia variables' \citep{balona:2015,balona:2020,mowlavi:2016}, although the supposed prototype of this class, the star Maia (20 Tau, HD\,23408), is itself an HgMn star showing rotational modulation due to chemical spots and no pulsations \citep{white:2017}. It is beyond the scope of our study to address this confusing nomenclature. In the remainder of this section we will refer to the low-frequency pulsational signals as `g-modes' and the higher frequency ones as `p-modes' without an attempt to formally classify corresponding pulsational variability. In this study we also searched for rapid oscillations with frequencies above 50 d$^{-1}$ typical of roAp stars \citep[e.g.][]{cunha:2019}, but no examples of such variability were found.

Seven stars included in our surveys show light curves dominated by multi-periodic pulsational variability. Examples of TESS data for these stars and corresponding periodograms are shown in Fig.~\ref{fig:puls}.

\smallskip

\begin{figure*}
\centering
%
\fivps{0.129\vsize}{0}{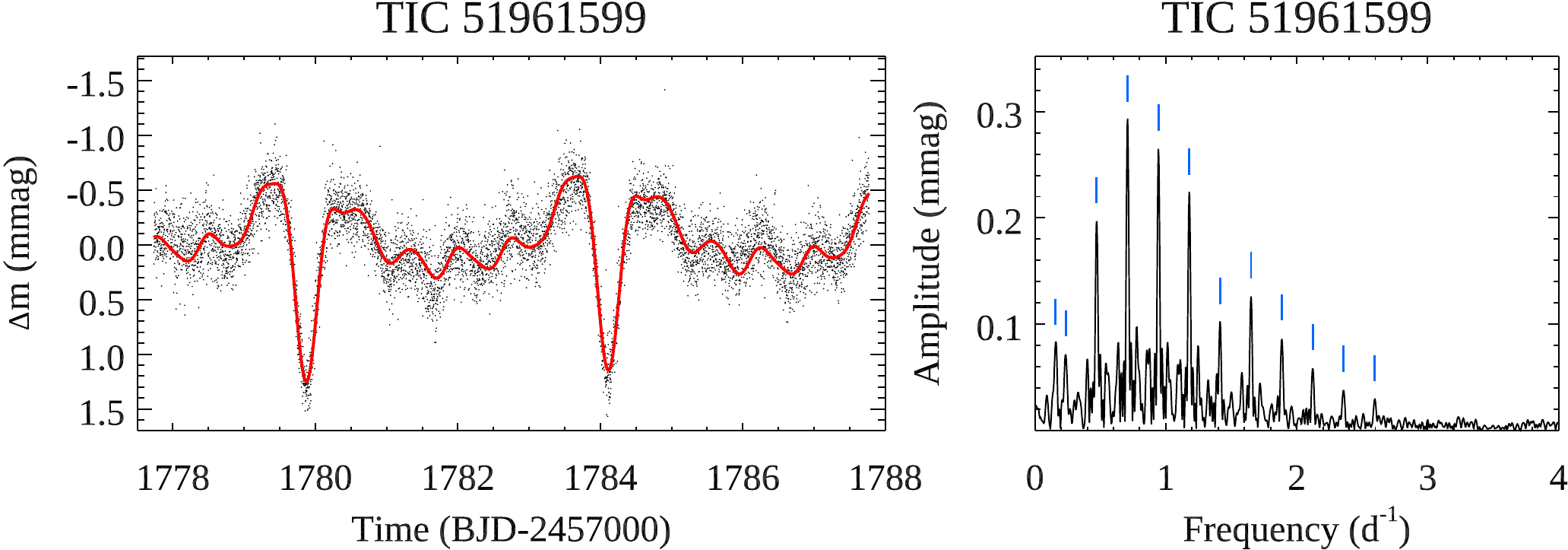}
\fivps{0.129\vsize}{0}{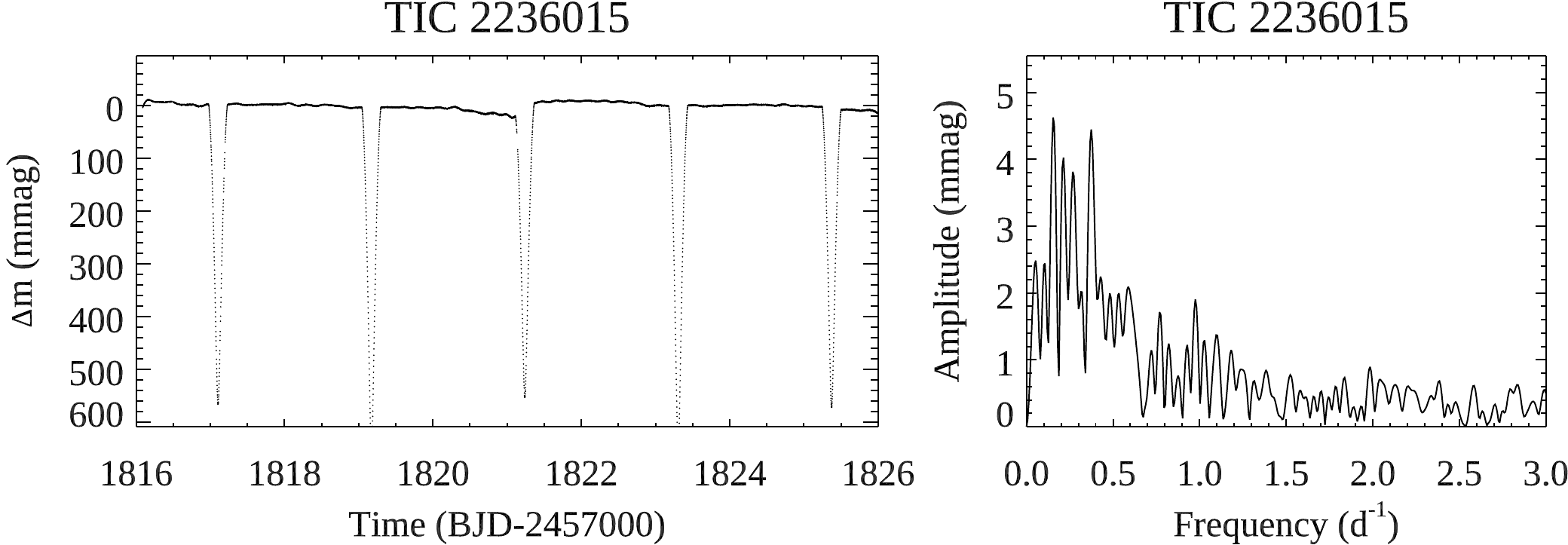}\vspace*{2mm}
\fivps{0.129\vsize}{0}{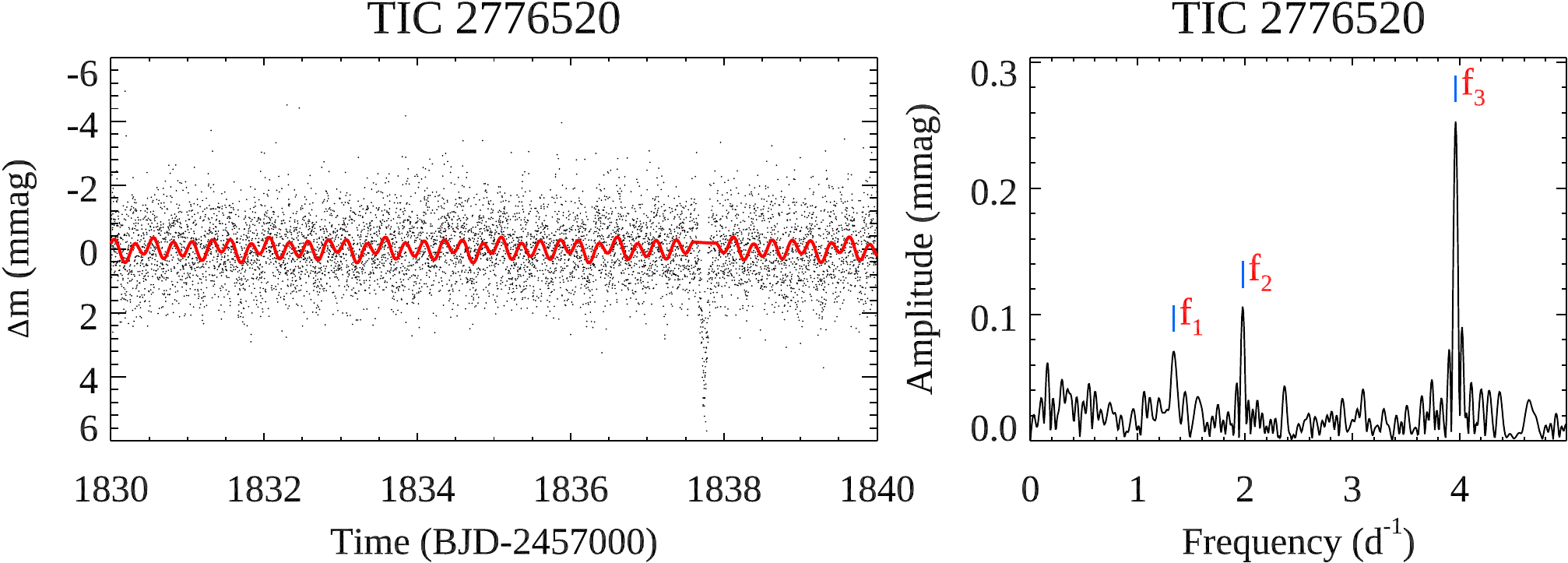}
\fivps{0.129\vsize}{0}{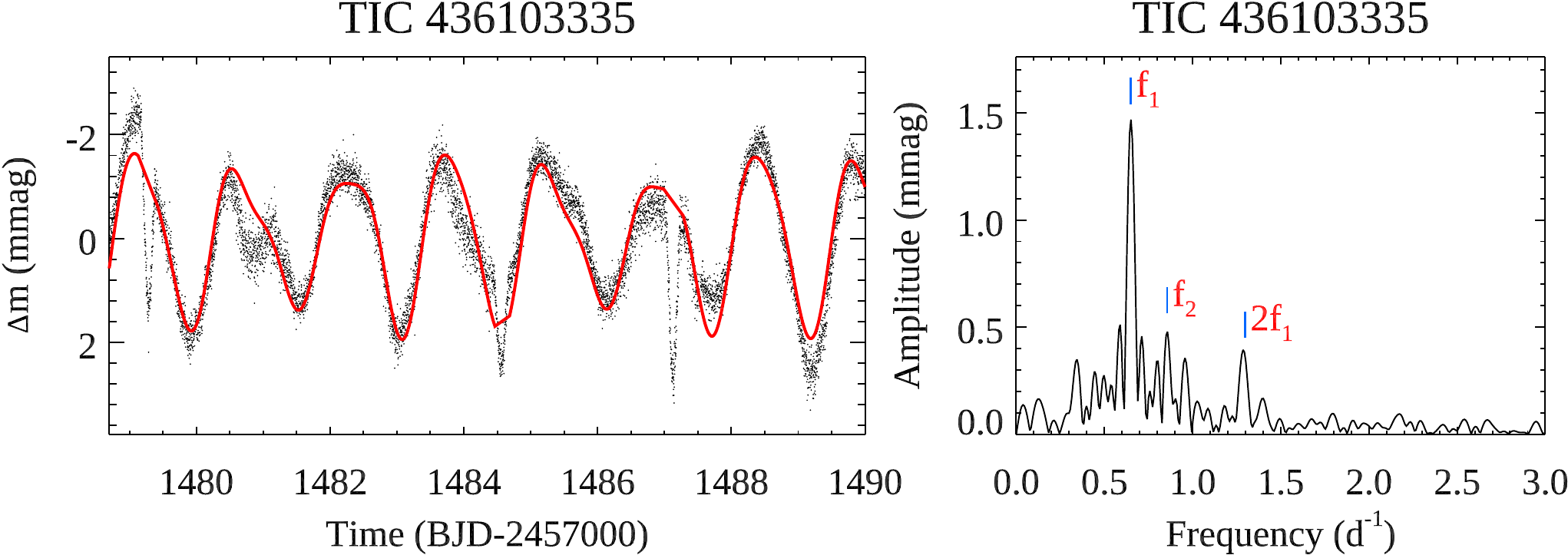}\vspace*{2mm}
\fivps{0.129\vsize}{0}{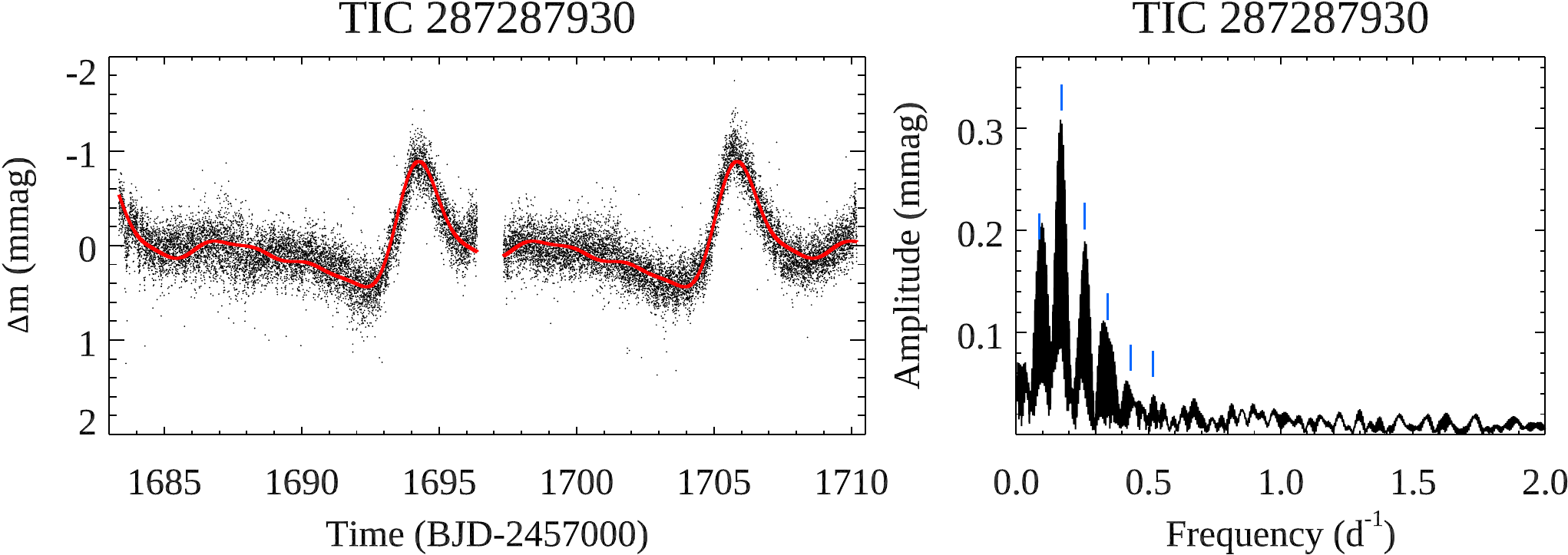}
\fivps{0.129\vsize}{0}{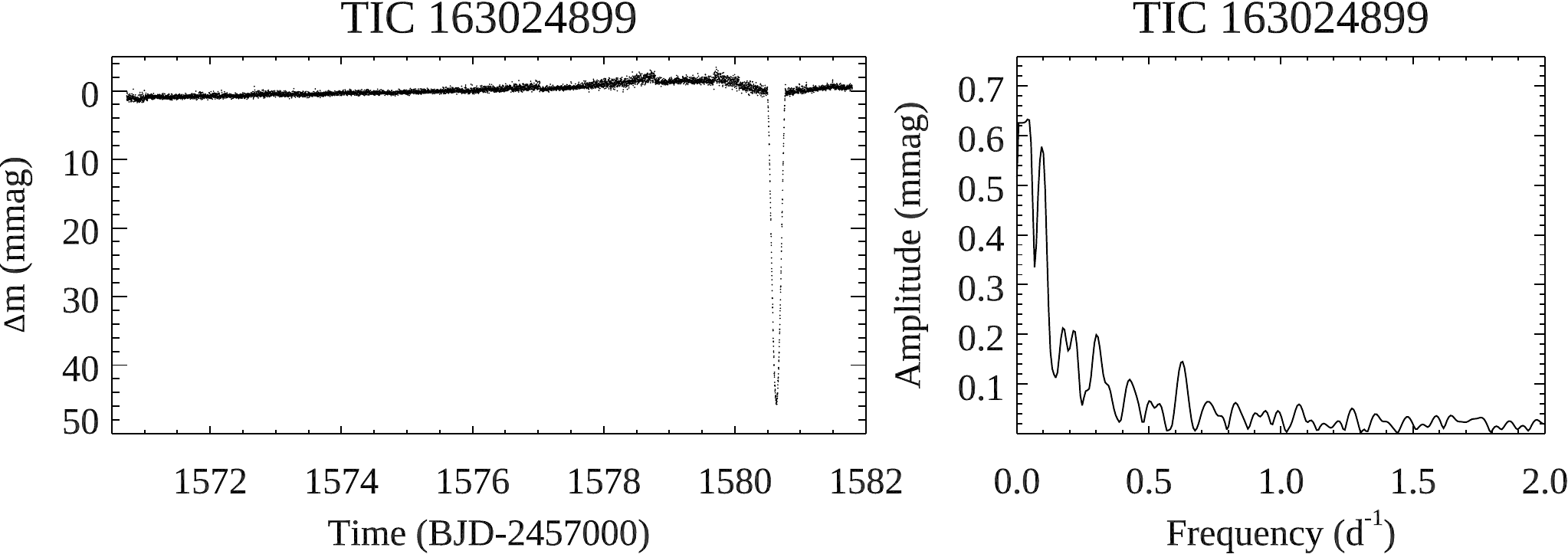}
\caption{Same as Fig.~\ref{fig:rot1}, but for stars with different types of variability related to binarity and/or companions. Periodograms of eclipsing binaries (all targets shown here except TIC 51961599 and 287287930) were computed excluding eclipses.}
\label{fig:binary}
\end{figure*}

TIC~245936817 (HD~29589, 93~Tau, Fig.~\ref{fig:puls}) is a well-established HgMn star \citep{kochukhov:2005b}, known to have a faint companion \citep{hubrig:2001a} and showing spectroscopic variability \citep{hubrig:2011b}. This star was found to be constant by \citet{paunzen:2013}. On the other hand, the TESS data reveals a rich spectrum of g-modes with periods between 0.27 and 0.75~d. It is plausible that the lowest frequency peak found in our study, corresponding to $P=1.567$~d, is due to stellar rotation.

TIC~391421890 (HD~93549, HR~4220, Fig.~\ref{fig:puls}) is part of an SB1 system with $P_{\rm orb}=5.4999$~d \citep{pourbaix:2004}. The TESS periodogram shows multiple peaks corresponding to periods in the 1.45--5.65~d range. This variability may be due to a combination of g-modes and tidally induced oscillations. The strongest signal at $P_{\rm rot}=0.874$~d is accompanied by a harmonic and is almost certainly due to rotation. This star has a moderate $v_{\rm e}\sin i$ of 67~\kms\ \citep{zorec:2012}, implying that the inclination angle has to be relatively small to be compatible with the radius typical of a late-B star.

TIC~219480018 (HD~145389, $\varphi$~Her, Fig.~\ref{fig:puls}) is a long-period ($P_{\rm orb}=565$~d) SB2 system \citep{zavala:2007} with a narrow-line primary and broad-line secondary. A relatively short rotation period of 3.708~d established by our analysis of the TESS light curve is compatible with $v_{\rm e}\sin i=8$~\kms\ measured for the primary if its rotational axis is aligned with the orbital axis. The latter is known to be nearly perpendicular to the plane of the sky \citep[$i_{\rm orb}=12\degr$,][]{zavala:2007}. In addition to the rotational modulation signal, there is a cluster of p-mode frequencies around 20~d$^{-1}$. This p-mode pulsational variability can originate either in the primary or secondary.

TIC~198388195 (HD~156127, Fig.~\ref{fig:puls}) is a recently identified and poorly studied HgMn star that was observed in 12 TESS sectors. Owing to this abundance of data, rotational variability with $P_{\rm rot}=2.459$~d and an amplitude of only 0.11~mmag is firmly established. In addition, this star shows a puzzling group of frequencies at 1.45~d$^{-1}$ as well as second and third harmonics. The main frequency is a quadruplet or, possibly, a quintuplet with the strongest component at $P=0.6889$~d. The quadruplet components are separated by 0.005365~d$^{-1}$, implying a modulation with a 186~d period. We have no explanation of this unique variability.

TIC~27697099 (HD~171301, HR~6968, Fig.~\ref{fig:puls}) is another HgMn star that, similar to 93~Tau, exhibits multi-periodic g-mode pulsations. The 3.104~d period is likely to be a signature of rotational modulation. At least three significant shorter periods, ranging between 0.63 and 0.95~d, are present in the TESS light curve of this star.

TIC~383676357 (HD~172728, HR~7018, Fig.~\ref{fig:puls}) rotates with $P_{\rm rot}=4.498$~d and shows at least five g-mode pulsation signals with periods in the range of 1.26--2.76~d. This star is not known to be a binary and, thus, represents yet another example of an HgMn star with significant pulsational variability.

TIC~392569763 (HD~173524, 46 Dra, Fig.~\ref{fig:puls}) is an SB2 system with nearly identical components and $P_{\rm orb}=9.811$~d \citep{catanzaro:2004}. Both components have chemical anomalies characteristic of HgMn stars (\citealt*{adelman:1998d}; \citealt{tsymbal:1998}). The TESS data for this system shows a clear rotational modulation with a period of 9.827~d. Considering the high precision of the orbital radial velocity and photometric period measurements, the discrepancy with $P_{\rm orb}$ is statistically significant. Nevertheless, this variability is still compatible with rotational modulation produced by one of the (nearly) synchronously rotating HgMn components. In addition, there are three higher frequency signals corresponding to periods in the range of 0.72--1.45~d, indicating that one or both of the components of 46~Dra are g-mode pulsators.

\subsection{Binarity}
\label{ssec:bin}

In this section we summarise time series analysis results for the six targets with variability related to their binary nature. This includes four eclipsing binaries as well as two objects showing heartbeat variation. The light curves and periodograms of these stars are shown in Fig.~\ref{fig:binary}.

\smallskip

TIC~51961599 (HD~5408, HR~266, Fig.~\ref{fig:binary}) is a quadruple system showing the presence of three components in its spectra \citep{cole:1992}. The broad-lined B7IV primary is orbited by a short-period ($P_{\rm orb}=4.241$~d), eccentric ($e=0.415$) binary comprising narrow-lined B9IV and A1V components. The B9IV object is classified as an HgMn star. The TESS light curve of this system reveals a complex periodic variability with up to eleven harmonics of the 4.241~d period. This signal is thus coming from the tight binary containing the HgMn star, but is inconsistent with a typical eclipse light curve or rotational modulation. Instead, this system appears to show the `heartbeat' variation caused by dynamic tidal distortions and tidally induced pulsations occurring in close eccentric binary systems, such as those discussed by \citet{thompson:2012}. In particular, the TESS light curve of HR~266 bears a striking resemblance to the variation of KIC 5034333 discussed in that paper.

TIC~2236015 (HD~34364, AR Aur, Fig.~\ref{fig:binary}) is a well-known double-line eclipsing binary with nearly identical components \citep[e.g.][]{folsom:2010,hubrig:2012} and a low-mass unseen tertiary on a wide orbit. The binary has an orbital period of 4.135~d \citep{albayrak:2003} with the components rotating synchronously with the orbital motion. The primary shows spectral peculiarities typical of HgMn stars whereas the secondary appears to be a weak Am star \citep{folsom:2010}. Spectral variability in the lines of several heavy elements was detected for the primary (\citealt*{zverko:1997}; \citealt{hubrig:2006,hubrig:2010,folsom:2010}). The TESS light curve of AR Aur shows prominent eclipses, from which we derive an orbital period of 4.134~d consistent with the previous determination. After eclipses are removed, no clear periodic variation can be identified in the residual photometry.

TIC~2776520 (HD~34923, Fig.~\ref{fig:binary}) shows three peaks in its TESS periodogram, indicating variation with periods from 0.252 to 0.749~d. This variability is likely to be caused by g-mode pulsations. In addition, a single low-amplitude transit event is observed at BJD=24571837.8, suggesting that HD~34923 is orbited by a low-mass companion.

TIC~436103335 (HD~36881, HR~1883, Fig.~\ref{fig:binary}) is known to be a long-period ($P_{\rm orb}=1857$~d), single-line spectroscopic binary \citep{dworetsky:1982}. According to our analysis, the HgMn primary exhibits a rotational variation with $P_{\rm rot}=1.541$~d. Furthermore, the TESS light curve shows evidence of two groups of eclipses, distinguished by different eclipse shapes. From both groups we measured a consistent period of 7.86~d. This leads to the conclusion that the secondary of HR~1883 is an eclipsing binary, consistent with one of the hypotheses considered by \citet{dworetsky:1982} to explain the unusually large mass function derived from the SB1 orbit.

TIC~287287930 (HD~89822, HR~4072, Fig.~\ref{fig:binary}) is an SB2 system with an HgMn primary and an Am secondary \citep{adelman:1994c}. The orbital period is $P_{\rm orb}=11.579$~d \citep{pourbaix:2004}. The TESS light curve of this star displays a coherent variation with $P=11.581$~d and a complex shape. The periodogram exhibits up to six harmonics, which is considerably more complex than the signal observed for any rotationally modulated TESS light curves of HgMn stars. Taking into account that the orbit of this system is eccentric ($e=0.26$), it is plausible that the observed behaviour of HR~4072 is another manifestation of the heartbeat variability phenomenon. 

TIC~163024899 (HD~99803, HR~4423, Fig.~\ref{fig:binary}) shows a single prominent eclipse in its TESS light curve. HR~4423 is known to be a visual binary with a wide enough separation for the components to be resolved by ground-based observations. The secondary is a 2.5~mag fainter Am star \citep[HD~99803B, HIP 56001][]{corbally:1984} which is also contributing to the TESS light curve. The Bright Star Catalogue \citep{hoffleit:1991} entry of HR 4423 mentions a spectroscopic binary nature of the primary. Thus, it is possible that the HgMn component of HR~4423 is a previously unrecognised eclipsing binary.

\section{Discussion}
\label{sec:disc}

We have carried out a systematic search of photometric variability among HgMn stars observed by the TESS satellite at 2-min cadence. Light curves of 65 stars classified as HgMn in the literature and observed during cycles 1 and 2 of the mission were examined. Variability consistent with rotational modulation was identified in 55 stars. One of these objects (HD\,168733, TIC\,323999777) has conflicting peculiarity classifications in the literature and is likely to be a magnetic Bp star. It is distinguished from the rest of the sample by its high variability amplitude. Excluding this target, we derive an 84 per cent (54 out of 64 stars) incidence of rotational modulation in HgMn stars. This shows that such variability is nearly ubiquitous among this class of CP stars and that the majority of them possess low-contrast chemical spots on their surfaces.

Rotational periods measured in our study range from 0.34 to 12.4~d. The median period is 3.77~d. Our period determinations generally agree with the rotational periods deduced in studies of line profile variability using high-resolution time series spectra. This is the case, for instance, for HD\,358, HD\,11753, HD\,32964, HD\,53244, and HD\,221507. This consistency confirms that both types of variability are linked to the same underlying surface inhomogeneity phenomenon.

Six rapid rotators with periods below 1.2~d were found in our survey (HD\,14228, HD\,37519, HD\,93549, HD\,101391, HD\,145842, and HD\,169027). Such a rapid rotation is not expected for HgMn stars. It was estimated that meridional circulation currents associated with equatorial rotational velocity in excess of $\approx$\,90~\kms\ would destroy the build up of chemical elements by atomic diffusion \citep{michaud:1982}. This equatorial rotation speed corresponds to a period of 1.4--1.6~d assuming a stellar radius of 2.5--2.9 $R_\odot$ as appropriate for B7--B9 main sequence stars \citep{pecaut:2013}. Our results challenge the hypothesis of a sharp rotational cutoff of the HgMn phenomenon, but should be verified by confirmation of the HgMn nature of each of these rapid rotators based on high-resolution spectroscopic observations and detailed abundance analysis. This confirmation has already been obtained for HD\,169027 by \citet{woolf:1999} and for HD\,93549 and HD\,145842 by \citet{gonzalez:2021}.

The star HD\,14228 (TIC\,354671857, $\varphi$~Eri) is particularly puzzling. This is an ultra-fast rotator with $P_{\rm rot}=0.344$~d, $v_{\rm e}\sin i=240$~\kms\ and a strongly distorted surface \citep{van-belle:2012}. Can this really be an HgMn star? Even if it is not confirmed to be a member of this CP-star class, the implications of its prominent rotational variability are still very intriguing. This would mean that at least some normal late-B stars exhibit rotational modulation similar to the behaviour of HgMn stars and hint that non-CP stars also possess a non-uniform surface structure \citep{balona:2019b}.

For clarifying if the possible rotational periods found in this study are physical,  we have generated the Hertzsprung-Russell diagram of the target star sample. For the $\log T_\mathrm{eff}$ calibrations, we used the Johnson $UBV$, Geneva 7-colour, and Str{\"o}mgren-Crawford $uvby\beta$ photometric systems. The individual values were taken from \citet{paunzen:2015} and the General Catalogue of Photometric data\footnote{\url{http://gcpd.physics.muni.cz/}}. \citet{netopil:2008} introduced calibrations for CP stars using individual corrections for the temperature domain and the CP subclass, which are summarised in their Table 2. Here we follow their approach. For the derivation of  the final effective temperatures, all calibrated values were averaged and the standard deviations were calculated. 
(Table~\ref{tab:params}).

To calibrate the luminosity, the geometric distances from \citet{bailer-jones:2021} on the basis of the {\it Gaia} EDR3 \citep{riello:2021} were taken. For HD 358 and HD 106625 the distances from the {\it Hipparcos} mission \citep{van-leeuwen:2007} were applied. For the bolometric correction, the calibration published by \citet{flower:1996} was applied. It is independent of the luminosity and the metallicity. For the bolometric magnitude of the Sun, a value of 4.75\,mag \citep{cayrel-de-strobel:1996} was used.
 
\begin{figure}
\centering
\figps{\hsize}{0}{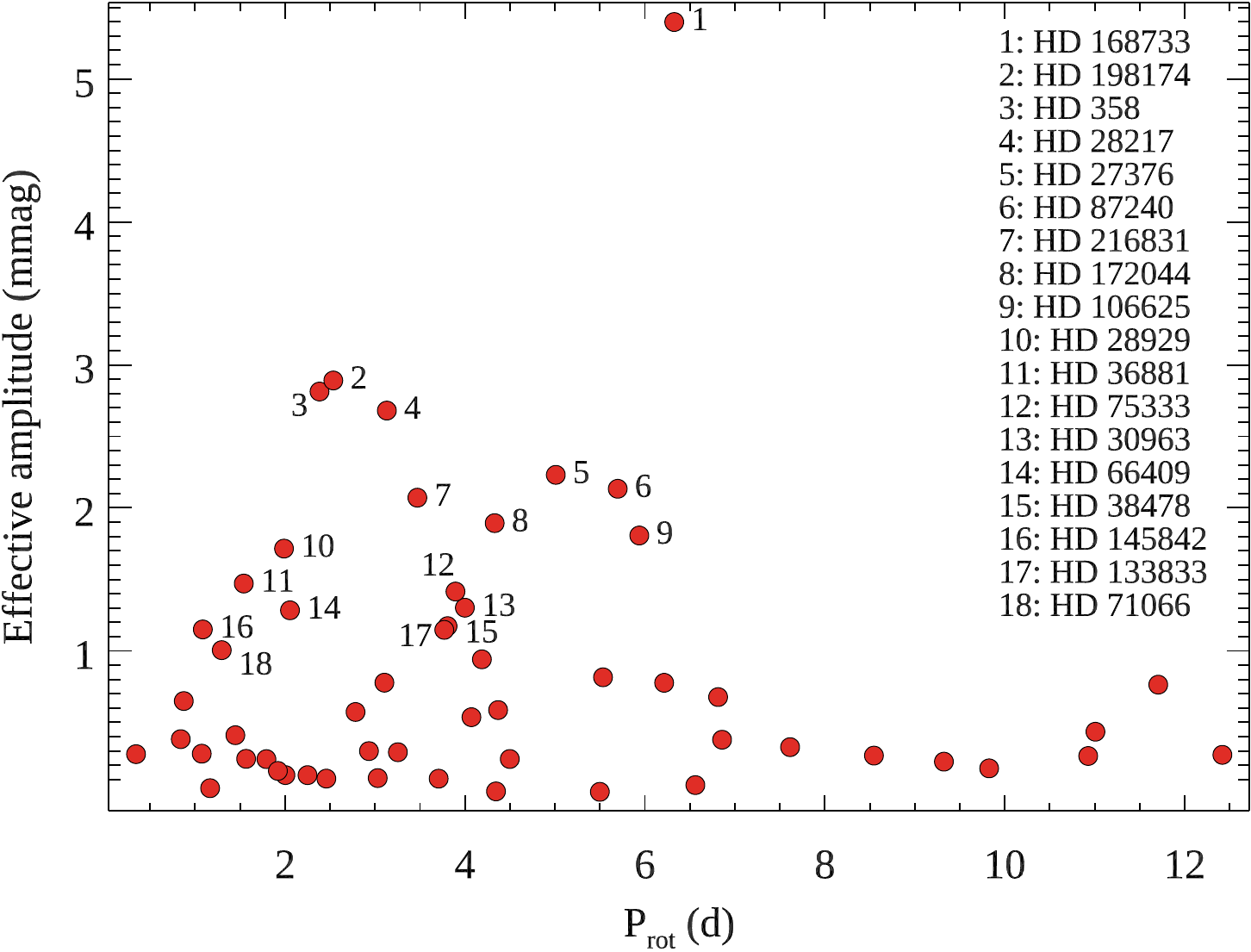}
\caption{Effective amplitude of the rotational modulation as a function of rotation period. Stars with the largest amplitude of photometric variability are identified in the legend.}
\label{fig:res1}
\end{figure}

The commonly employed dereddening procedures in the Str{\"o}mgren-Crawford $uvby\beta$ photometric system published by \citet*{napiwotzki:1993} was supplemented by the reddening map\footnote{\url{http://argonaut.skymaps.info/}} from \citet{green:2019}, which is based on {\it Gaia} parallaxes and stellar photometry from Pan-STARRS 1 and 2MASS. For the calibrations of the different photometric systems, we used the following relations \citep*{paunzen:2006}:
\begin{equation}
A_V = 3.1E(B-V) = 4.3E(b-y) = 4.95E(B2-V1).
\end{equation}
An error value of 0.01\,mag was adopted for all objects.

To derive equatorial velocities $v_\mathrm{eq}$ we used the following equation \citep{netopil:2017}:
\begin{equation}
v_\mathrm{eq} = 50.579R/P_\mathrm{rot},
\end{equation}
where $v_\mathrm{eq}$ is in \kms, $R$ is in solar units and $P_\mathrm{rot}$ is in days. This relation requires us to
estimate the individual stellar radii. For this we employed the Stellar Isochrone Fitting Tool\footnote{\url{https://github.com/Johaney-s/StIFT}} using the stellar isochrones by \citet{bressan:2012} for [Z]\,=\,0.014. To finally get the ratio of the equatorial to the critical equatorial ($v_\mathrm{crit}$), the latter were taken from \citet{georgy:2013} for the same metallicity as the isochrones.

\citet{netopil:2017} have shown that the majority of the magnetic CP stars show a ratio $v_\mathrm{eq}$/$v_\mathrm{crit}$ of less than 30 per cent with a peak at about 5 per cent. Only a few stars in their sample have values as large as 70 per cent. From our target star sample, 48 out of 55 objects (87 per cent) the ratio falls below 30 per cent. Only one object (HD 36881) has a ratio larger than one, indicating that its period is not due to rotation. Two other stars, HD 14228 and HD 37519, are rotating close to critical velocity ($v_\mathrm{eq}$/$v_\mathrm{crit}$ is 87 and 97 per cent respectively), which is unusual for CP stars. 

In order to get further insights into the statistical characteristics of the rotational variability of HgMn stars, we computed the total effective amplitude by adding in quadrature the amplitudes of all detected harmonics of the rotational frequency \citep{mikulasek:2007}. This effective amplitude is found to be in the range from 0.014 and 2.89 mmag. This parameter is shown as a function of rotational period in Fig.~\ref{fig:res1}. The object with the largest amplitude in this plot, HD\,168733, is likely to be a magnetic Bp star. The bona fide HgMn stars with the highest photometric variability are HD\,358, HD\,28217, and HD\,198174. There is no obvious dependence of amplitude on rotational period, but stars with the highest amplitude in the TESS band tend to cluster between 2 and 6-d periods.

\begin{figure}
\centering
\figps{\hsize}{0}{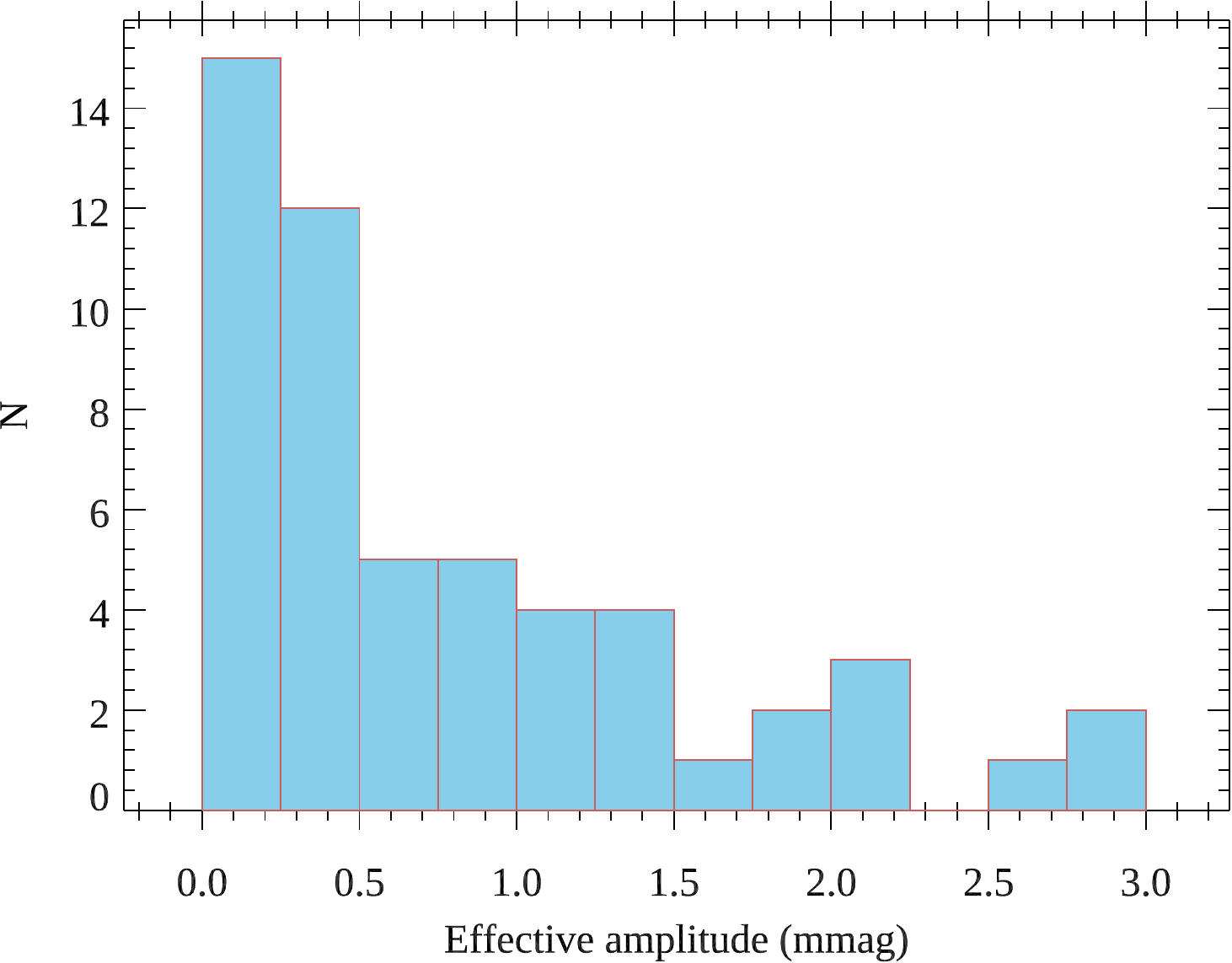}
\caption{Distribution of the effective variability amplitude for bona fide HgMn stars showing photometric rotational modulation.}
\label{fig:res2}
\end{figure}

The distribution of effective rotational modulation amplitude (Fig.~\ref{fig:res2}) is exponential, with half of the sample stars having an amplitude below 0.54 mmag. This variability is challenging to detect from the ground, which explains why HgMn stars were considered photometrically constant before the advent of space photometry missions \citep[e.g.][]{adelman:1994b,adelman:1998b,adelman:2002a}. From the comparison of the effective amplitude with the amplitudes of individual harmonics of rotational frequency we can conclude that the shapes of photometric phase curves of HgMn stars are mostly sinusoidal. Only 22 stars (41 per cent of the sample exhibiting rotationally modulated light curves) show deviation from a sinusoidal variability at the level of 30 per cent or more as judged by the difference between the effective amplitude and that of the first harmonic. For 6 stars (11 per cent) the second harmonic dominates over the first one, indicating a double-wave photometric variation. These morphological characteristics of HgMn light curves can be compared to those of magnetic CP (mCP) stars as summarised by \citet{jagelka:2019}. These authors found that 67 per cent of mCP stars exhibit single-wave photometric phase curves and 29 per cent have some form of double-wave curves. The latter is somewhat higher than the fraction of double-wave stars derived here, suggesting that HgMn stars have simpler spots distributions than mCP stars. The observed simple photometric phase behaviour of HgMn stars is incompatible with the hypothesis that their surfaces are covered by a complex network of small magnetic spots discussed in the context of magnetic studies of these stars \citep{hubrig:2012,kochukhov:2013a}.

Although our investigation has focused on the analysis of rotational modulation, TESS observations allowed us to assess other types of stellar variability. We found 8 HgMn stars (7 objects discussed in Sect.~\ref{ssec:puls} and HD\,34923) with multi-periodic pulsations which have amplitude higher or comparable to that of rotational modulation. The 3 stars with 2--5 mmag g-mode oscillations (HD\,29589, HD\,93549, HD\,171301) are particularly interesting for follow up studies and detailed asteroseismic analyses. All three objects have been studied with high-resolution spectra (\citealt{hubrig:2011b}; \citealt{gonzalez:2021}; \citealt*{adelman:2017}) and their HgMn classification is indisputable. Two other well-studied HgMn stars, HD\,172728 \citep*{adelman:2001c} and HD\,173524 \citep{adelman:1998d}, have the longest TESS data sets (12 and 10 TESS sectors respectively). Both stars exhibit g-mode pulsations with amplitudes below 1~mmag. Similar variability is observed in HD\,34923. A few other HgMn stars in our survey appear to show weak periodic signal consistent with g-mode pulsation (e.g. HD\,11921, 27376, 110073) in addition to stronger rotational modulation. It is not always clear which of these signals can be attributed to pulsations in HgMn stars and which are caused by blending by nearby faint objects. In any case, we find that high-amplitude g-mode pulsations are rare among HgMn stars, with about 10 per cent of the targets showing this type of variability. The single candidate p-mode pulsator found in this study, HD\,145389 ($\varphi$~Her), is the primary component in a binary system with a late-A star \citep{zavala:2007}. It is likely that the high-frequency pulsational signal is coming from the secondary, which is a $\delta$~Sct star. Thus, we find no definite cases of high-frequency oscillation in an HgMn star.

Binarity is another phenomenon responsible for photometric variability of HgMn stars. This includes well-understood eclipse and transit variation in binary systems seen edge-on. In this study we reported discovery of a possible eclipsing binary with an HgMn component (HD\,99803) and another HgMn star (HD\,36881) orbiting an eclipsing binary in a hierarchical triple system. Furthermore, the precision of space photometric data enabled detection of new types of variability related to binarity other than eclipses. Here we identified two HgMn stars, HD\,5408 and HD\,89822, with heartbeat photometric variability caused by the tidal distortion and tidally induced oscillations excited in the components of an eccentric close binary \citep[e.g.][]{thompson:2012}. Only one case of heartbeat variability in an HgMn star, corresponding to the eclipsing system V680~Mon, was previously known \citep{paunzen:2021a}.

HgMn stars identified in this work as rotational variables are good candidates to search for magnetic fields. It has already been shown that strong magnetic fields like those found in Bp stars are not present in HgMn stars. However, ultra-weak fields could be present. Detecting ultra-weak fields requires very deep spectropolarimetric observations, with an error bar on the longitudinal field measurement of the order of 0.1~G. Such a sensitivity can only be reached with current instrumentation for stars satisfying specific criteria: very bright stars, relatively cool stars (as the spectral type impacts the number and depth of lines that can be used in the calculations), low $v\sin i$ (to increase the amplitude of the Zeeman signature in the spectral line), long rotation period (to accumulate more easily measurements in a given rotation phase bin), and no additional variability (such as pulsations or binarity that would blur the magnetic signal). These criteria have already been successfully used to detect ultra-weak fields in Am stars \citep{blazere:2016}. Moreover, it is expected that magnetic field strength decreases at the surface of stars as they evolve, due to magnetic flux conservation \citep{neiner:2018}. Therefore magnetic field detections require a lower precision for main sequence stars than for very evolved stars. As a consequence we identify HD\,358 ($\alpha$\,And), HD\,33904 ($\mu$\,Lep), and HD\,106625 ($\gamma$\,Crv) as the best candidates to search for ultra-weak fields in HgMn stars. These targets will be the goal of a future spectropolarimetric study.

\section*{Acknowledgements}
O.\,Kochukhov acknowledges support by the Swedish Research Council, the Royal Swedish Academy of Sciences and the Swedish National Space Agency.
V.\,Khalack and O.\,Kobzar acknowledge support from the Natural Sciences and Engineering Research Council of Canada (NSERC) and from the Facult\'{e} des \'{E}tudes Sup\'{e}rieures et de la Recherch de l'Universit\'{e} de Moncton.
E.\,Paunzen acknowledges support by the Erasmus+ programme of the European Union under grant number 2020-1-CZ01-KA203-078200.
A.\,David-Uraz acknowledges support by NASA under award number 80GSFC21M0002.
This research has made extensive use of the SIMBAD database, operated at CDS, Strasbourg, France.
This work has made use of data from the European Space Agency (ESA) mission
{\it Gaia} (\url{https://www.cosmos.esa.int/gaia}), processed by the {\it Gaia}
Data Processing and Analysis Consortium (DPAC,
\url{https://www.cosmos.esa.int/web/gaia/dpac/consortium}). Funding for the DPAC
has been provided by national institutions, in particular the institutions
participating in the {\it Gaia} Multilateral Agreement.

\section*{Data availability}
The TESS light curve data underlying this article can be obtained from the Mikulski Archive for Space Telescopes (MAST).


\begin{thebibliography}{}
\makeatletter
\relax
\def\mn@urlcharsother{\let\do\@makeother \do\$\do\&\do\#\do\^\do\_\do\%\do\~}
\def\mn@doi{\begingroup\mn@urlcharsother \@ifnextchar [ {\mn@doi@}
  {\mn@doi@[]}}
\def\mn@doi@[#1]#2{\def\@tempa{#1}\ifx\@tempa\@empty \href
  {http://dx.doi.org/#2} {doi:#2}\else \href {http://dx.doi.org/#2} {#1}\fi
  \endgroup}
\def\mn@eprint#1#2{\mn@eprint@#1:#2::\@nil}
\def\mn@eprint@arXiv#1{\href {http://arxiv.org/abs/#1} {{\tt arXiv:#1}}}
\def\mn@eprint@dblp#1{\href {http://dblp.uni-trier.de/rec/bibtex/#1.xml}
  {dblp:#1}}
\def\mn@eprint@#1:#2:#3:#4\@nil{\def\@tempa {#1}\def\@tempb {#2}\def\@tempc
  {#3}\ifx \@tempc \@empty \let \@tempc \@tempb \let \@tempb \@tempa \fi \ifx
  \@tempb \@empty \def\@tempb {arXiv}\fi \@ifundefined
  {mn@eprint@\@tempb}{\@tempb:\@tempc}{\expandafter \expandafter \csname
  mn@eprint@\@tempb\endcsname \expandafter{\@tempc}}}

\bibitem[\protect\citeauthoryear{{Abt}, {Levato}  \& {Grosso}}{{Abt}
  et~al.}{2002}]{abt:2002}
{Abt} H.~A.,  {Levato} H.,   {Grosso} M.,  2002, \mn@doi [\apj]
  {10.1086/340590}, \href {http://adsabs.harvard.edu/abs/2002ApJ...573..359A}
  {573, 359}

\bibitem[\protect\citeauthoryear{{Adelman}}{{Adelman}}{1994b}]{adelman:1994b}
{Adelman} S.~J.,  1994b, \mnras, \href
  {http://adsabs.harvard.edu/abs/1994MNRAS.266...97A} {266, 97}

\bibitem[\protect\citeauthoryear{{Adelman}}{{Adelman}}{1994a}]{adelman:1994c}
{Adelman} S.~J.,  1994a, \mn@doi [\mnras] {10.1093/mnras/266.1.97}, \href
  {https://ui.adsabs.harvard.edu/abs/1994MNRAS.266...97A} {266, 97}

\bibitem[\protect\citeauthoryear{{Adelman}}{{Adelman}}{1998}]{adelman:1998b}
{Adelman} S.~J.,  1998, \mn@doi [\aaps] {10.1051/aas:1998361}, \href
  {http://esoads.eso.org/abs/1998A%26AS..132...93A} {132, 93}

\bibitem[\protect\citeauthoryear{{Adelman} \& {Brunhouse}}{{Adelman} \&
  {Brunhouse}}{1998}]{adelman:1998c}
{Adelman} S.~J.,  {Brunhouse} E.~F.,  1998, \mn@doi [\pasp] {10.1086/316250},
  \href {https://ui.adsabs.harvard.edu/abs/1998PASP..110.1304A} {110, 1304}

\bibitem[\protect\citeauthoryear{{Adelman} \& {Meadows}}{{Adelman} \&
  {Meadows}}{2002}]{adelman:2002a}
{Adelman} S.~J.,  {Meadows} S.~A.,  2002, \mn@doi [\aap]
  {10.1051/0004-6361:20020761}, \href
  {https://ui.adsabs.harvard.edu/abs/2002A&A...390.1023A} {390, 1023}

\bibitem[\protect\citeauthoryear{{Adelman}, {Ryabchikova}  \&
  {Davydova}}{{Adelman} et~al.}{1998}]{adelman:1998d}
{Adelman} S.~J.,  {Ryabchikova} T.~A.,   {Davydova} E.~S.,  1998, \mn@doi
  [\mnras] {10.1046/j.1365-8711.1998.01344.x}, \href
  {https://ui.adsabs.harvard.edu/abs/1998MNRAS.297....1A} {297, 1}

\bibitem[\protect\citeauthoryear{{Adelman}, {Gulliver}  \& {Rayle}}{{Adelman}
  et~al.}{2001}]{adelman:2001c}
{Adelman} S.~J.,  {Gulliver} A.~F.,   {Rayle} K.~E.,  2001, \mn@doi [\aap]
  {10.1051/0004-6361:20010007}, \href
  {https://ui.adsabs.harvard.edu/abs/2001A&A...367..597A} {367, 597}

\bibitem[\protect\citeauthoryear{{Adelman}, {Gulliver}, {Kochukhov}  \&
  {Ryabchikova}}{{Adelman} et~al.}{2002}]{adelman:2002}
{Adelman} S.~J.,  {Gulliver} A.~F.,  {Kochukhov} O.~P.,   {Ryabchikova} T.~A.,
  2002, \mn@doi [\apj] {10.1086/341140}, \href
  {http://adsabs.harvard.edu/abs/2002ApJ...575..449A} {575, 449}

\bibitem[\protect\citeauthoryear{{Adelman}, {Gulliver}  \& {Gucella}}{{Adelman}
  et~al.}{2017}]{adelman:2017}
{Adelman} S.~J.,  {Gulliver} A.~F.,   {Gucella} L.~J.,  2017, \mn@doi
  [Astronomische Nachrichten] {10.1002/asna.201613214}, \href
  {https://ui.adsabs.harvard.edu/abs/2017AN....338..584A} {338, 584}

\bibitem[\protect\citeauthoryear{{Aerts}, {Christensen-Dalsgaard}  \&
  {Kurtz}}{{Aerts} et~al.}{2010}]{aerts:2010}
{Aerts} C.,  {Christensen-Dalsgaard} J.,   {Kurtz} D.~W.,  2010,
  {Asteroseismology}.
Astronomy and Astrophysics Library, Springer-Verlag, Heidelberg

\bibitem[\protect\citeauthoryear{{Albayrak}, {Ak}  \& {Elmasli}}{{Albayrak}
  et~al.}{2003}]{albayrak:2003}
{Albayrak} B.,  {Ak} T.,   {Elmasli} A.,  2003, \mn@doi [Astronomische
  Nachrichten] {10.1002/asna.200310161}, \href
  {https://ui.adsabs.harvard.edu/abs/2003AN....324..523A} {324, 523}

\bibitem[\protect\citeauthoryear{{Alecian}}{{Alecian}}{2015}]{alecian:2015}
{Alecian} G.,  2015, \mn@doi [\mnras] {10.1093/mnras/stv2205}, \href
  {http://adsabs.harvard.edu/abs/2015MNRAS.454.3143A} {454, 3143}

\bibitem[\protect\citeauthoryear{{Alecian}, {Gebran}, {Auvergne}, {Richard},
  {Samadi}, {Weiss}  \& {Baglin}}{{Alecian} et~al.}{2009}]{alecian:2009a}
{Alecian} G.,  {Gebran} M.,  {Auvergne} M.,  {Richard} O.,  {Samadi} R.,
  {Weiss} W.~W.,   {Baglin} A.,  2009, \mn@doi [\aap]
  {10.1051/0004-6361/200911678}, \href
  {http://adsabs.harvard.edu/abs/2009A%26A...506...69A} {506, 69}

\bibitem[\protect\citeauthoryear{{Alecian}, {Stift}  \& {Dorfi}}{{Alecian}
  et~al.}{2011}]{alecian:2011}
{Alecian} G.,  {Stift} M.~J.,   {Dorfi} E.~A.,  2011, \mn@doi [\mnras]
  {10.1111/j.1365-2966.2011.19547.x}, \href
  {http://adsabs.harvard.edu/abs/2011MNRAS.418..986A} {418, 986}

\bibitem[\protect\citeauthoryear{{Alonso}, {L{\'o}pez-Garc{\'\i}a}, {Malaroda}
  \& {Leone}}{{Alonso} et~al.}{2003}]{alonso:2003}
{Alonso} M.~S.,  {L{\'o}pez-Garc{\'\i}a} Z.,  {Malaroda} S.,   {Leone} F.,
  2003, \mn@doi [\aap] {10.1051/0004-6361:20030222}, \href
  {https://ui.adsabs.harvard.edu/abs/2003A&A...402..331A} {402, 331}

\bibitem[\protect\citeauthoryear{{Auri{\`e}re} et~al.,}{{Auri{\`e}re}
  et~al.}{2010}]{auriere:2010a}
{Auri{\`e}re} M.,  et~al., 2010, \mn@doi [\aap] {10.1051/0004-6361/201014848},
  \href {http://adsabs.harvard.edu/abs/2010A%26A...523A..40A} {523, A40}

\bibitem[\protect\citeauthoryear{{Babel} \& {Michaud}}{{Babel} \&
  {Michaud}}{1991}]{babel:1991}
{Babel} J.,  {Michaud} G.,  1991, \mn@doi [\apj] {10.1086/169591}, \href
  {http://adsabs.harvard.edu/abs/1991ApJ...366..560B} {366, 560}

\bibitem[\protect\citeauthoryear{{Bagnulo}, {Landstreet}, {Fossati}  \&
  {Kochukhov}}{{Bagnulo} et~al.}{2012}]{bagnulo:2012}
{Bagnulo} S.,  {Landstreet} J.~D.,  {Fossati} L.,   {Kochukhov} O.,  2012,
  \mn@doi [\aap] {10.1051/0004-6361/201118098}, \href
  {http://adsabs.harvard.edu/abs/2012A%26A...538A.129B} {538, A129}

\bibitem[\protect\citeauthoryear{{Bagnulo}, {Fossati}, {Landstreet}  \&
  {Izzo}}{{Bagnulo} et~al.}{2015}]{bagnulo:2015}
{Bagnulo} S.,  {Fossati} L.,  {Landstreet} J.~D.,   {Izzo} C.,  2015, \mn@doi
  [\aap] {10.1051/0004-6361/201526497}, \href
  {https://ui.adsabs.harvard.edu/abs/2015A&A...583A.115B} {583, A115}

\bibitem[\protect\citeauthoryear{{Bailer-Jones}, {Rybizki}, {Fouesneau},
  {Demleitner}  \& {Andrae}}{{Bailer-Jones} et~al.}{2021}]{bailer-jones:2021}
{Bailer-Jones} C.~A.~L.,  {Rybizki} J.,  {Fouesneau} M.,  {Demleitner} M.,
  {Andrae} R.,  2021, \mn@doi [\aj] {10.3847/1538-3881/abd806}, \href
  {https://ui.adsabs.harvard.edu/abs/2021AJ....161..147B} {161, 147}

\bibitem[\protect\citeauthoryear{{Balona}}{{Balona}}{2019}]{balona:2019b}
{Balona} L.~A.,  2019, \mn@doi [\mnras] {10.1093/mnras/stz2808}, \href
  {https://ui.adsabs.harvard.edu/abs/2019MNRAS.490.2112B} {490, 2112}

\bibitem[\protect\citeauthoryear{{Balona} \& {Ozuyar}}{{Balona} \&
  {Ozuyar}}{2020}]{balona:2020}
{Balona} L.~A.,  {Ozuyar} D.,  2020, \mn@doi [\mnras] {10.1093/mnras/staa670},
  \href {https://ui.adsabs.harvard.edu/abs/2020MNRAS.493.5871B} {493, 5871}

\bibitem[\protect\citeauthoryear{{Balona} et~al.,}{{Balona}
  et~al.}{2011}]{balona:2011a}
{Balona} L.~A.,  et~al., 2011, \mn@doi [\mnras]
  {10.1111/j.1365-2966.2011.18311.x}, \href
  {https://ui.adsabs.harvard.edu/abs/2011MNRAS.413.2403B} {413, 2403}

\bibitem[\protect\citeauthoryear{{Balona}, {Baran}, {Daszy{\'n}ska-Daszkiewicz}
   \& {De Cat}}{{Balona} et~al.}{2015}]{balona:2015}
{Balona} L.~A.,  {Baran} A.~S.,  {Daszy{\'n}ska-Daszkiewicz} J.,   {De Cat} P.,
   2015, \mn@doi [\mnras] {10.1093/mnras/stv1017}, \href
  {https://ui.adsabs.harvard.edu/abs/2015MNRAS.451.1445B} {451, 1445}

\bibitem[\protect\citeauthoryear{{Balona} et~al.,}{{Balona}
  et~al.}{2019}]{balona:2019a}
{Balona} L.~A.,  et~al., 2019, \mn@doi [\mnras] {10.1093/mnras/stz586}, \href
  {https://ui.adsabs.harvard.edu/abs/2019MNRAS.485.3457B} {485, 3457}

\bibitem[\protect\citeauthoryear{{Blaz{\`e}re} et~al.,}{{Blaz{\`e}re}
  et~al.}{2016}]{blazere:2016}
{Blaz{\`e}re} A.,  et~al., 2016, \mn@doi [\aap] {10.1051/0004-6361/201527556},
  \href {https://ui.adsabs.harvard.edu/abs/2016A&A...586A..97B} {586, A97}

\bibitem[\protect\citeauthoryear{{Breger} et~al.,}{{Breger}
  et~al.}{1993}]{breger:1993}
{Breger} M.,  et~al., 1993, \aap, \href
  {https://ui.adsabs.harvard.edu/abs/1993A&A...271..482B} {271, 482}

\bibitem[\protect\citeauthoryear{{Bressan}, {Marigo}, {Girardi}, {Salasnich},
  {Dal Cero}, {Rubele}  \& {Nanni}}{{Bressan} et~al.}{2012}]{bressan:2012}
{Bressan} A.,  {Marigo} P.,  {Girardi} L.,  {Salasnich} B.,  {Dal Cero} C.,
  {Rubele} S.,   {Nanni} A.,  2012, \mn@doi [\mnras]
  {10.1111/j.1365-2966.2012.21948.x}, \href
  {https://ui.adsabs.harvard.edu/abs/2012MNRAS.427..127B} {427, 127}

\bibitem[\protect\citeauthoryear{{Briquet}, {Hubrig}, {De Cat}, {Aerts},
  {North}  \& {Sch{\"o}ller}}{{Briquet} et~al.}{2007}]{briquet:2007}
{Briquet} M.,  {Hubrig} S.,  {De Cat} P.,  {Aerts} C.,  {North} P.,
  {Sch{\"o}ller} M.,  2007, \mn@doi [\aap] {10.1051/0004-6361:20066940}, \href
  {http://adsabs.harvard.edu/abs/2007A%26A...466..269B} {466, 269}

\bibitem[\protect\citeauthoryear{{Briquet}, {Korhonen}, {Gonz{\'a}lez},
  {Hubrig}  \& {Hackman}}{{Briquet} et~al.}{2010}]{briquet:2010}
{Briquet} M.,  {Korhonen} H.,  {Gonz{\'a}lez} J.~F.,  {Hubrig} S.,   {Hackman}
  T.,  2010, \mn@doi [\aap] {10.1051/0004-6361/200913775}, \href
  {http://adsabs.harvard.edu/abs/2010A%26A...511A..71B} {511, A71}

\bibitem[\protect\citeauthoryear{{Burssens} et~al.,}{{Burssens}
  et~al.}{2020}]{burssens:2020}
{Burssens} S.,  et~al., 2020, \mn@doi [\aap] {10.1051/0004-6361/202037700},
  \href {https://ui.adsabs.harvard.edu/abs/2020A&A...639A..81B} {639, A81}

\bibitem[\protect\citeauthoryear{{Buysschaert}, {Neiner}, {Briquet}  \&
  {Aerts}}{{Buysschaert} et~al.}{2017}]{buysschaert:2017}
{Buysschaert} B.,  {Neiner} C.,  {Briquet} M.,   {Aerts} C.,  2017, \mn@doi
  [\aap] {10.1051/0004-6361/201731012}, \href
  {https://ui.adsabs.harvard.edu/abs/2017A&A...605A.104B} {605, A104}

\bibitem[\protect\citeauthoryear{{Castelli} \& {Hubrig}}{{Castelli} \&
  {Hubrig}}{2004}]{castelli:2004}
{Castelli} F.,  {Hubrig} S.,  2004, \mn@doi [\aap]
  {10.1051/0004-6361:20041011}, \href
  {http://cdsads.u-strasbg.fr/abs/2004A%26A...425..263C} {425, 263}

\bibitem[\protect\citeauthoryear{{Catanzaro} \& {Leto}}{{Catanzaro} \&
  {Leto}}{2004}]{catanzaro:2004}
{Catanzaro} G.,  {Leto} P.,  2004, \mn@doi [\aap] {10.1051/0004-6361:20034445},
  \href {http://cdsads.u-strasbg.fr/abs/2004A%26A...416..661C} {416, 661}

\bibitem[\protect\citeauthoryear{{Catanzaro}, {Frasca}, {Molenda-{\.Z}akowicz}
  \& {Marilli}}{{Catanzaro} et~al.}{2010}]{catanzaro:2010}
{Catanzaro} G.,  {Frasca} A.,  {Molenda-{\.Z}akowicz} J.,   {Marilli} E.,
  2010, \mn@doi [\aap] {10.1051/0004-6361/201014189}, \href
  {https://ui.adsabs.harvard.edu/abs/2010A&A...517A...3C} {517, A3}

\bibitem[\protect\citeauthoryear{{Catanzaro}, {Giarrusso}, {Leone}, {Munari},
  {Scalia}, {Sparacello}  \& {Scuderi}}{{Catanzaro}
  et~al.}{2016}]{catanzaro:2016}
{Catanzaro} G.,  {Giarrusso} M.,  {Leone} F.,  {Munari} M.,  {Scalia} C.,
  {Sparacello} E.,   {Scuderi} S.,  2016, \mn@doi [\mnras]
  {10.1093/mnras/stw923}, \href
  {https://ui.adsabs.harvard.edu/abs/2016MNRAS.460.1999C} {460, 1999}

\bibitem[\protect\citeauthoryear{{Catanzaro}, {Giarrusso}, {Munari}  \&
  {Leone}}{{Catanzaro} et~al.}{2020}]{catanzaro:2020}
{Catanzaro} G.,  {Giarrusso} M.,  {Munari} M.,   {Leone} F.,  2020, \mn@doi
  [\mnras] {10.1093/mnras/staa3108}, \href
  {https://ui.adsabs.harvard.edu/abs/2020MNRAS.499.3720C} {499, 3720}

\bibitem[\protect\citeauthoryear{{Cayrel de Strobel}}{{Cayrel de
  Strobel}}{1996}]{cayrel-de-strobel:1996}
{Cayrel de Strobel} G.,  1996, \mn@doi [\aapr] {10.1007/s001590050006}, \href
  {https://ui.adsabs.harvard.edu/abs/1996A&ARv...7..243C} {7, 243}

\bibitem[\protect\citeauthoryear{{Chojnowski}, {Hubrig}, {Hasselquist},
  {Beaton}, {Majewski}, {Garc{\'\i}a-Hern{\'a}ndez}  \&
  {DeColibus}}{{Chojnowski} et~al.}{2020}]{chojnowski:2020}
{Chojnowski} S.~D.,  {Hubrig} S.,  {Hasselquist} S.,  {Beaton} R.~L.,
  {Majewski} S.~R.,  {Garc{\'\i}a-Hern{\'a}ndez} D.~A.,   {DeColibus} D.,
  2020, \mn@doi [\mnras] {10.1093/mnras/staa1527}, \href
  {https://ui.adsabs.harvard.edu/abs/2020MNRAS.496..832C} {496, 832}

\bibitem[\protect\citeauthoryear{{Cole}, {Fekel}, {Hartkopf}, {McAlister}  \&
  {Tomkin}}{{Cole} et~al.}{1992}]{cole:1992}
{Cole} W.~A.,  {Fekel} F.~C.,  {Hartkopf} W.~I.,  {McAlister} H.~A.,   {Tomkin}
  J.,  1992, \mn@doi [\aj] {10.1086/116149}, \href
  {https://ui.adsabs.harvard.edu/abs/1992AJ....103.1357C} {103, 1357}

\bibitem[\protect\citeauthoryear{{Collado} \&
  {L{\'o}pez-Garc{\'\i}a}}{{Collado} \&
  {L{\'o}pez-Garc{\'\i}a}}{2009}]{collado:2009}
{Collado} A.,  {L{\'o}pez-Garc{\'\i}a} Z.,  2009, \rmxaa, \href
  {https://ui.adsabs.harvard.edu/abs/2009RMxAA..45...95C} {45, 95}

\bibitem[\protect\citeauthoryear{{Corbally}}{{Corbally}}{1984}]{corbally:1984}
{Corbally} C.~J.,  1984, \mn@doi [\apjs] {10.1086/190973}, \href
  {https://ui.adsabs.harvard.edu/abs/1984ApJS...55..657C} {55, 657}

\bibitem[\protect\citeauthoryear{{Cowley}, {Hubrig}, {Palmeri}, {Quinet},
  {Bi{\'e}mont}, {Wahlgren}, {Sch{\"u}tz}  \& {Gonz{\'a}lez}}{{Cowley}
  et~al.}{2010}]{cowley:2010}
{Cowley} C.~R.,  {Hubrig} S.,  {Palmeri} P.,  {Quinet} P.,  {Bi{\'e}mont}
  {\'E}.,  {Wahlgren} G.~M.,  {Sch{\"u}tz} O.,   {Gonz{\'a}lez} J.~F.,  2010,
  \mn@doi [\mnras] {10.1111/j.1365-2966.2010.16529.x}, \href
  {http://cdsads.u-strasbg.fr/abs/2010MNRAS.405.1271C} {405, 1271}

\bibitem[\protect\citeauthoryear{{Cucchiaro}, {Macau-Hercot}, {Jaschek}  \&
  {Jaschek}}{{Cucchiaro} et~al.}{1977}]{cucchiaro:1977}
{Cucchiaro} A.,  {Macau-Hercot} D.,  {Jaschek} M.,   {Jaschek} C.,  1977,
  \aaps, \href {https://ui.adsabs.harvard.edu/abs/1977A&AS...30...71C} {30, 71}

\bibitem[\protect\citeauthoryear{{Cunha} et~al.,}{{Cunha}
  et~al.}{2019}]{cunha:2019}
{Cunha} M.~S.,  et~al., 2019, \mn@doi [\mnras] {10.1093/mnras/stz1332}, \href
  {https://ui.adsabs.harvard.edu/abs/2019MNRAS.487.3523C} {487, 3523}

\bibitem[\protect\citeauthoryear{{David-Uraz} et~al.,}{{David-Uraz}
  et~al.}{2021}]{david-uraz:2021}
{David-Uraz} A.,  et~al., 2021, \mn@doi [\mnras] {10.1093/mnras/stab899}, \href
  {https://ui.adsabs.harvard.edu/abs/2021MNRAS.504.4841D} {504, 4841}

\bibitem[\protect\citeauthoryear{{Dolk}, {Wahlgren}  \& {Hubrig}}{{Dolk}
  et~al.}{2003}]{dolk:2003}
{Dolk} L.,  {Wahlgren} G.~M.,   {Hubrig} S.,  2003, \mn@doi [\aap]
  {10.1051/0004-6361:20030213}, \href
  {http://adsabs.harvard.edu/abs/2003A%26A...402..299D} {402, 299}

\bibitem[\protect\citeauthoryear{{Dworetsky}}{{Dworetsky}}{1982}]{dworetsky:1982}
{Dworetsky} M.~M.,  1982, \mn@doi [\mnras] {10.1093/mnras/199.2.303}, \href
  {https://ui.adsabs.harvard.edu/abs/1982MNRAS.199..303D} {199, 303}

\bibitem[\protect\citeauthoryear{{Flower}}{{Flower}}{1996}]{flower:1996}
{Flower} P.~J.,  1996, \mn@doi [\apj] {10.1086/177785}, \href
  {http://adsabs.harvard.edu/abs/1996ApJ...469..355F} {469, 355}

\bibitem[\protect\citeauthoryear{{Folsom}, {Kochukhov}, {Wade}, {Silvester}  \&
  {Bagnulo}}{{Folsom} et~al.}{2010}]{folsom:2010}
{Folsom} C.~P.,  {Kochukhov} O.,  {Wade} G.~A.,  {Silvester} J.,   {Bagnulo}
  S.,  2010, \mn@doi [\mnras] {10.1111/j.1365-2966.2010.17057.x}, \href
  {http://adsabs.harvard.edu/abs/2010MNRAS.407.2383F} {407, 2383}

\bibitem[\protect\citeauthoryear{{Folsom}, {Wade}  \& {Johnson}}{{Folsom}
  et~al.}{2013}]{folsom:2013}
{Folsom} C.~P.,  {Wade} G.~A.,   {Johnson} N.~M.,  2013, \mn@doi [\mnras]
  {10.1093/mnras/stt1003}, \href
  {http://adsabs.harvard.edu/abs/2013MNRAS.433.3336F} {433, 3336}

\bibitem[\protect\citeauthoryear{{Gaia Collaboration} et~al.,}{{Gaia
  Collaboration} et~al.}{2021}]{gaia-collaboration:2021}
{Gaia Collaboration} et~al., 2021, \mn@doi [\aap]
  {10.1051/0004-6361/202039657}, \href
  {https://ui.adsabs.harvard.edu/abs/2021A&A...649A...1G} {649, A1}

\bibitem[\protect\citeauthoryear{{Georgy}, {Ekstr{\"o}m}, {Granada}, {Meynet},
  {Mowlavi}, {Eggenberger}  \& {Maeder}}{{Georgy} et~al.}{2013}]{georgy:2013}
{Georgy} C.,  {Ekstr{\"o}m} S.,  {Granada} A.,  {Meynet} G.,  {Mowlavi} N.,
  {Eggenberger} P.,   {Maeder} A.,  2013, \mn@doi [\aap]
  {10.1051/0004-6361/201220558}, \href
  {https://ui.adsabs.harvard.edu/abs/2013A&A...553A..24G} {553, A24}

\bibitem[\protect\citeauthoryear{{Gerbaldi}, {Floquet}  \& {Hauck}}{{Gerbaldi}
  et~al.}{1985}]{gerbaldi:1985}
{Gerbaldi} M.,  {Floquet} M.,   {Hauck} B.,  1985, \aap, \href
  {http://adsabs.harvard.edu/abs/1985A%26A...146..341G} {146, 341}

\bibitem[\protect\citeauthoryear{{Ghazaryan} \& {Alecian}}{{Ghazaryan} \&
  {Alecian}}{2016}]{ghazaryan:2016}
{Ghazaryan} S.,  {Alecian} G.,  2016, \mn@doi [\mnras] {10.1093/mnras/stw911},
  \href {http://adsabs.harvard.edu/abs/2016MNRAS.460.1912G} {460, 1912}

\bibitem[\protect\citeauthoryear{{Ghazaryan}, {Alecian}  \&
  {Hakobyan}}{{Ghazaryan} et~al.}{2018}]{ghazaryan:2018}
{Ghazaryan} S.,  {Alecian} G.,   {Hakobyan} A.~A.,  2018, \mn@doi [\mnras]
  {10.1093/mnras/sty1912}, \href
  {https://ui.adsabs.harvard.edu/abs/2018MNRAS.480.2953G} {480, 2953}

\bibitem[\protect\citeauthoryear{{Glagolevskij}, {Panov}  \&
  {Chunakova}}{{Glagolevskij} et~al.}{1985}]{glagolevskij:1985}
{Glagolevskij} Y.~V.,  {Panov} K.,   {Chunakova} N.~M.,  1985, Pisma v
  Astronomicheskii Zhurnal, \href
  {https://ui.adsabs.harvard.edu/abs/1985PAZh...11..749G} {11, 749}

\bibitem[\protect\citeauthoryear{{Gonz{\'a}lez}, {Nu{\~n}ez}, {Saffe}, {Alejo},
  {Veramendi}  \& {Collado}}{{Gonz{\'a}lez} et~al.}{2021}]{gonzalez:2021}
{Gonz{\'a}lez} J.~F.,  {Nu{\~n}ez} N.~E.,  {Saffe} C.,  {Alejo} A.~D.,
  {Veramendi} M.~E.,   {Collado} A.,  2021, \mn@doi [\mnras]
  {10.1093/mnras/staa3401}, \href
  {https://ui.adsabs.harvard.edu/abs/2021MNRAS.502.3670G} {502, 3670}

\bibitem[\protect\citeauthoryear{{Green}, {Schlafly}, {Zucker}, {Speagle}  \&
  {Finkbeiner}}{{Green} et~al.}{2019}]{green:2019}
{Green} G.~M.,  {Schlafly} E.,  {Zucker} C.,  {Speagle} J.~S.,   {Finkbeiner}
  D.,  2019, \mn@doi [\apj] {10.3847/1538-4357/ab5362}, \href
  {https://ui.adsabs.harvard.edu/abs/2019ApJ...887...93G} {887, 93}

\bibitem[\protect\citeauthoryear{{Hoffleit} \& {Warren}}{{Hoffleit} \&
  {Warren}}{1991}]{hoffleit:1991}
{Hoffleit} D.,  {Warren} Jr. W.~H.,  1991, {VizieR Online Data Catalog: Bright
  Star Catalogue, 5th Revised Ed.}

\bibitem[\protect\citeauthoryear{{Hubrig}, {Le Mignant}, {North}  \&
  {Krautter}}{{Hubrig} et~al.}{2001}]{hubrig:2001a}
{Hubrig} S.,  {Le Mignant} D.,  {North} P.,   {Krautter} J.,  2001, \mn@doi
  [\aap] {10.1051/0004-6361:20010452}, \href
  {https://ui.adsabs.harvard.edu/abs/2001A&A...372..152H} {372, 152}

\bibitem[\protect\citeauthoryear{{Hubrig}, {Gonz{\'a}lez}, {Savanov},
  {Sch{\"o}ller}, {Ageorges}, {Cowley}  \& {Wolff}}{{Hubrig}
  et~al.}{2006}]{hubrig:2006}
{Hubrig} S.,  {Gonz{\'a}lez} J.~F.,  {Savanov} I.,  {Sch{\"o}ller} M.,
  {Ageorges} N.,  {Cowley} C.~R.,   {Wolff} B.,  2006, \mn@doi [\mnras]
  {10.1111/j.1365-2966.2006.10863.x}, \href
  {http://adsabs.harvard.edu/abs/2006MNRAS.371.1953H} {371, 1953}

\bibitem[\protect\citeauthoryear{{Hubrig} et~al.,}{{Hubrig}
  et~al.}{2010}]{hubrig:2010}
{Hubrig} S.,  et~al., 2010, \mn@doi [\mnras]
  {10.1111/j.1745-3933.2010.00928.x}, \href
  {http://adsabs.harvard.edu/abs/2010MNRAS.408L..61H} {408, L61}

\bibitem[\protect\citeauthoryear{{Hubrig} et~al.,}{{Hubrig}
  et~al.}{2011}]{hubrig:2011b}
{Hubrig} S.,  et~al., 2011, \mn@doi [Astronomische Nachrichten]
  {10.1002/asna.201111599}, \href
  {https://ui.adsabs.harvard.edu/abs/2011AN....332..998H} {332, 998}

\bibitem[\protect\citeauthoryear{{Hubrig} et~al.,}{{Hubrig}
  et~al.}{2012}]{hubrig:2012}
{Hubrig} S.,  et~al., 2012, \mn@doi [\aap] {10.1051/0004-6361/201219778}, \href
  {http://adsabs.harvard.edu/abs/2012A%26A...547A..90H} {547, A90}

\bibitem[\protect\citeauthoryear{{Hubrig} et~al.,}{{Hubrig}
  et~al.}{2014}]{hubrig:2014}
{Hubrig} S.,  et~al., 2014, \mn@doi [\mnras] {10.1093/mnras/stu1122}, \href
  {https://ui.adsabs.harvard.edu/abs/2014MNRAS.442.3604H} {442, 3604}

\bibitem[\protect\citeauthoryear{{Hubrig}, {J{\"a}rvinen}, {Korhonen}, {Ilyin},
  {Sch{\"o}ller}, {Niemczura}  \& {Chojnowski}}{{Hubrig}
  et~al.}{2020}]{hubrig:2020}
{Hubrig} S.,  {J{\"a}rvinen} S.~P.,  {Korhonen} H.,  {Ilyin} I.,
  {Sch{\"o}ller} M.,  {Niemczura} E.,   {Chojnowski} S.~D.,  2020, \mn@doi
  [\mnras] {10.1093/mnrasl/slaa064}, \href
  {https://ui.adsabs.harvard.edu/abs/2020MNRAS.495L..97H} {495, L97}

\bibitem[\protect\citeauthoryear{{Hummel}, {Sch{\"o}ller}, {Duvert}  \&
  {Hubrig}}{{Hummel} et~al.}{2017}]{hummel:2017}
{Hummel} C.~A.,  {Sch{\"o}ller} M.,  {Duvert} G.,   {Hubrig} S.,  2017, \mn@doi
  [\aap] {10.1051/0004-6361/201730414}, \href
  {https://ui.adsabs.harvard.edu/abs/2017A&A...600L...5H} {600, L5}

\bibitem[\protect\citeauthoryear{{H{\"u}mmerich}, {Niemczura}, {Walczak},
  {Paunzen}, {Bernhard}, {Murphy}  \& {Drobek}}{{H{\"u}mmerich}
  et~al.}{2018}]{hummerich:2018}
{H{\"u}mmerich} S.,  {Niemczura} E.,  {Walczak} P.,  {Paunzen} E.,  {Bernhard}
  K.,  {Murphy} S.~J.,   {Drobek} D.,  2018, \mn@doi [\mnras]
  {10.1093/mnras/stx2974}, \href
  {https://ui.adsabs.harvard.edu/abs/2018MNRAS.474.2467H} {474, 2467}

\bibitem[\protect\citeauthoryear{{Hutchings}, {Nemec}  \&
  {Cassidy}}{{Hutchings} et~al.}{1979}]{hutchings:1979}
{Hutchings} J.~B.,  {Nemec} J.~M.,   {Cassidy} J.,  1979, \mn@doi [\pasp]
  {10.1086/130490}, \href
  {https://ui.adsabs.harvard.edu/abs/1979PASP...91..313H} {91, 313}

\bibitem[\protect\citeauthoryear{{Ilyin}}{{Ilyin}}{2000}]{ilyin:2000}
{Ilyin} I.~V.,  2000, PhD thesis, {Astronomy Division, Department of Physical
  Sciences, University of Oulu, Finland}

\bibitem[\protect\citeauthoryear{{Jagelka}, {Mikul{\'a}{\v s}ek},
  {H{\"u}mmerich}  \& {Paunzen}}{{Jagelka} et~al.}{2019}]{jagelka:2019}
{Jagelka} M.,  {Mikul{\'a}{\v s}ek} Z.,  {H{\"u}mmerich} S.,   {Paunzen} E.,
  2019, \mn@doi [\aap] {10.1051/0004-6361/201833482}, \href
  {http://adsabs.harvard.edu/abs/2019A%26A...622A.199J} {622, A199}

\bibitem[\protect\citeauthoryear{{Jenkins} et~al.,}{{Jenkins}
  et~al.}{2016}]{jenkins:2016}
{Jenkins} J.~M.,  et~al., 2016, in {Chiozzi} G.,  {Guzman} J.~C.,  eds,
  Society of Photo-Optical Instrumentation Engineers (SPIE) Conference Series
  Vol. 9913, Software and Cyberinfrastructure for Astronomy IV. p. 99133E

\bibitem[\protect\citeauthoryear{{Kammerer}, {Ireland}, {Martinache}  \&
  {Girard}}{{Kammerer} et~al.}{2019}]{kammerer:2019}
{Kammerer} J.,  {Ireland} M.~J.,  {Martinache} F.,   {Girard} J.~H.,  2019,
  \mn@doi [\mnras] {10.1093/mnras/stz882}, \href
  {https://ui.adsabs.harvard.edu/abs/2019MNRAS.486..639K} {486, 639}

\bibitem[\protect\citeauthoryear{{Kochukhov} \& {Bagnulo}}{{Kochukhov} \&
  {Bagnulo}}{2006}]{kochukhov:2006}
{Kochukhov} O.,  {Bagnulo} S.,  2006, \mn@doi [\aap]
  {10.1051/0004-6361:20054596}, \href
  {http://adsabs.harvard.edu/abs/2006A%26A...450..763K} {450, 763}

\bibitem[\protect\citeauthoryear{{Kochukhov}, {Drake}, {Piskunov}  \& {de la
  Reza}}{{Kochukhov} et~al.}{2004}]{kochukhov:2004e}
{Kochukhov} O.,  {Drake} N.~A.,  {Piskunov} N.,   {de la Reza} R.,  2004,
  \mn@doi [\aap] {10.1051/0004-6361:20040517}, \href
  {http://adsabs.harvard.edu/abs/2004A%26A...424..935K} {424, 935}

\bibitem[\protect\citeauthoryear{{Kochukhov}, {Piskunov}, {Sachkov}  \&
  {Kudryavtsev}}{{Kochukhov} et~al.}{2005}]{kochukhov:2005b}
{Kochukhov} O.,  {Piskunov} N.,  {Sachkov} M.,   {Kudryavtsev} D.,  2005,
  \mn@doi [\aap] {10.1051/0004-6361:20053123}, \href
  {http://adsabs.harvard.edu/abs/2005A%26A...439.1093K} {439, 1093}

\bibitem[\protect\citeauthoryear{{Kochukhov}, {Adelman}, {Gulliver}  \&
  {Piskunov}}{{Kochukhov} et~al.}{2007}]{kochukhov:2007b}
{Kochukhov} O.,  {Adelman} S.~J.,  {Gulliver} A.~F.,   {Piskunov} N.,  2007,
  \mn@doi [Nature Physics] {10.1038/nphys648}, \href
  {http://esoads.eso.org/abs/2007NatPh...3..526K} {3, 526}

\bibitem[\protect\citeauthoryear{{Kochukhov} et~al.,}{{Kochukhov}
  et~al.}{2011}]{kochukhov:2011b}
{Kochukhov} O.,  et~al., 2011, \mn@doi [\aap] {10.1051/0004-6361/201117970},
  \href {http://cdsads.u-strasbg.fr/abs/2011A%26A...534L..13K} {534, L13}

\bibitem[\protect\citeauthoryear{{Kochukhov} et~al.,}{{Kochukhov}
  et~al.}{2013}]{kochukhov:2013a}
{Kochukhov} O.,  et~al., 2013, \mn@doi [\aap] {10.1051/0004-6361/201321467},
  \href {http://adsabs.harvard.edu/abs/2013A%26A...554A..61K} {554, A61}

\bibitem[\protect\citeauthoryear{{Kochukhov}, {L{\"u}ftinger}, {Neiner},
  {Alecian}  \& {MiMeS Collaboration}}{{Kochukhov}
  et~al.}{2014}]{kochukhov:2014}
{Kochukhov} O.,  {L{\"u}ftinger} T.,  {Neiner} C.,  {Alecian} E.,   {MiMeS
  Collaboration} 2014, \mn@doi [\aap] {10.1051/0004-6361/201423472}, \href
  {http://adsabs.harvard.edu/abs/2014A%26A...565A..83K} {565, A83}

\bibitem[\protect\citeauthoryear{{Kochukhov}, {Shultz}  \&
  {Neiner}}{{Kochukhov} et~al.}{2019}]{kochukhov:2019}
{Kochukhov} O.,  {Shultz} M.,   {Neiner} C.,  2019, \mn@doi [\aap]
  {10.1051/0004-6361/201834279}, \href
  {http://adsabs.harvard.edu/abs/2019A%26A...621A..47K} {621, A47}

\bibitem[\protect\citeauthoryear{{Kochukhov}, {Johnston}, {Labadie-Bartz},
  {Shetye}, {Ryabchikova}, {Tkachenko}  \& {Shultz}}{{Kochukhov}
  et~al.}{2021}]{kochukhov:2021a}
{Kochukhov} O.,  {Johnston} C.,  {Labadie-Bartz} J.,  {Shetye} S.,
  {Ryabchikova} T.~A.,  {Tkachenko} A.,   {Shultz} M.~E.,  2021, \mn@doi
  [\mnras] {10.1093/mnras/staa3472}, \href
  {https://ui.adsabs.harvard.edu/abs/2021MNRAS.500.2577K} {500, 2577}

\bibitem[\protect\citeauthoryear{{Korhonen} et~al.,}{{Korhonen}
  et~al.}{2013}]{korhonen:2013}
{Korhonen} H.,  et~al., 2013, \mn@doi [\aap] {10.1051/0004-6361/201220951},
  \href {http://adsabs.harvard.edu/abs/2013A%26A...553A..27K} {553, A27}

\bibitem[\protect\citeauthoryear{{Krti{\v{c}}ka} et~al.,}{{Krti{\v{c}}ka}
  et~al.}{2020}]{krticka:2020}
{Krti{\v{c}}ka} J.,  et~al., 2020, \mn@doi [\aap]
  {10.1051/0004-6361/202037953}, \href
  {https://ui.adsabs.harvard.edu/abs/2020A&A...639A...8K} {639, A8}

\bibitem[\protect\citeauthoryear{{Kuschnig}, {Weiss}, {Gruber}, {Bely}  \&
  {Jenkner}}{{Kuschnig} et~al.}{1997}]{kuschnig:1997}
{Kuschnig} R.,  {Weiss} W.~W.,  {Gruber} R.,  {Bely} P.~Y.,   {Jenkner} H.,
  1997, \aap, \href {https://ui.adsabs.harvard.edu/abs/1997A&A...328..544K}
  {328, 544}

\bibitem[\protect\citeauthoryear{{Le Bouquin}, {Beust}, {Duvert}, {Berger},
  {M{\'e}nard}  \& {Zins}}{{Le Bouquin} et~al.}{2013}]{le-bouquin:2013}
{Le Bouquin} J.~B.,  {Beust} H.,  {Duvert} G.,  {Berger} J.~P.,  {M{\'e}nard}
  F.,   {Zins} G.,  2013, \mn@doi [\aap] {10.1051/0004-6361/201220454}, \href
  {https://ui.adsabs.harvard.edu/abs/2013A&A...551A.121L} {551, A121}

\bibitem[\protect\citeauthoryear{{LeBlanc}, {Monin}, {Hui-Bon-Hoa}  \&
  {Hauschildt}}{{LeBlanc} et~al.}{2009}]{leblanc:2009}
{LeBlanc} F.,  {Monin} D.,  {Hui-Bon-Hoa} A.,   {Hauschildt} P.~H.,  2009,
  \mn@doi [\aap] {10.1051/0004-6361:200810848}, \href
  {http://adsabs.harvard.edu/abs/2009A%26A...495..937L} {495, 937}

\bibitem[\protect\citeauthoryear{{Makaganiuk} et~al.,}{{Makaganiuk}
  et~al.}{2011a}]{makaganiuk:2011a}
{Makaganiuk} V.,  et~al., 2011a, \mn@doi [\aap] {10.1051/0004-6361/201015666},
  \href {http://adsabs.harvard.edu/abs/2011A%26A...525A..97M} {525, A97}

\bibitem[\protect\citeauthoryear{{Makaganiuk} et~al.,}{{Makaganiuk}
  et~al.}{2011b}]{makaganiuk:2011}
{Makaganiuk} V.,  et~al., 2011b, \mn@doi [\aap] {10.1051/0004-6361/201016302},
  \href {http://adsabs.harvard.edu/abs/2011A%26A...529A.160M} {529, A160}

\bibitem[\protect\citeauthoryear{{Makaganiuk} et~al.,}{{Makaganiuk}
  et~al.}{2012}]{makaganiuk:2012}
{Makaganiuk} V.,  et~al., 2012, \mn@doi [\aap] {10.1051/0004-6361/201118167},
  \href {http://adsabs.harvard.edu/abs/2012A%26A...539A.142M} {539, A142}

\bibitem[\protect\citeauthoryear{{Markwardt}}{{Markwardt}}{2009}]{markwardt:2009}
{Markwardt} C.~B.,  2009, in {D.~A.~Bohlender, D.~Durand, \& P.~Dowler} ed.,
  Astronomical Society of the Pacific Conference Series Vol. 411, Astronomical
  Society of the Pacific Conference Series. pp 251--254

\bibitem[\protect\citeauthoryear{{Martin} et~al.,}{{Martin}
  et~al.}{2018}]{martin:2018}
{Martin} A.~J.,  et~al., 2018, \mn@doi [\mnras] {10.1093/mnras/stx3264}, \href
  {https://ui.adsabs.harvard.edu/abs/2018MNRAS.475.1521M} {475, 1521}

\bibitem[\protect\citeauthoryear{{Mathys}}{{Mathys}}{1994}]{mathys:1994}
{Mathys} G.,  1994, \aaps, \href
  {http://adsabs.harvard.edu/abs/1994A%26AS..108..547M} {108, 547}

\bibitem[\protect\citeauthoryear{{Mathys} \& {Hubrig}}{{Mathys} \&
  {Hubrig}}{1997}]{mathys:1997a}
{Mathys} G.,  {Hubrig} S.,  1997, \aaps, \href
  {http://adsabs.harvard.edu/abs/1997A%26AS..124..475M} {124, 475}

\bibitem[\protect\citeauthoryear{{Michaud}}{{Michaud}}{1982}]{michaud:1982}
{Michaud} G.,  1982, \mn@doi [\apj] {10.1086/160083}, \href
  {https://ui.adsabs.harvard.edu/abs/1982ApJ...258..349M} {258, 349}

\bibitem[\protect\citeauthoryear{{Michaud}, {Charland}  \&
  {Megessier}}{{Michaud} et~al.}{1981}]{michaud:1981}
{Michaud} G.,  {Charland} Y.,   {Megessier} C.,  1981, \aap, \href
  {http://adsabs.harvard.edu/abs/1981A%26A...103..244M} {103, 244}

\bibitem[\protect\citeauthoryear{{Michaud}, {Alecian}  \& {Richer}}{{Michaud}
  et~al.}{2015}]{michaud:2015}
{Michaud} G.,  {Alecian} G.,   {Richer} J.,  2015, {Atomic Diffusion in Stars},
  \mn@doi{10.1007/978-3-319-19854-5.
}

\bibitem[\protect\citeauthoryear{{Mikul{\'a}{\v{s}}ek}, {Jan{\'\i}k}, {Zverko},
  {{\v{Z}}i{\v{z}}{\v{n}}ovsk{\'y}}, {Zejda}, {Netolick{\'y}}  \&
  {Va{\v{n}}ko}}{{Mikul{\'a}{\v{s}}ek} et~al.}{2007}]{mikulasek:2007}
{Mikul{\'a}{\v{s}}ek} Z.,  {Jan{\'\i}k} J.,  {Zverko} J.,
  {{\v{Z}}i{\v{z}}{\v{n}}ovsk{\'y}} J.,  {Zejda} M.,  {Netolick{\'y}} M.,
  {Va{\v{n}}ko} M.,  2007, \mn@doi [Astronomische Nachrichten]
  {10.1002/asna.200610705}, \href
  {https://ui.adsabs.harvard.edu/abs/2007AN....328...10M} {328, 10}

\bibitem[\protect\citeauthoryear{{Monier}, {Gebran}  \& {Royer}}{{Monier}
  et~al.}{2015}]{monier:2015}
{Monier} R.,  {Gebran} M.,   {Royer} F.,  2015, \mn@doi [\aap]
  {10.1051/0004-6361/201526106}, \href
  {http://adsabs.harvard.edu/abs/2015A%26A...577A..96M} {577, A96}

\bibitem[\protect\citeauthoryear{{Monier}, {Griffin}, {Gebran},
  {K{\i}l{\i}{\c{c}}o{\u{g}}lu}, {Merle}  \& {Royer}}{{Monier}
  et~al.}{2019}]{monier:2019}
{Monier} R.,  {Griffin} E.,  {Gebran} M.,  {K{\i}l{\i}{\c{c}}o{\u{g}}lu} T.,
  {Merle} T.,   {Royer} F.,  2019, \mn@doi [\aj] {10.3847/1538-3881/ab3b59},
  \href {https://ui.adsabs.harvard.edu/abs/2019AJ....158..157M} {158, 157}

\bibitem[\protect\citeauthoryear{{Morel} et~al.,}{{Morel}
  et~al.}{2014}]{morel:2014}
{Morel} T.,  et~al., 2014, \mn@doi [\aap] {10.1051/0004-6361/201322289}, \href
  {http://adsabs.harvard.edu/abs/2014A%26A...561A..35M} {561, A35}

\bibitem[\protect\citeauthoryear{{Mowlavi}, {Saesen}, {Semaan}, {Eggenberger},
  {Barblan}, {Eyer}, {Ekstr{\"o}m}  \& {Georgy}}{{Mowlavi}
  et~al.}{2016}]{mowlavi:2016}
{Mowlavi} N.,  {Saesen} S.,  {Semaan} T.,  {Eggenberger} P.,  {Barblan} F.,
  {Eyer} L.,  {Ekstr{\"o}m} S.,   {Georgy} C.,  2016, \mn@doi [\aap]
  {10.1051/0004-6361/201629175}, \href
  {https://ui.adsabs.harvard.edu/abs/2016A&A...595L...1M} {595, L1}

\bibitem[\protect\citeauthoryear{{Napiwotzki}, {Schoenberner}  \&
  {Wenske}}{{Napiwotzki} et~al.}{1993}]{napiwotzki:1993}
{Napiwotzki} R.,  {Schoenberner} D.,   {Wenske} V.,  1993, \aap, \href
  {http://adsabs.harvard.edu/abs/1993A%26A...268..653N} {268, 653}

\bibitem[\protect\citeauthoryear{{Neiner}, {Martin}, {Wade}  \&
  {Oksala}}{{Neiner} et~al.}{2018}]{neiner:2018}
{Neiner} C.,  {Martin} A.,  {Wade} G.,   {Oksala} M.,  2018, in {Di Matteo} P.,
   {Billebaud} F.,  {Herpin} F.,  {Lagarde} N.,  {Marquette} J.~B.,  {Robin}
  A.,   {Venot} O.,  eds, SF2A-2018: Proceedings of the Annual meeting of the
  French Society of Astronomy and Astrophysics. p.~Di (\mn@eprint {arXiv}
  {1811.05258})

\bibitem[\protect\citeauthoryear{{Netopil}, {Paunzen}, {Maitzen}, {North}  \&
  {Hubrig}}{{Netopil} et~al.}{2008}]{netopil:2008}
{Netopil} M.,  {Paunzen} E.,  {Maitzen} H.~M.,  {North} P.,   {Hubrig} S.,
  2008, \mn@doi [\aap] {10.1051/0004-6361:200810325}, \href
  {https://ui.adsabs.harvard.edu/abs/2008A&A...491..545N} {491, 545}

\bibitem[\protect\citeauthoryear{{Netopil}, {Paunzen}, {H{\"u}mmerich}  \&
  {Bernhard}}{{Netopil} et~al.}{2017}]{netopil:2017}
{Netopil} M.,  {Paunzen} E.,  {H{\"u}mmerich} S.,   {Bernhard} K.,  2017,
  \mn@doi [\mnras] {10.1093/mnras/stx674}, \href
  {https://ui.adsabs.harvard.edu/abs/2017MNRAS.468.2745N} {468, 2745}

\bibitem[\protect\citeauthoryear{{Nu{\~n}ez}, {Gonz{\'a}lez}  \&
  {Hubrig}}{{Nu{\~n}ez} et~al.}{2010}]{nunez:2010}
{Nu{\~n}ez} N.~E.,  {Gonz{\'a}lez} J.~F.,   {Hubrig} S.,  2010, in
  {Kudryavtsev} D.~O.,  {Romanyuk} I.~I.,   {Zyazeva} A.~V.,  eds, Magnetic
  Stars. pp 109--118

\bibitem[\protect\citeauthoryear{{Paunzen}}{{Paunzen}}{2015}]{paunzen:2015}
{Paunzen} E.,  2015, \mn@doi [\aap] {10.1051/0004-6361/201526413}, \href
  {https://ui.adsabs.harvard.edu/abs/2015A&A...580A..23P} {580, A23}

\bibitem[\protect\citeauthoryear{{Paunzen}, {Schnell}  \& {Maitzen}}{{Paunzen}
  et~al.}{2006}]{paunzen:2006}
{Paunzen} E.,  {Schnell} A.,   {Maitzen} H.~M.,  2006, \mn@doi [\aap]
  {10.1051/0004-6361:20064889}, \href
  {https://ui.adsabs.harvard.edu/abs/2006A&A...458..293P} {458, 293}

\bibitem[\protect\citeauthoryear{{Paunzen}, {Wraight}, {Fossati}, {Netopil},
  {White}  \& {Bewsher}}{{Paunzen} et~al.}{2013}]{paunzen:2013}
{Paunzen} E.,  {Wraight} K.~T.,  {Fossati} L.,  {Netopil} M.,  {White} G.~J.,
  {Bewsher} D.,  2013, \mn@doi [\mnras] {10.1093/mnras/sts318}, \href
  {https://ui.adsabs.harvard.edu/abs/2013MNRAS.429..119P} {429, 119}

\bibitem[\protect\citeauthoryear{{Paunzen}, {Huemmerich}, {Fedurco},
  {Bernhard}, {Komzik}  \& {Vanko}}{{Paunzen} et~al.}{2021a}]{paunzen:2021a}
{Paunzen} E.,  {Huemmerich} S.,  {Fedurco} M.,  {Bernhard} K.,  {Komzik} R.,
  {Vanko} M.,  2021a, arXiv e-prints, \href
  {https://ui.adsabs.harvard.edu/abs/2021arXiv210407627P} {p. arXiv:2104.07627}

\bibitem[\protect\citeauthoryear{{Paunzen}, {H{\"u}mmerich}  \&
  {Bernhard}}{{Paunzen} et~al.}{2021b}]{paunzen:2021}
{Paunzen} E.,  {H{\"u}mmerich} S.,   {Bernhard} K.,  2021b, \mn@doi [\aap]
  {10.1051/0004-6361/202038847}, \href
  {https://ui.adsabs.harvard.edu/abs/2021A&A...645A..34P} {645, A34}

\bibitem[\protect\citeauthoryear{{Pecaut} \& {Mamajek}}{{Pecaut} \&
  {Mamajek}}{2013}]{pecaut:2013}
{Pecaut} M.~J.,  {Mamajek} E.~E.,  2013, \mn@doi [\apjs]
  {10.1088/0067-0049/208/1/9}, \href
  {https://ui.adsabs.harvard.edu/abs/2013ApJS..208....9P} {208, 9}

\bibitem[\protect\citeauthoryear{{Pedersen} et~al.,}{{Pedersen}
  et~al.}{2019}]{pedersen:2019}
{Pedersen} M.~G.,  et~al., 2019, \mn@doi [\apjl] {10.3847/2041-8213/ab01e1},
  \href {https://ui.adsabs.harvard.edu/abs/2019ApJ...872L...9P} {872, L9}

\bibitem[\protect\citeauthoryear{{Petit} et~al.,}{{Petit}
  et~al.}{2010}]{petit:2010}
{Petit} P.,  et~al., 2010, \mn@doi [\aap] {10.1051/0004-6361/201015307}, \href
  {http://cdsads.u-strasbg.fr/abs/2010A%26A...523A..41P} {523, A41}

\bibitem[\protect\citeauthoryear{{Petit} et~al.,}{{Petit}
  et~al.}{2011}]{petit:2011}
{Petit} P.,  et~al., 2011, \mn@doi [\aap] {10.1051/0004-6361/201117573}, \href
  {http://adsabs.harvard.edu/abs/2011A%26A...532L..13P} {532, L13}

\bibitem[\protect\citeauthoryear{{Pope} et~al.,}{{Pope}
  et~al.}{2019}]{pope:2019}
{Pope} B. J.~S.,  et~al., 2019, \mn@doi [\apjs] {10.3847/1538-4365/ab3d29},
  \href {https://ui.adsabs.harvard.edu/abs/2019ApJS..245....8P} {245, 8}

\bibitem[\protect\citeauthoryear{{Pourbaix} et~al.,}{{Pourbaix}
  et~al.}{2004}]{pourbaix:2004}
{Pourbaix} D.,  et~al., 2004, \mn@doi [\aap] {10.1051/0004-6361:20041213},
  \href {http://adsabs.harvard.edu/abs/2004A%26A...424..727P} {424, 727}

\bibitem[\protect\citeauthoryear{{Pourbaix}, {Boffin}, {Chini}  \&
  {Dembsky}}{{Pourbaix} et~al.}{2013}]{pourbaix:2013}
{Pourbaix} D.,  {Boffin} H.~M.~J.,  {Chini} R.,   {Dembsky} T.,  2013, \mn@doi
  [\aap] {10.1051/0004-6361/201321699}, \href
  {https://ui.adsabs.harvard.edu/abs/2013A&A...556A..45P} {556, A45}

\bibitem[\protect\citeauthoryear{{Preston}}{{Preston}}{1974}]{preston:1974}
{Preston} G.~W.,  1974, \mn@doi [\araa] {10.1146/annurev.aa.12.090174.001353},
  \href {http://adsabs.harvard.edu/abs/1974ARA%26A..12..257P} {12, 257}

\bibitem[\protect\citeauthoryear{{Prv{\'a}k}, {Krti{\v{c}}ka}  \&
  {Korhonen}}{{Prv{\'a}k} et~al.}{2020}]{prvak:2020}
{Prv{\'a}k} M.,  {Krti{\v{c}}ka} J.,   {Korhonen} H.,  2020, \mn@doi [\mnras]
  {10.1093/mnras/stz3564}, \href
  {https://ui.adsabs.harvard.edu/abs/2020MNRAS.492.1834P} {492, 1834}

\bibitem[\protect\citeauthoryear{{Renson} \& {Manfroid}}{{Renson} \&
  {Manfroid}}{2009}]{renson:2009}
{Renson} P.,  {Manfroid} J.,  2009, \mn@doi [\aap]
  {10.1051/0004-6361/200810788}, \href
  {http://adsabs.harvard.edu/abs/2009A%26A...498..961R} {498, 961}

\bibitem[\protect\citeauthoryear{{Ricker} et~al.,}{{Ricker}
  et~al.}{2015}]{ricker:2015}
{Ricker} G.~R.,  et~al., 2015, \mn@doi [Journal of Astronomical Telescopes,
  Instruments, and Systems] {10.1117/1.JATIS.1.1.014003}, \href
  {http://adsabs.harvard.edu/abs/2015JATIS...1a4003R} {1, 014003}

\bibitem[\protect\citeauthoryear{{Riello} et~al.,}{{Riello}
  et~al.}{2021}]{riello:2021}
{Riello} M.,  et~al., 2021, \mn@doi [\aap] {10.1051/0004-6361/202039587}, \href
  {https://ui.adsabs.harvard.edu/abs/2021A&A...649A...3R} {649, A3}

\bibitem[\protect\citeauthoryear{{Roby} \& {Lambert}}{{Roby} \&
  {Lambert}}{1990}]{roby:1990}
{Roby} S.~W.,  {Lambert} D.~L.,  1990, \mn@doi [\apjs] {10.1086/191440}, \href
  {http://adsabs.harvard.edu/abs/1990ApJS...73...67R} {73, 67}

\bibitem[\protect\citeauthoryear{{Ryabchikova}, {Zakharova}  \&
  {Adelman}}{{Ryabchikova} et~al.}{1996}]{ryabchikova:1996a}
{Ryabchikova} T.~A.,  {Zakharova} L.~A.,   {Adelman} S.~J.,  1996, \mn@doi
  [\mnras] {10.1093/mnras/283.4.1115}, \href
  {https://ui.adsabs.harvard.edu/abs/1996MNRAS.283.1115R} {283, 1115}

\bibitem[\protect\citeauthoryear{{Ryabchikova}, {Piskunov}, {Stempels}, {Kupka}
   \& {Weiss}}{{Ryabchikova} et~al.}{1999}]{ryabchikova:1999b}
{Ryabchikova} T.~A.,  {Piskunov} N.~E.,  {Stempels} H.~C.,  {Kupka} F.,
  {Weiss} W.~W.,  1999, \mn@doi [Physica Scripta Volume T]
  {10.1238/Physica.Topical.083a00162}, \href
  {http://adsabs.harvard.edu/abs/1999PhST...83..162R} {83, 162}

\bibitem[\protect\citeauthoryear{{Schneider}}{{Schneider}}{1981}]{schneider:1981}
{Schneider} H.,  1981, \aaps, \href
  {https://ui.adsabs.harvard.edu/abs/1981A&AS...44..137S} {44, 137}

\bibitem[\protect\citeauthoryear{{Sch{\"o}ller}, {Correia}, {Hubrig}  \&
  {Ageorges}}{{Sch{\"o}ller} et~al.}{2010}]{scholler:2010}
{Sch{\"o}ller} M.,  {Correia} S.,  {Hubrig} S.,   {Ageorges} N.,  2010, \mn@doi
  [\aap] {10.1051/0004-6361/201014246}, \href
  {http://adsabs.harvard.edu/abs/2010A%26A...522A..85S} {522, A85}

\bibitem[\protect\citeauthoryear{{Shorlin}, {Wade}, {Donati}, {Landstreet},
  {Petit}, {Sigut}  \& {Strasser}}{{Shorlin} et~al.}{2002}]{shorlin:2002}
{Shorlin} S.~L.~S.,  {Wade} G.~A.,  {Donati} J.-F.,  {Landstreet} J.~D.,
  {Petit} P.,  {Sigut} T.~A.~A.,   {Strasser} S.,  2002, \mn@doi [\aap]
  {10.1051/0004-6361:20021192}, \href
  {http://adsabs.harvard.edu/abs/2002A%26A...392..637S} {392, 637}

\bibitem[\protect\citeauthoryear{{Shultz}, {Rivinius}, {Wade}, {Kochukhov},
  {Alecian}, {David-Uraz}, {Sikora}  \& {MiMeS Collaboration}}{{Shultz}
  et~al.}{2021}]{shultz:2021}
{Shultz} M.~E.,  {Rivinius} T.,  {Wade} G.~A.,  {Kochukhov} O.,  {Alecian} E.,
  {David-Uraz} A.,  {Sikora} J.,   {MiMeS Collaboration} 2021, \mn@doi [\mnras]
  {10.1093/mnras/staa3158}, \href
  {https://ui.adsabs.harvard.edu/abs/2021MNRAS.tmp..908S} {}

\bibitem[\protect\citeauthoryear{{Sikora}, {Wade}  \& {Rowe}}{{Sikora}
  et~al.}{2020}]{sikora:2020}
{Sikora} J.,  {Wade} G.~A.,   {Rowe} J.,  2020, \mn@doi [\mnras]
  {10.1093/mnras/staa2444}, \href
  {https://ui.adsabs.harvard.edu/abs/2020MNRAS.498.2456S} {498, 2456}

\bibitem[\protect\citeauthoryear{{Stassun} et~al.,}{{Stassun}
  et~al.}{2019}]{stassun:2019}
{Stassun} K.~G.,  et~al., 2019, \mn@doi [\aj] {10.3847/1538-3881/ab3467}, \href
  {https://ui.adsabs.harvard.edu/abs/2019AJ....158..138S} {158, 138}

\bibitem[\protect\citeauthoryear{{Strassmeier}, {Granzer}, {Mallonn}, {Weber}
  \& {Weingrill}}{{Strassmeier} et~al.}{2017}]{strassmeier:2017}
{Strassmeier} K.~G.,  {Granzer} T.,  {Mallonn} M.,  {Weber} M.,   {Weingrill}
  J.,  2017, \mn@doi [\aap] {10.1051/0004-6361/201629150}, \href
  {http://adsabs.harvard.edu/abs/2017A%26A...597A..55S} {597, A55}

\bibitem[\protect\citeauthoryear{{Strassmeier} et~al.,}{{Strassmeier}
  et~al.}{2020}]{strassmeier:2020}
{Strassmeier} K.~G.,  et~al., 2020, \mn@doi [\aap]
  {10.1051/0004-6361/202039310}, \href
  {https://ui.adsabs.harvard.edu/abs/2020A&A...644A.104S} {644, A104}

\bibitem[\protect\citeauthoryear{{Thompson} et~al.,}{{Thompson}
  et~al.}{2012}]{thompson:2012}
{Thompson} S.~E.,  et~al., 2012, \mn@doi [\apj] {10.1088/0004-637X/753/1/86},
  \href {https://ui.adsabs.harvard.edu/abs/2012ApJ...753...86T} {753, 86}

\bibitem[\protect\citeauthoryear{{Tsymbal}, {Kotchukhov}, {Khokhlova}  \&
  {Lambert}}{{Tsymbal} et~al.}{1998}]{tsymbal:1998}
{Tsymbal} V.~V.,  {Kotchukhov} O.~P.,  {Khokhlova} V.~L.,   {Lambert} D.~L.,
  1998, Astronomy Letters, \href
  {https://ui.adsabs.harvard.edu/abs/1998AstL...24...90T} {24, 90}

\bibitem[\protect\citeauthoryear{{Wade} et~al.,}{{Wade}
  et~al.}{2006}]{wade:2006}
{Wade} G.~A.,  et~al., 2006, \mn@doi [\aap] {10.1051/0004-6361:20054502}, \href
  {http://adsabs.harvard.edu/abs/2006A%26A...451..293W} {451, 293}

\bibitem[\protect\citeauthoryear{{White} et~al.,}{{White}
  et~al.}{2017}]{white:2017}
{White} T.~R.,  et~al., 2017, \mn@doi [\mnras] {10.1093/mnras/stx1050}, \href
  {https://ui.adsabs.harvard.edu/abs/2017MNRAS.471.2882W} {471, 2882}

\bibitem[\protect\citeauthoryear{{Woolf} \& {Lambert}}{{Woolf} \&
  {Lambert}}{1999}]{woolf:1999}
{Woolf} V.~M.,  {Lambert} D.~L.,  1999, \mn@doi [\apj] {10.1086/307551}, \href
  {http://adsabs.harvard.edu/abs/1999ApJ...521..414W} {521, 414}

\bibitem[\protect\citeauthoryear{{Y{\"u}ce} \& {Adelman}}{{Y{\"u}ce} \&
  {Adelman}}{2014}]{yuce:2014}
{Y{\"u}ce} K.,  {Adelman} S.~J.,  2014, \mn@doi [\pasp] {10.1086/676335}, \href
  {https://ui.adsabs.harvard.edu/abs/2014PASP..126..345Y} {126, 345}

\bibitem[\protect\citeauthoryear{{Yushchenko}, {Gopka}, {Khokhlova}, {Musaev}
  \& {Bikmaev}}{{Yushchenko} et~al.}{1999}]{yushchenko:1999}
{Yushchenko} A.~V.,  {Gopka} V.~F.,  {Khokhlova} V.~L.,  {Musaev} F.~A.,
  {Bikmaev} I.~F.,  1999, Astronomy Letters, \href
  {https://ui.adsabs.harvard.edu/abs/1999AstL...25..453Y} {25, 453}

\bibitem[\protect\citeauthoryear{{Zavala} et~al.,}{{Zavala}
  et~al.}{2007}]{zavala:2007}
{Zavala} R.~T.,  et~al., 2007, \mn@doi [\apj] {10.1086/510108}, \href
  {https://ui.adsabs.harvard.edu/abs/2007ApJ...655.1046Z} {655, 1046}

\bibitem[\protect\citeauthoryear{{Zechmeister} \& {K{\"u}rster}}{{Zechmeister}
  \& {K{\"u}rster}}{2009}]{zechmeister:2009}
{Zechmeister} M.,  {K{\"u}rster} M.,  2009, \mn@doi [\aap]
  {10.1051/0004-6361:200811296}, \href
  {https://ui.adsabs.harvard.edu/abs/2009A&A...496..577Z} {496, 577}

\bibitem[\protect\citeauthoryear{{Zorec} \& {Royer}}{{Zorec} \&
  {Royer}}{2012}]{zorec:2012}
{Zorec} J.,  {Royer} F.,  2012, \mn@doi [\aap] {10.1051/0004-6361/201117691},
  \href {https://ui.adsabs.harvard.edu/abs/2012A&A...537A.120Z} {537, A120}

\bibitem[\protect\citeauthoryear{{Zverko}, {Ziznovsky}  \&
  {Khokhlova}}{{Zverko} et~al.}{1997}]{zverko:1997}
{Zverko} J.,  {Ziznovsky} J.,   {Khokhlova} V.~L.,  1997, Contributions of the
  Astronomical Observatory Skalnate Pleso, \href
  {https://ui.adsabs.harvard.edu/abs/1997CoSka..27...41Z} {27, 41}

\bibitem[\protect\citeauthoryear{{van Belle}}{{van
  Belle}}{2012}]{van-belle:2012}
{van Belle} G.~T.,  2012, \mn@doi [\aapr] {10.1007/s00159-012-0051-2}, \href
  {https://ui.adsabs.harvard.edu/abs/2012A&ARv..20...51V} {20, 51}

\bibitem[\protect\citeauthoryear{{van Leeuwen}}{{van
  Leeuwen}}{2007}]{van-leeuwen:2007}
{van Leeuwen} F.,  2007, \mn@doi [\aap] {10.1051/0004-6361:20078357}, \href
  {https://ui.adsabs.harvard.edu/abs/2007A&A...474..653V} {474, 653}

\makeatother
\end{thebibliography}

\appendix

\section{Time series analysis results}

\begin{table*}
\caption{Results of time series analysis of TESS light curves of HgMn stars. The columns give HD number for every target, other commonly used name, the TIC identification number, $V$ magnitude, analysed sectors, periods and amplitudes of variability. Harmonic signals are indicated by ``/'' followed by the harmonic number. Uncertainties in the last significant digits are indicated in parentheses for periods and amplitudes. The last column highlights periods attributed to rotational modulation or binarity. \label{tab:results}}
\begin{tabular}{llllllll}
\hline
 HD         & Other id.     & TIC         &$V$ (mag)& Sectors            & $P$ (d)       & $A$ (mmag) & Comment \\
\hline
   358      & $\alpha$~And  &  427733653  &  2.070  &  17                & 2.38337(17)   & 2.8136(41) & ROT         \\[1mm]  
   1009     &               &  83803744   &  8.400  &  17,18             & 2.24978(96)   & 0.0697(42) & ROT         \\       
            &               &             &         &                    & /2            & 0.1106(42) &             \\
            &               &             &         &                    & 0.0808487(67) & 0.0426(41) &             \\
            &               &             &         &                    & 0.0461809(33) & 0.0283(41) &             \\
            &               &             &         &                    & 0.0449932(36) & 0.0245(41) &             \\[1mm]
   1279     & HR 62         &  440076466  &  5.858  &  17                & 7.614(11)     & 0.3021(50) & ROT         \\       
            &               &             &         &                    & /2            & 0.1118(32) &             \\         
            &               &             &         &                    & /3            & 0.0542(31) &             \\[1mm]         
   4382     & 23 Cas        &  275361674  &  5.422  &  18,19             & 11.0079(76)   & 0.4343(28) & ROT?        \\       
            &               &             &         &                    & 6.1938(39)    & 0.2544(23) &             \\[1mm]
   5408     & HR 266        &  51961599   &  5.562  &  17,18             & 6.357(12)     & 0.0724(20) &             \\       
            &               &             &         &                    & 4.24074(16)   & 0.0611(21) & heartbeat   \\       
            &               &             &         &                    & /2            & 0.1837(19) &             \\
            &               &             &         &                    & /3            & 0.2748(19) &             \\
            &               &             &         &                    & /4            & 0.2456(19) &             \\
            &               &             &         &                    & /5            & 0.2115(19) &             \\
            &               &             &         &                    & /6            & 0.0949(19) &             \\
            &               &             &         &                    & /7            & 0.1146(19) &             \\
            &               &             &         &                    & /8            & 0.0819(19) &             \\
            &               &             &         &                    & /9            & 0.0526(19) &             \\
            &               &             &         &                    & /10           & 0.0307(19) &             \\
            &               &             &         &                    & /11           & 0.0196(19) &             \\[1mm]
   11291    & 2 Per         &  285425945  &  5.702  &  18                & 6.384(17)     & 0.1865(34) &             \\       
            &               &             &         &                    & 3.9369(46)    & 0.2769(33) &             \\
            &               &             &         &                    & 2.78350(99)   & 0.5725(32) & ROT?        \\[1mm]
   11753    & $\varphi$~Phe &  229099027  &  5.120  &  3                 & 9.3238(82)    & 0.0833(18) & ROT         \\       
            &               &             &         &                    & /2            & 0.2031(18) &             \\       
            &               &             &         &                    & /3            & 0.0446(18) &             \\
            &               &             &         &                    & /4            & 0.0227(18) &             \\[1mm]
   14228    & $\varphi$~Eri &  354671857  &  3.560  &  2,3               & 0.461391(57)  & 0.03292(88)&             \\       
            &               &             &         &                    & 0.454053(59)  & 0.03081(88)&             \\
            &               &             &         &                    & 0.3437583(35) & 0.27766(82)& ROT         \\[1mm]
   16727    & 11 Per        &  245758891  &  5.761  &  18                & 11.706(14)    & 0.7451(92) & ROT         \\       
            &               &             &         &                    & /2            & 0.1669(45) &             \\[1mm]
   19400    & $\theta$~Hyi  &  280051467  &  5.510  &  1,2,13            & 4.368500(45)  & 0.5826(13) & ROT         \\       
            &               &             &         &                    & /2            & 0.0626(13) &             \\[1mm]
   27376    & 41 Eri        &  168847194  &  3.550  &  4,5               & 5.00966(19)   & 2.1838(17) & ROT         \\       
            &               &             &         &                    & /2            & 0.4563(17) &             \\
            &               &             &         &                    & 0.773022(16)  & 0.6515(17) &             \\[1mm]
   28217    & HR 1402       &  373026963  &  5.872  &  5                 & 3.13164(30)   & 2.6615(39) & ROT         \\       
            &               &             &         &                    & /2            & 0.3186(39) &             \\
            &               &             &         &                    & 2.3540(11)    & 0.4335(39) &             \\[1mm]
   28929    & HR 1445       &  268507411  &  5.894  &  19                & 1.98864(15)   & 1.6861(31) & ROT         \\       
            &               &             &         &                    & /2            & 0.3144(31) &             \\[1mm]
   29589    & 93 Tau        &  245936817  &  5.455  &  5                 & 1.56746(77)   & 0.2448(37) & ROT?        \\       
            &               &             &         &                    & 0.754386(90)  & 0.5583(37) &             \\       
            &               &             &         &                    & 0.703415(26)  & 1.9520(40) &             \\
            &               &             &         &                    & 0.67346(11)   & 0.4339(42) &             \\
            &               &             &         &                    & 0.643669(21)  & 1.8721(41) &             \\
            &               &             &         &                    & 0.610245(48)  & 0.6492(39) &             \\
            &               &             &         &                    & 0.546276(18)  & 1.3110(37) &             \\
            &               &             &         &                    & 0.344503(35)  & 0.2621(37) &             \\
            &               &             &         &                    & 0.329194(48)  & 0.1759(37) &             \\
            &               &             &         &                    & 0.272684(15)  & 0.3886(37) &             \\
\hline
\end{tabular}
\end{table*}
\begin{table*}
\contcaption{}
\begin{tabular}{llllllll}
\hline
 HD         & Other id.     & TIC         &$V$ (mag)& Sectors            & $P$ (d)       & $A$ (mmag) & Comment \\
\hline                                                                                                                    
   30963    &               &  9355205    &  7.265  &  5                 & 3.99912(90)   & 1.2976(36) & ROT         \\       
            &               &             &         &                    & /2            & 0.1054(36) &             \\[1mm]
   31373    & HR 1576       &  436723855  &  5.791  &  5                 & 2.6438(35)    & 0.0876(22) &             \\       
            &               &             &         &                    & 1.44772(19)   & 0.3817(21) & ROT         \\
            &               &             &         &                    & /2            & 0.1503(21) &             \\[1mm]
   32964    & 66 Eri        &  248396674  &  5.097  &  5                 & 5.5347(10)    & 0.6415(21) & ROT?        \\       
            &               &             &         &                    & /2            & 0.4928(20) &             \\
            &               &             &         &                    & /3            & 0.0977(20) &             \\
            &               &             &         &                    & /4            & 0.0269(20) &             \\[1mm]
   33904    & $\mu$~Lep     &  169534653  &  3.290  &  5                 & 2.9327(11)    & 0.2903(21) & ROT         \\       
            &               &             &         &                    & /2            & 0.0712(21) &             \\
            &               &             &         &                    & 0.46771(19)   & 0.0501(21) &             \\[1mm]
   34364    & AR Aur        &  2236015    &  6.147  &  19                & 4.134         &$\approx$600& eclipses    \\[1mm]  
   34923    &               &  2776520    &  9.734  &  19                & 0.7485(17)    & 0.072(11)  & single transit \\    
            &               &             &         &                    & 0.50543(53)   & 0.109(11)  &             \\
            &               &             &         &                    & 0.252364(56)  & 0.255(10)  &             \\[1mm]
   36881    & HR 1883       &  436103335  &  5.604  &  6                 & 1.54148(20)   & 1.4404(53) & ROT         \\       
            &               &             &         &                    & /2            & 0.2978(52) &             \\
            &               &             &         &                    & 1.16825(53)   & 0.3384(53) &             \\
            &               &             &         &                    & 7.86          &            & eclipses    \\[1mm]
                                                                           
   37437    &               &  116154521  &  8.129  &  19                & 6.858(10)     & 0.3289(58) & ROT         \\       
            &               &             &         &                    & /2            & 0.1663(58) &             \\
            &               &             &         &                    & /3            & 0.0863(57) &             \\[1mm]
   37519    & HR 1938       &  116333643  &  6.040  &  19                & 0.840782(85)  & 0.3295(29) & ROT         \\       
            &               &             &         &                    & /2            & 0.1885(29) &             \\
            &               &             &         &                    & /3            & 0.0410(29) &             \\
            &               &             &         &                    & /4            & 0.0218(29) &             \\[1mm]
   37886    &               &  11411002   &  8.979  &  6                 & 4.0719(69)    & 0.5361(87) & ROT?        \\       
            &               &             &         &                    & 2.5140(46)    & 0.3021(86) &             \\[1mm]
   38478    & 129 Tau       &  247638066  &  6.010  &  6                 & 3.8065(12)    & 1.1621(35) & ROT         \\       
            &               &             &         &                    & /2            & 0.1579(35) &             \\[1mm]
   49606    & 33 Gem        &  155095401  &  5.855  &  6                 & 8.546(13)     & 0.1808(32) & ROT         \\       
            &               &             &         &                    & /2            & 0.1972(31) &             \\[1mm]
   53244    & $\gamma$~CMa  &  148109427  &  4.110  &  7                 & 7.8042(90)    & 0.3702(17) &             \\       
            &               &             &         &                    & 6.2140(27)    & 0.7720(16) & ROT         \\       
            &               &             &         &                    & /2            & 0.0846(15) &             \\[1mm]  
   65950    &               &  372913233  &  6.860  &  1,2,7-11          & $>$0.5        & $<$0.0463  &             \\[1mm]  
   66409    &               &  410451677  &  8.373  &  1,7-11            & 2.055116(15)  & 1.2783(26) & ROT         \\       
            &               &             &         &                    & /2            & 0.1174(26) &             \\[1mm]
   69028    &               &  139271386  &  7.990  &  20                & 6.8137(46)    & 0.4711(60) & ROT         \\       
            &               &             &         &                    & /2            & 0.4738(58) &             \\
            &               &             &         &                    & /3            & 0.1086(57) &             \\[1mm]
   71066    & $\kappa^2$~Vol&  307291308  &  5.630  &  5,6,10-13         & 1.2963766(63) & 1.0050(19) & ROT         \\[1mm]  
   75333    & 14 Hya        &  62480591   &  5.291  &  8                 & 3.89443(75)   & 1.4147(35) & ROT         \\[1mm]  
   76728    & c Car         &  356299294  &  3.840  &  10                & 4.1880(11)    & 0.9409(26) & ROT         \\[1mm]  
   77350    & $\nu$ Cnc     &  126349649  &  5.456  &  21                & $>$0.5        & $<$0.2462  &             \\[1mm]  
   80078    &               &  357981882  &  8.203  &  9-11              & 10.9274(58)   & 0.1642(38) & ROT         \\       
            &               &             &         &                    & /2            & 0.2076(32) &             \\[1mm]
   87240    &               &  462162948  &  9.650  &  9,10              & 5.6980(15)    & 2.1335(95) & ROT         \\[1mm]  
   89822    & HR 4072       &  287287930  &  4.950  &  14,21             & 11.58105(53)  & 0.2638(39) & hearbeat    \\       
            &               &             &         &                    & /2            & 0.2939(25) &             \\
            &               &             &         &                    & /3            & 0.2091(24) &             \\
            &               &             &         &                    & /4            & 0.0980(24) &             \\
            &               &             &         &                    & /5            & 0.0561(23) &             \\
            &               &             &         &                    & /6            & 0.0385(23) &             \\
\hline
\end{tabular}
\end{table*}
\begin{table*}
\contcaption{}
\begin{tabular}{llllllll}
\hline
 HD         & Other id.    & TIC         &$V$ (mag)& Sectors            & $P$ (d)       & $A$ (mmag) & Comment \\
\hline
   93549    & HR 4220      &  391421890  &  5.230  &  10,11             & 5.6537(35)    & 0.1932(22) &             \\        
            &              &             &         &                    & 3.72367(54)   & 0.5346(21) &             \\
            &              &             &         &                    & 2.89581(46)   & 0.3709(21) &             \\
            &              &             &         &                    & 1.93136(29)   & 0.2526(20) &             \\
            &              &             &         &                    & 1.44737(24)   & 0.1708(20) &             \\
            &              &             &         &                    & 0.874401(23)  & 0.6476(20) & ROT         \\
            &              &             &         &                    & /2            & 0.0429(20) &             \\[1mm]
   99803    & HR 4423      &  163024899  &  5.140  &  10                &               & $\approx$46  & single eclipse \\[1mm] 
   101189   & HR 4487      &  319809753  &  5.150  &  10,11             & 2.00407(53)   & 0.1312(18) & ROT         \\[1mm]   
   101391   & HR 4493      &  137752707  &  6.353  &  21                & 1.07428(20)   & 0.2787(25) & ROT         \\        
            &              &             &         &                    & /2            & 0.0332(25) &             \\
            &              &             &         &                    & /3            & 0.0097(25) &             \\[1mm]
   106625   & $\gamma$~Crv &  348987372  &  2.580  &  10                & 5.9377(15)    & 1.8000(39) & ROT         \\        
            &              &             &         &                    & /2            & 0.1572(38) &             \\[1mm]
   110073   & HR 4817      &  144395071  &  4.630  &  10                & 5.903(11)     & 0.1122(17) &             \\        
            &              &             &         &                    & 3.0301(27)    & 0.1066(17) & ROT?        \\
            &              &             &         &                    & /2            & 0.0280(16) &             \\
            &              &             &         &                    & 2.3479(31)    & 0.0599(16) &             \\
            &              &             &         &                    & 1.3414(12)    & 0.0534(16) &             \\[1mm]
   133833   &              &  121161014  &  8.225  &  12                & 3.7700(16)    & 1.1373(72) & ROT         \\        
            &              &             &         &                    & /2            & 0.1477(70) &             \\[1mm]
   141556   & $\chi$~Lup   &  442652828  &  3.970  &  12                & 1.79286(54)   & 0.2429(19) & ROT         \\[1mm]   
   143807   & $\iota$~CrB  &  356010327  &  4.978  &  24,25             & $>$0.5        & $<$0.0313  &             \\[1mm]
   144206   &$\upsilon$~Her&  417582433  &  4.723  &  23-25             & $>$0.5        & $<$0.0221  &             \\[1mm]
   145389   &$\varphi$~Her &  219480018  &  4.237  &  25                & 4.446(16)     & 0.0424(13) &             \\
            &              &             &         &                    & 3.7076(29)    & 0.0990(14) & ROT         \\        
            &              &             &         &                    & /2            & 0.0049(13) &             \\        
            &              &             &         &                    & /3            & 0.0381(13) &             \\
            &              &             &         &                    & 0.0546655(55) & 0.0144(13) &             \\
            &              &             &         &                    & 0.0527289(33) & 0.0222(13) &             \\
            &              &             &         &                    & 0.05097180(75)& 0.0924(13) &             \\
            &              &             &         &                    & 0.0493830(42) & 0.0156(13) &             \\
            &              &             &         &                    & 0.0465250(51) & 0.0112(13) &             \\
            &              &             &         &                    & 0.0448131(49) & 0.0109(13) &             \\[1mm]
   145842   & $\theta$~Nor &  51612589   &  5.130  &  12                & 1.08588(13)   & 1.1444(60) & ROT         \\        
            &              &             &         &                    & /2            & 0.1119(60) &             \\[1mm]
   156127   &              &  198388195  &  8.300  &  14-26             & 2.45915(14)   & 0.1065(16) & ROT         \\        
            &              &             &         &                    & 0.694363(50)  & 0.0237(16) &             \\
            &              &             &         &                    & 0.688940(20)  & 0.0395(16) &             \\
            &              &             &         &                    & /2            & 0.0126(16) &             \\
            &              &             &         &                    & /3            & 0.0113(16) &             \\
            &              &             &         &                    & 0.683537(44)  & 0.0262(16) &             \\
            &              &             &         &                    & 0.678268(64)  & 0.0177(16) &             \\[1mm]
   156678   &              &  198409372  &  8.632  &  14-25             & 4.3448(34)    & 0.0172(19) & ROT?        \\        
            &              &             &         &                    & 0.388511(21)  & 0.0223(18) &             \\[1mm]
   169027   &  38 Dra      &  236770286  &  6.784  &  14-26             & 1.21850(19)   & 0.01119(97)&             \\        
            &              &             &         &                    & 1.167052(50)  & 0.03959(97)& ROT?        \\[1mm]
   168733   &  HR 6870     &  323999777  &  5.330  &  13                & 6.32604(88)   & 5.3980(55) & ROT         \\[1mm]   
   171301   &  HR 6968     &  27697099   &  5.478  &  26                & 3.10449(96)   & 0.7775(35) & ROT?        \\        %
            &              &             &         &                    & 0.947931(35)  & 1.9107(34) &             \\        
            &              &             &         &                    & 0.795271(11)  & 4.1444(34) &             \\
            &              &             &         &                    & 0.633192(60)  & 0.4981(34) &             \\
\hline
\end{tabular}
\end{table*}
\begin{table*}
\contcaption{}
\begin{tabular}{llllllll}
\hline
 HD         & Other id.    & TIC         &$V$ (mag)& Sectors            & $P$ (d)       & $A$ (mmag) & Comment \\
\hline
   172044   &  HR 6997     &   27901267  &  5.409  &  26                & 4.33023(84)   & 1.7691(42) & ROT         \\        
            &              &             &         &                    & /2            & 0.6662(40) &             \\
            &              &             &         &                    & /3            & 0.0910(39) &             \\
            &              &             &         &                    & /4            & 0.0579(39) &             \\
            &              &             &         &                    & 3.6797(74)    & 0.2179(40) &             \\
            &              &             &         &                    & 1.00444(69)   & 0.1264(39) &             \\[1mm]
   172728   & HR 7018      &  383676357  &  5.746  &  14-26             & 4.498374(95)  & 0.24364(76)& ROT         \\        
            &              &             &         &                    & 2.76518(16)   & 0.05519(75)&             \\
            &              &             &         &                    & 2.606514(75)  & 0.10438(76)&             \\
            &              &             &         &                    & 2.533836(88)  & 0.09821(79)&             \\
            &              &             &         &                    & 2.512467(99)  & 0.08684(79)&             \\
            &              &             &         &                    & 1.258558(37)  & 0.04831(75)&             \\[1mm]
   173524   & 46 Dra       &  392569763  &  5.030  &  15,16,18-23,25,26 & 9.82663(47)   & 0.15617(78)& ROT         \\        
            &              &             &         &                    & /2            & 0.07340(72)&             \\
            &              &             &         &                    & /3            & 0.04465(71)&             \\
            &              &             &         &                    & 1.447225(18)  & 0.14437(72)&             \\
            &              &             &         &                    & 1.0131045(67) & 0.18481(71)&             \\
            &              &             &         &                    & 0.722427(16)  & 0.03865(71)&             \\[1mm]
   174933   & 112 Her      &  347160134  &  5.430  &  26                & 12.419(17)    & 0.1153(66) & ROT         \\        
            &              &             &         &                    & /2            & 0.2385(36) &             \\
            &              &             &         &                    & /3            & 0.0652(27) &             \\[1mm]
   198174   & HR 7961      &  270070443  &  5.860  &  1                 & 2.53650(21)   & 2.8910(46) & ROT         \\[1mm]   
   202033   &              &  76960308   &  8.470  &  15                & $>$0.5        & $<$0.0685  &             \\[1mm]   
   212093   &              &  422012928  &  8.266  &  16,17             & 6.561(24)     & 0.0422(47) & ROT?        \\        
            &              &             &         &                    & /2            & 0.0443(45) &             \\[1mm]
   216831   & HR 8723      &  129533458  &  5.743  &  16                & 3.4709(13)    & 2.0713(96) & ROT         \\[1mm]   
   219485   & HR 8844      &  417703090  &  5.899  &  17,19,24,25       & 5.4993(41)    & 0.0141(11) & ROT?        \\        
            &              &             &         &                    & 3.9488(24)    & 0.0122(11) &             \\
            &              &             &         &                    & 2.5276(11)    & 0.0102(10) &             \\[1mm] 
   221507   & $\beta$\,Scl &  224244458  &  4.380  &  2                 & 1.92060(45)   & 0.15928(99)& ROT         \\        
            &              &             &         &                    & /2            & 0.01832(99)&             \\        
            &              &             &         &                    & 0.65573(12)   & 0.05919(99)& ROT companion\\       
            &              &             &         &                    & /2            & 0.01677(99)&             \\[1mm]
   225289   & HR 9110      &  359033491  &  5.783  &  17,18             & 3.25423(85)   & 0.2733(27) & ROT         \\        
            &              &             &         &                    & /2            & 0.0931(27) &             \\
            &              &             &         &                    & /3            & 0.0384(27) &             \\
\hline
\end{tabular}
\end{table*}

\section{Stellar parameters}

\begin{table*}
\caption{Essential data for HgMn stars with rotation periods measured in our study. The columns denote: (1) HD number, (2) $V$ magnitude, 
(3) distance, (4) logarithmic effective temperature, (5) bolometric correction,  (6) logarithmic luminosity, (7) radius, 
(8) observed period, (9) calibrated equatorial velocity, (10) critical velocity, (11) ratio of the calibrated equatorial velocity
and critical velocity. \label{tab:params}}
\begin{tabular}{lcccccccccc}
\hline
(1) & (2) & (3) & (4) & (5) & (6) & (7) & (8) & (9) & (10) & (11)  \\
\hline
HD	& $V$ & $D$ & $\log T_\mathrm{eff}$ & BC & $\log L/L_\odot$ & $R$ & $P_\mathrm{rot}$ & $v_\mathrm{eq}$ & $v_\mathrm{crit}$ & $v_\mathrm{eq}$/$v_\mathrm{crit}$\\
& (mag) & (pc) & (K) & (mag) & & ($R_\odot$) & (d) & (km\,s$^{-1}$) & (km\,s$^{-1}$) & \\
\hline
358	&	2.070 & $29.7   \pm 0.3  $ & $4.129 \pm 0.004$ & -0.975 & $2.429 \pm 0.007$ & 3.031  & 2.383   & 64      & 383 & 0.17    \\
1009	&	8.400 & $554.0  \pm 6.2  $ & $4.081 \pm 0.002$ & -0.695 & $2.670 \pm 0.007$ & 4.980  & 2.250   & 112     & 324 & 0.35    \\
1279	&	5.858 & $347.4  \pm 6.7  $ & $4.114 \pm 0.004$ & -0.887 & $3.044 \pm 0.009$ & 6.572  & 7.614   & 44      & 348 & 0.13    \\
4382	&	5.422 & $239.9  \pm 2.9  $ & $4.110 \pm 0.006$ & -0.862 & $2.901 \pm 0.007$ & 5.678  & 11.008  & 26      & 303 & 0.09    \\
11291	&	5.702 & $145.9  \pm 2.3  $ & $4.067 \pm 0.005$ & -0.614 & $2.227 \pm 0.008$ & 3.189  & 2.784   & 58      & 362 & 0.16    \\
11753	&	5.120 & $98.2   \pm 1.8  $ & $4.027 \pm 0.013$ & -0.385 & $1.990 \pm 0.009$ & 2.922  & 9.324   & 16      & 330 & 0.05    \\
14228	&	3.560 & $46.1   \pm 0.5  $ & $4.102 \pm 0.004$ & -0.814 & $2.129 \pm 0.007$ & 2.422  & 0.344   & 356     & 409 & 0.87    \\
16727	&	5.761 & $128.2  \pm 1.2  $ & $4.144 \pm 0.005$ & -1.057 & $2.255 \pm 0.007$ & 2.312  & 11.706  & 10      & 420 & 0.02    \\
19400	&	5.510 & $153.8  \pm 1.4  $ & $4.133 \pm 0.003$ & -0.995 & $2.468 \pm 0.007$ & 3.108  & 4.369   & 36      & 359 & 0.10    \\
27376	&	3.550 & $54.5   \pm 0.7  $ & $4.100 \pm 0.001$ & -0.804 & $2.274 \pm 0.008$ & 2.898  & 5.010   & 29      & 407 & 0.07    \\
28217	&	5.872 & $321.1  \pm 54.2 $ & $4.127 \pm 0.002$ & -0.960 & $3.183 \pm 0.066$ & 7.269  & 3.132   & 117     & 296 & 0.40    \\
28929	&	5.894 & $138.3  \pm 0.9  $ & $4.103 \pm 0.001$ & -0.820 & $2.226 \pm 0.006$ & 2.708  & 1.989   & 69      & 366 & 0.19    \\
29589	&	5.455 & $120.6  \pm 1.8  $ & $4.147 \pm 0.005$ & -1.076 & $2.311 \pm 0.008$ & 2.435  & 1.567   & 79      & 394 & 0.20    \\
30963	&	7.265 & $310.9  \pm 3.0  $ & $4.069 \pm 0.001$ & -0.623 & $2.289 \pm 0.007$ & 3.400  & 3.999   & 43      & 338 & 0.13    \\
31373	&	5.791 & $140.2  \pm 2.5  $ & $4.130 \pm 0.003$ & -0.980 & $2.304 \pm 0.009$ & 2.615  & 1.448   & 91      & 396 & 0.23    \\
32964	&	5.097 & $93.1   \pm 1.0  $ & $4.046 \pm 0.005$ & -0.492 & $1.999 \pm 0.007$ & 2.707  & 5.535   & 25      & 395 & 0.06    \\
33904	&	3.290 & $52.0   \pm 0.9  $ & $4.097 \pm 0.005$ & -0.788 & $2.333 \pm 0.009$ & 3.138  & 2.933   & 54      & 380 & 0.14    \\
36881	&	5.604 & $500.8  \pm 25.6 $ & $4.039 \pm 0.015$ & -0.451 & $3.526 \pm 0.021$ & 16.171 & 1.541   & 531     & 200 & 2.65    \\
37437	&	8.129 & $532.0  \pm 16.1 $ & $4.133 \pm 0.006$ & -0.995 & $2.571 \pm 0.013$ & 3.500  & 6.858   & 26      & 380 & 0.07    \\
37519	&	6.040 & $239.2  \pm 2.8  $ & $4.062 \pm 0.006$ & -0.584 & $2.656 \pm 0.007$ & 5.343  & 0.841   & 321     & 333 & 0.97    \\
37886	&	8.979 & $365.0  \pm 4.9  $ & $4.092 \pm 0.002$ & -0.755 & $1.814 \pm 0.008$ & 1.800  & 4.072   & 22      & 459 & 0.05    \\
38478	&	6.010 & $270.5  \pm 5.8  $ & $4.116 \pm 0.003$ & -0.895 & $2.758 \pm 0.010$ & 4.684  & 3.807   & 62      & 367 & 0.17    \\
49606	&	5.855 & $204.5  \pm 3.8  $ & $4.136 \pm 0.003$ & -1.015 & $2.591 \pm 0.009$ & 3.530  & 8.546   & 21      & 374 & 0.06    \\
53244	&	4.110 & $130.1  \pm 3.2  $ & $4.122 \pm 0.005$ & -0.933 & $2.873 \pm 0.011$ & 5.205  & 6.214   & 42      & 329 & 0.13    \\
66409	&	8.373 & $424.8  \pm 8.7  $ & $4.103 \pm 0.003$ & -0.824 & $2.291 \pm 0.010$ & 2.907  & 2.055   & 72      & 399 & 0.18    \\
69028	&	7.990 & $436.0  \pm 9.8  $ & $4.039 \pm 0.002$ & -0.453 & $2.274 \pm 0.010$ & 3.829  & 6.814   & 28      & 333 & 0.09    \\
71066	&	5.630 & $126.2  \pm 1.0  $ & $4.077 \pm 0.004$ & -0.670 & $2.118 \pm 0.006$ & 2.696  & 1.296   & 105     & 357 & 0.29    \\
75333	&	5.291 & $131.0  \pm 1.8  $ & $4.081 \pm 0.004$ & -0.693 & $2.339 \pm 0.008$ & 3.402  & 3.894   & 44      & 331 & 0.13    \\
76728	&	3.840 & $89.3   \pm 2.3  $ & $4.120 \pm 0.003$ & -0.918 & $2.649 \pm 0.011$ & 4.057  & 4.188   & 49      & 398 & 0.12    \\
80078	&	8.203 & $479.1  \pm 5.0  $ & $4.110 \pm 0.007$ & -0.864 & $2.350 \pm 0.007$ & 3.022  & 10.927  & 14      & 372 & 0.04    \\
87240	&	9.650 & $1341.7 \pm 134.0$ & $4.108 \pm 0.008$ & -0.850 & $2.694 \pm 0.039$ & 4.520  & 5.698   & 40      & 327 & 0.12    \\
93549	&	5.230 & $149.8  \pm 2.2  $ & $4.148 \pm 0.006$ & -1.083 & $2.657 \pm 0.008$ & 3.606  & 0.874   & 209     & 398 & 0.52    \\
101189	&	5.150 & $96.1   \pm 1.3  $ & $4.045 \pm 0.008$ & -0.488 & $2.038 \pm 0.008$ & 2.848  & 2.004   & 72      & 379 & 0.19    \\
101391	&	6.353 & $153.3  \pm 0.9  $ & $4.087 \pm 0.003$ & -0.728 & $2.024 \pm 0.006$ & 2.305  & 1.074   & 109     & 378 & 0.29    \\
106625	&	2.580 & $47.1   \pm 0.4  $ & $4.076 \pm 0.004$ & -0.661 & $2.479 \pm 0.007$ & 4.087  & 5.938   & 35      & 306 & 0.11    \\
110073	&	4.630 & $101.8  \pm 2.1  $ & $4.107 \pm 0.004$ & -0.842 & $2.454 \pm 0.010$ & 3.446  & 3.030   & 58      & 332 & 0.17    \\
133833	&	8.225 & $332.8  \pm 50.3 $ & $4.084 \pm 0.002$ & -0.712 & $1.963 \pm 0.059$ & 2.185  & 3.770   & 29      & 403 & 0.07    \\
141556	&	3.970 & $62.6   \pm 1.5  $ & $4.030 \pm 0.009$ & -0.402 & $2.069 \pm 0.011$ & 3.161  & 1.793   & 89      & 349 & 0.26    \\
145389	&	4.237 & $73.7   \pm 2.6  $ & $4.065 \pm 0.006$ & -0.599 & $2.222 \pm 0.015$ & 3.203  & 3.708   & 44      & 363 & 0.12    \\
145842	&	5.130 & $111.7  \pm 1.3  $ & $4.099 \pm 0.003$ & -0.800 & $2.264 \pm 0.007$ & 2.879  & 1.086   & 134     & 410 & 0.33    \\
156127	&	8.300 & $417.9  \pm 4.4  $ & $4.025 \pm 0.010$ & -0.376 & $2.070 \pm 0.007$ & 3.237  & 2.459   & 67      & 345 & 0.19    \\
156678	&	8.632 & $330.4  \pm 2.3  $ & $3.977 \pm 0.011$ & -0.145 & $1.580 \pm 0.006$ & 2.290  & 4.345   & 27      & 330 & 0.08    \\
168733	&	6.784 & $197.2  \pm 3.9  $ & $4.120 \pm 0.011$ & -0.920 & $2.164 \pm 0.010$ & 2.332  & 1.167   & 101     & 396 & 0.26    \\
169027	&	5.330 & $178.7  \pm 1.0  $ & $4.057 \pm 0.006$ & -0.556 & $2.495 \pm 0.006$ & 4.546  & 6.326   & 36      & 303 & 0.12    \\
171301	&	5.478 & $106.9  \pm 0.7  $ & $4.086 \pm 0.002$ & -0.724 & $2.074 \pm 0.006$ & 2.461  & 3.104   & 40      & 355 & 0.11    \\
172044	&	5.409 & $162.8  \pm 1.8  $ & $4.143 \pm 0.003$ & -1.051 & $2.625 \pm 0.007$ & 3.556  & 4.330   & 42      & 410 & 0.10    \\
172728	&	5.746 & $128.2  \pm 1.4  $ & $4.035 \pm 0.010$ & -0.431 & $2.011 \pm 0.007$ & 2.884  & 4.498   & 32      & 389 & 0.08    \\
173524	&	5.030 & $110.3  \pm 1.0  $ & $4.061 \pm 0.005$ & -0.580 & $2.215 \pm 0.007$ & 3.240  & 9.827   & 17      & 325 & 0.05    \\
174933	&	5.430 & $127.3  \pm 1.2  $ & $4.109 \pm 0.003$ & -0.858 & $2.308 \pm 0.007$ & 2.888  & 12.419  & 12      & 391 & 0.03    \\
198174	&	5.860 & $149.9  \pm 3.7  $ & $4.114 \pm 0.004$ & -0.887 & $2.330 \pm 0.011$ & 2.901  & 2.537   & 58      & 382 & 0.15    \\
212093	&	8.266 & $317.9  \pm 11.3 $ & $4.090 \pm 0.007$ & -0.745 & $1.957 \pm 0.015$ & 2.106  & 6.561   & 16      & 407 & 0.04    \\
216831	&	5.743 & $217.8  \pm 4.6  $ & $4.107 \pm 0.005$ & -0.844 & $2.684 \pm 0.010$ & 4.488  & 3.471   & 65      & 334 & 0.20    \\
219485	&	5.899 & $136.3  \pm 1.0  $ & $3.999 \pm 0.012$ & -0.243 & $1.909 \pm 0.006$ & 3.025  & 5.499   & 28      & 324 & 0.09    \\
221507	&	4.380 & $55.9   \pm 0.7  $ & $4.093 \pm 0.007$ & -0.764 & $1.948 \pm 0.008$ & 2.054  & 1.921   & 54      & 411 & 0.13    \\
225289	&	5.783 & $209.9  \pm 2.1  $ & $4.104 \pm 0.005$ & -0.826 & $2.628 \pm 0.007$ & 4.270  & 3.254   & 66      & 350 & 0.19    \\
\hline
\end{tabular}
\end{table*}

\bsp
\label{lastpage}
\end{document}